\documentclass[useAMS,usenatbib]{mn2e}
\usepackage{graphicx} 
\usepackage{amsmath} 
\usepackage{subfigure}

\title[The Luminous Convolution Model as an alternative to dark matter in spiral galaxies]{The Luminous Convolution Model  as an alternative to dark matter in spiral galaxies}
\author[S. ~Cisneros et. al] {S.\,~Cisneros$^{1}$\thanks{E-mail:cisneros@mit.edu}, N.\,S.~Oblath$^{1}$, J.\,A.~Formaggio$^{1}$,  R.\,A.~Ott$^{1,3}$, D.~Chester$^{1,2}$, \\D.\,J.~Battaglia$^{1}$,  A.~Ashley$^{1}$, R.~Robinson$^{1}$,
and A.~Rodriguez$^{1}$\\
$^{1}$Laboratory for Nuclear Science, Massachusetts Institute of Technology, Cambridge, MA~02139, USA  \\
$^{2}$Present address: University of California at Los Angeles \\
$^{3}$Present address: University of California at Davis }
\begin{document}

\date{Submitted 2014 July 24 }

\pagerange{\pageref{firstpage}--\pageref{lastpage}} \pubyear{2014} 
\maketitle

\label{firstpage}

\begin{abstract}
The Luminous Convolution Model (LCM) demonstrates that it is possible to predict the   rotation curves of spiral galaxies directly from estimates of the luminous matter.  We  consider    two frame-dependent effects on the light   observed from other galaxies:     relative velocity  and     relative curvature.  With one  free  parameter,  we predict the rotation curves of twenty-three (23)   galaxies represented in forty-two (42) data sets. Relative curvature effects rely upon knowledge of both the gravitational potential from luminous mass of the emitting galaxy and the  receiving galaxy, and so each   emitter galaxy is compared to four (4) different Milky Way luminous mass models.      On average in this sample, the LCM is    more successful than either dark matter or modified gravity models in fitting the observed rotation curve data. 

   Implications of LCM constraints on populations synthesis modeling are discussed in this paper.  This paper substantially expands the results in arXiv:1309.7370.
 \end{abstract}

\begin{keywords}
cosmology: dark matter, theory; galaxies: distances and redshifts,  Dark matter - galaxies: Stellar dynamics and kinematics - galaxies: Modified theories of gravity: Spiral Galaxies
\end{keywords}
  \section[]{INTRODUCTION}

Flat rotation-curve observations from spiral galaxies   have been  the smoking gun for theories of  missing  mass      or    new laws of gravity since the 1970's ~\citep{1978Rubin,Bosma78}.  There are two fundamental observables involved  in the flat rotation-curve problem: photometry and spectra.  
  Photometry is the measurement of the total light, in a specific wavelength band, which is  interpreted as mass through a  population synthesis model.  The   visible mass then yields  orbital velocities by Newton's second law and the    expected Keplerian fall-off at large radii.  The second observable is    Doppler-shifted  spectra from   characteristic atomic transitions,  which   imply  the dynamical
 mass   by        the   relation between the      orbital velocities (from the Lorentz Doppler-shift formula) and  the   Poisson equation.   These two observables  predict  very different mass content, since  rotation curves from spectra do not  fall off in   a  Keplerian sense at large radii.  The most popular theories to explain the conflicting observations are missing mass or new laws of gravity.
 The  missing mass, commonly called dark matter, invokes a    new particle which does not interact electromagnetically, and hence is ``dark.''   New laws of physics  commonly work on the principle that our  understanding of gravity, via General Relativity, is incomplete on extra-galactic distance scales.\\
 
      	Examples of   dark matter   theories  include the psuedo-isothermal dark matter halo model (ISO)~\citep{ISO} and the  Navarro, Frenk \& White (NFW)~\citep{NFW} dark matter halo model.   Examples of   new    laws of gravity include    modified Newtonian dynamics (MOND)~\citep{Milgrom}, non-local gravity ~\citep{Rahvar} and conformal relativity ~\citep{Mannheim}.  For a full review of   NFW or MOND approaches see  ~\citet{SanMcGa,Gianfranco}.  MOND successfully predicts core densities, the baryonic Tully-Fisher relation, and rotation curves.  Dark matter theories are the standard for interpreting flat rotation-curves and form the basis of  a  comprehensive 
     cosmology.   But the fact remains that  neither class of theory, dark particles or deviations from General Relativity,  has been observed in decisive terrestrial experiments and neither has resulted in a   full and rigorously computed set of population synthesis models~\citep{Blok}.   Dark matter may have been observed in the DAMA ~\citep{DAMA}, CoGeNT ~\citep{CoGeNT}, and CDMS ~\citep{CDMS} experiments, although these results are in some conflict with the limits of XENON100 ~\citep{XENON}.  \\

 It is a common feature of rotation curves that   galaxies smaller than the Milky Way are  `dark-matter dominated' (ascending rotation curves) and   galaxies larger than the Milky Way have minimal dark matter halos  (descending rotation curves).  Classification of galaxies based on     the relative size of the  dark matter halo  is called the universal rotation curve~\citep{Rub,PS88,PS90}.   Since  a priori there appears to be no physical reason for the Milky Way to occupy the median point in this dark matter halo distribution, we interpret the universal rotation curve as indicating frame-dependent effects in rotation curve observations.  \\
 
  Currently the  Doppler-shifted spectra   are interpreted  to   imply  only     frame-dependent effects  from relative velocity.  However, it  is known from  classical electrodynamic theory    that  effects on light naturally divide mathematically between   those from relative velocity and those from relative   acceleration\footnote{of the emitter with respect to the receiver frame}~\citep[see][Eqs.~(14.13) \& (14.14)]{Jack}.   In the context of Relativistic kinematics we replace the word  ``acceleration''  with   ``curvature.''   The   Luminous Convolution Model (LCM) is an empirical formula for predicting flat rotation curves based  on   interpreting the  Doppler-shifted spectra as having     two contributions:   relative velocity \emph{and}    relative curvature.   LCM   curvature   effects  are quantified via the gravitational redshifts for  the respective luminous matter content of the emitter and receiver galaxies, and rephrased in kinematical terms using equivalent Doppler-shifts.\\ 

 A sample of  twenty-three $(23)$    galaxies from forty-two $(42)$ different  data sets, and      four $(4)$ different Milky Way luminous mass models will be used  to demonstrate the utility of the LCM to constrain luminous mass modeling.     Comparisons are made to reported mass-to-light ratios from dark matter and MOND models, and differing distance indicators are noted. The highest resolution rotation curve data are presented here along with the older standards in the field of spiral-galaxy rotation-curve studies. 
 The  LCM has  one  free parameter  $\tilde{a}$ that is dimensionless and does not yet have a physical interpretation.   The LCM     successfully     predicts rotation curves  across a broad range of galaxy sizes and morphologies. In this paper we allow the fits to scale the given  mass-to-light ratios and gas fractions reported by the reference data sets,  within   a fixed range.   Our analysis is  focused     on  the high  symmetry  case of spiral galaxies in the plane of the galactic disk.   While the    LCM  mapping can be   extended to    an arbitrary  metric~\citep{Cisneros:2013vha}  to predict other astrophysical observations, such studies are beyond the scope of the current paper.    On average for the  data studied in this paper, the   LCM  rotation curve fits are better than MOND or dark matter model fits, returning    reasonable estimates of stellar and gas masses. 

The paper is divided as follows:  section~\ref{sec:stationTotal}    describes the rotation curve formalism and associated mass modeling;  section~\ref{sec:DERIVE} derives the  LCM;      section~\ref{sec:RESULTS}  describes the sample and     LCM results,   and  section~\ref{sec:conclusion} gives our  conclusions.

  \section[]{ROTATION CURVES OF SPIRAL GALAXIES}
\label{sec:stationTotal} 
 \subsection{Velocity addition formulae} 
 
In dark matter halo theories  the total dynamical mass $M'$  as a function of radius 
is a sum of the  luminous mass $M_{l}$ and the 
dark   matter halo mass $M_{dm}$:
 \begin{equation}
    M'(r)=M_{l}(r)+M_{dm}(r).
         \label{eq:mass1}
 \end{equation}
 
 The mass components are then related to Poisson law forces for the appropriate geometry. 
 Stellar disks and gas distributions are generally modeled as exponential profiles and dark matter 
 halos and stellar bulges are modeled as spherical. The associated sum of forces then  
 yields a sum of centripetal accelerations,  $v^2/r$, attributed to each component.   Orbital velocities are generally treated 
as those  experienced by  a test particle  at some radius $r$ from the center of the  mass distribution, assumed to be   in a stable, circular orbit.  
  The total  velocity   is taken to be  a quadratic sum of the contributions from the luminous    $v^2_{l}$   and the dark halo  
$ v^2_{halo}$ velocities:
  \begin{equation}
v_{rot}^2 =  v_{l}^2 +  v_{halo}^2.
\label{eq:zonte1}
\end{equation} 
  All velocities are  functions of the radius.  The  predicted total velocity $v_{rot}$   is  fit to the observed rotation curve  $v_{obs}$. 
 
 For a rotation curve observation  made using a  well-known atomic transition,
 having a frequency $\omega_o$ in the observer's rest-frame  and
 Doppler-shifted frequency  $\omega'$, the observed rotation curve is: 
 \begin{equation}
 \frac{v_{obs}(r)}{c}=
\frac{  \frac{\omega'(r)}{\omega_o}- \frac{\omega_o}{\omega'(r)}}{ 
\frac{\omega'(r)}{\omega_o}+ \frac{\omega_o}{\omega'(r)}}.
\label{eq:dataLorentz}
\end{equation}
This is the standard   Lorentz Doppler-shift formula.
 
 \subsubsection{Assumption of spherical symmetry as a final approximation}
 \label{spherical}
Forces associated with each of the mass components  in Eq.~\ref{eq:mass1} (gas, stellar disk and bulge, and dark matter halo) come  individually from  the solution to a classical Poisson equation for a given geometry.  Dark matter halos and stellar bulges are generally modeled as spherical, whereas stellar disks and gas distributions are   modeled as exponential disks with Bessel functions. Since   the  mass sum  and  the resulting sum of centripetal accelerations are not functionally dependent on $r$ in the same way~\citep{Binney},  the quadratic velocity sum in Eq.~\ref{eq:zonte1} is a simplifying assumption of  spherical symmetry.  

Rotation curve derivations  based  on  concentric rings also employ the   assumption of spherical symmetry,  so that  the  Gaussian shell technique can be used~\citep{Frat}.    Luminous   mass modeling     relies  on  the assumption of spherical symmetry as a final approximation as well, to  sum the contributions to the   gravitational  potential from the terms calculated in their respective  geometries~\citep{Xue,Klypin}.   The logic  of invoking this assumption     is 
 that the error introduced  by assuming  spherical geometry for  gravitational potentials calculated with disk geometries is     a   difference in magnitudes, not functional line shape with respect to radius~\citep{Chatterjee}.  
 
 The spherical assumption is   beneficial   computationally and conceptually.   Since   the flat rotation-curve problem   is clearly a  functional difference at large radii, this approximation cannot be responsible for the rotation curve velocities. Since the assumption of spherical symmetry is commonly used as a final approximation in evaluating spiral galaxies,    we    will use  this   assumption  in the LCM derivation (section~\ref{sec:DERIVE}) in the same ways .
 
 \subsubsection{LCM velocity addition formula}
 \label{LCMfunc}
In the LCM, the   observed Doppler-shifted frequencies  $\omega'$  in Eq.~\ref{eq:dataLorentz} are   posited to include two contributions: relative velocity and relative curvature.   Since these frequency shifts are   reported as the total rotation velocity, $v_{obs}$,  we   phrase our arguments kinematically,  including contributions to the relative curvature  in terms of equivalent Doppler-shifts.  It is clear from the work of ~\citet{Cisn} and ~\citet{Radosz} that 
 curvature effects on light can be cast  in purely kinematical terms.   
 
  Since dark matter theories are based on  classical Newtonian and   Relativistic gravity, we co-op the dark matter halo velocity addition formula in  Eq.~\ref{eq:zonte1},  to represent the contributions from relative velocity and relative curvature to the total frequency shifts by replacing the dark matter contribution $v^2_{dm}$  with a relative curvature term $\tilde{v}^2_{lcm}$:
  \begin{equation}
v_{rot}^2 =  v_{l}^2+ \tilde{a} \tilde{v}_{lcm}^2 ,
\label{eq:zonteLCM}
\end{equation}  
 for $ v_{l}^2$,    the  relative velocity  contribution;  $\tilde{a}$ is 
a dimensionless, free  parameter (see Fig.~\ref{fig:alphaDistrib1}), and $v_{rot}$  is the total rotation  velocity.

 Relative velocity and relative curvature effects are calculated  using only the luminous matter as   reported  in the cited literature, consistent with photometry and population synthesis modeling. In this way, by recasting the curvature contributions, using 
the associated  gravitational redshifts as a function of radius to get the equivalent  Doppler-shifts,  the relative velocity and relative curvature are  represented by the kinematics of the Lorentz group. 

 \subsection{Luminous mass modeling}
\label{sec:POTENTIALS}

\begin{figure*}
 \centering
\subfigure[]{\includegraphics[width=0.3\textwidth]{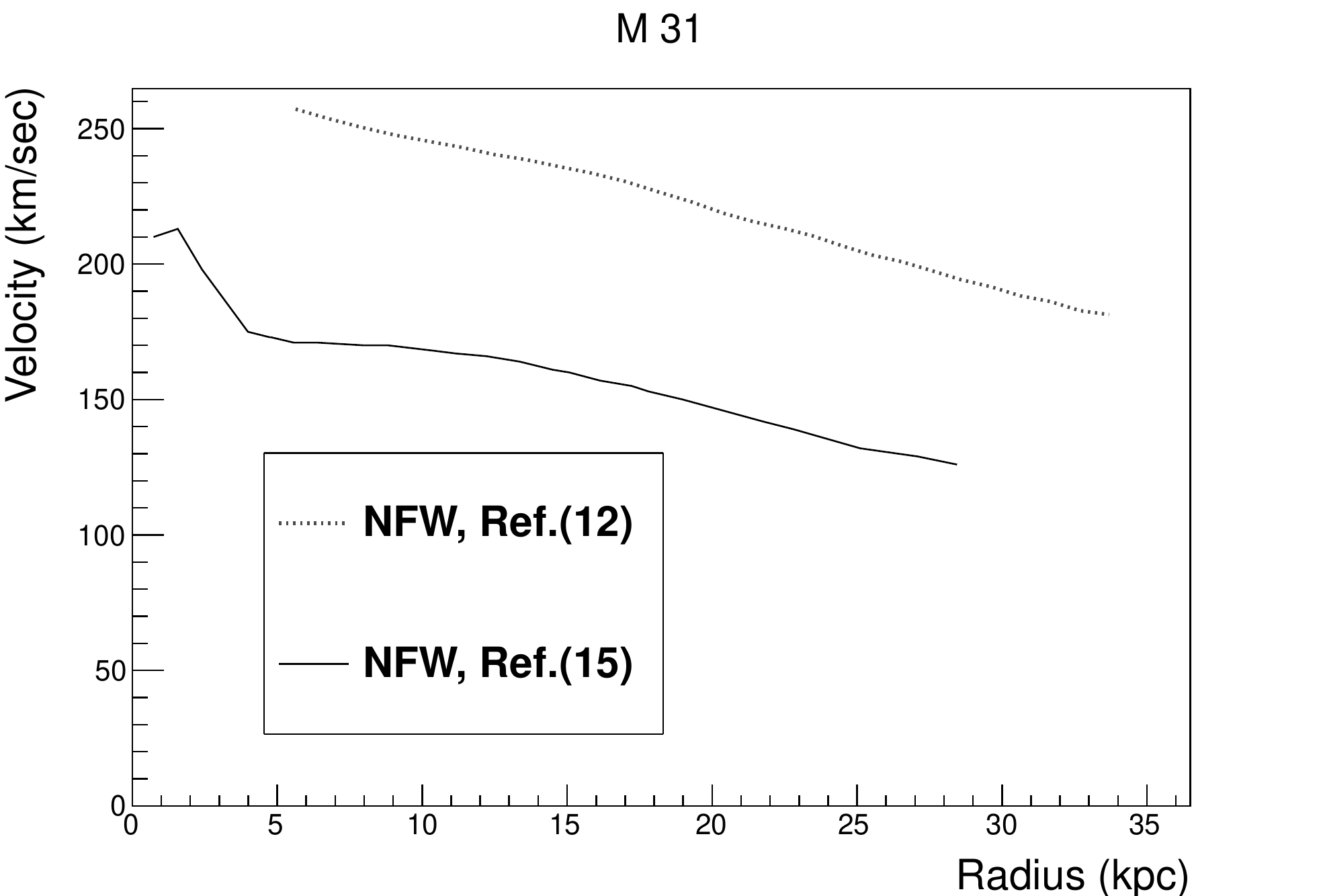}}
 \subfigure[]{\includegraphics[width=0.3\textwidth]{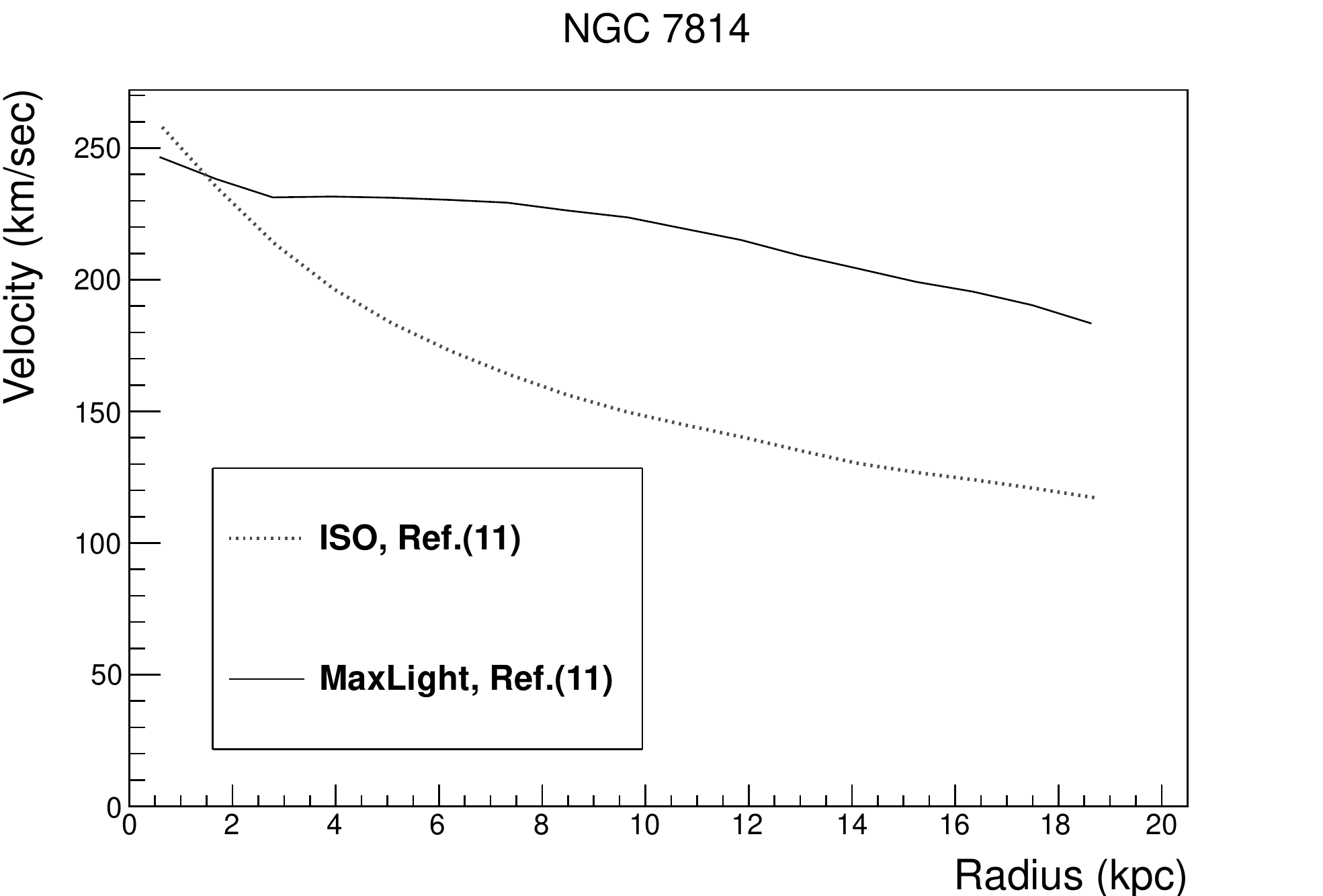}}
 \subfigure[]{\includegraphics[width=0.3\textwidth]{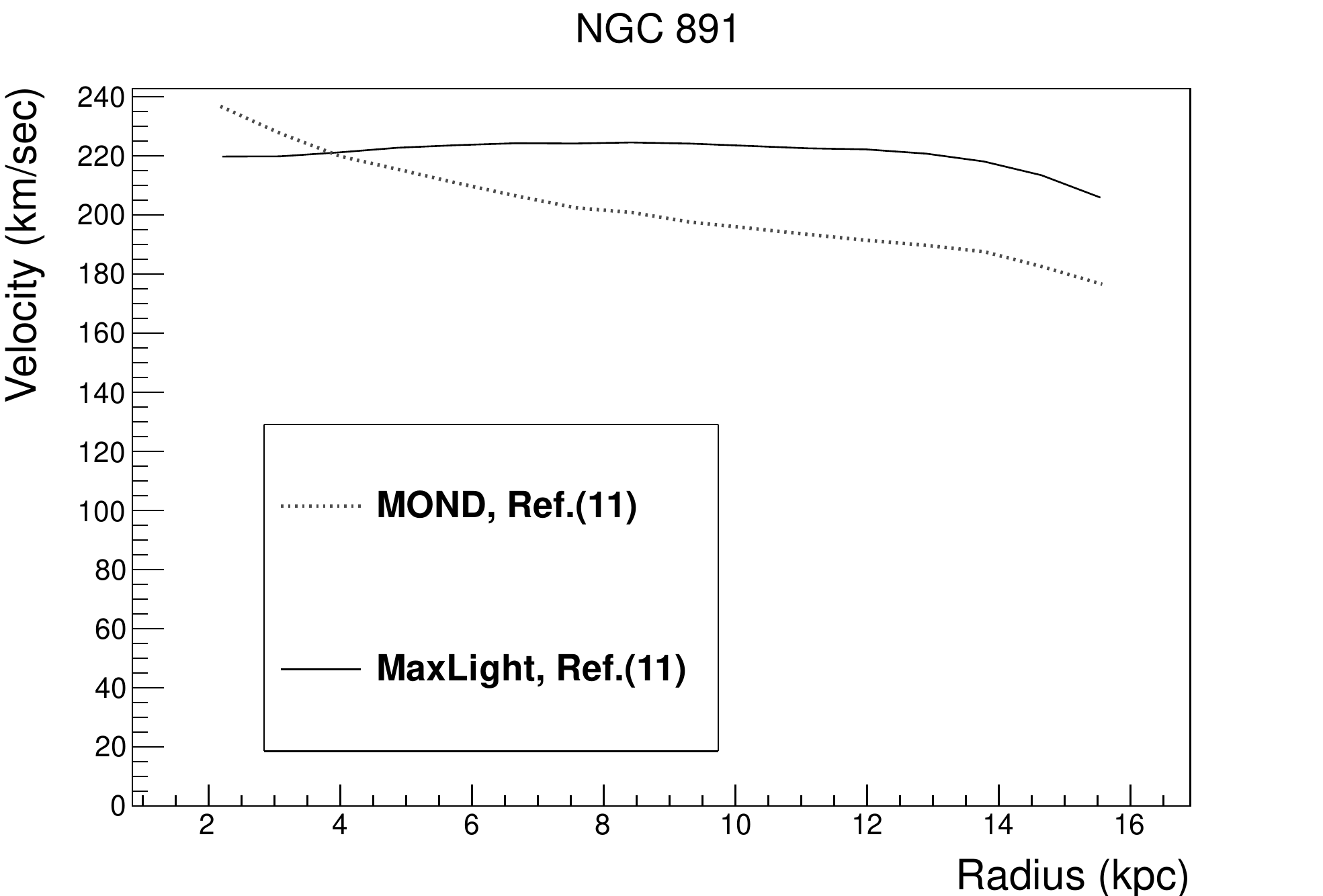}}\\
 
 \subfigure[]{\includegraphics[width=0.3\textwidth]{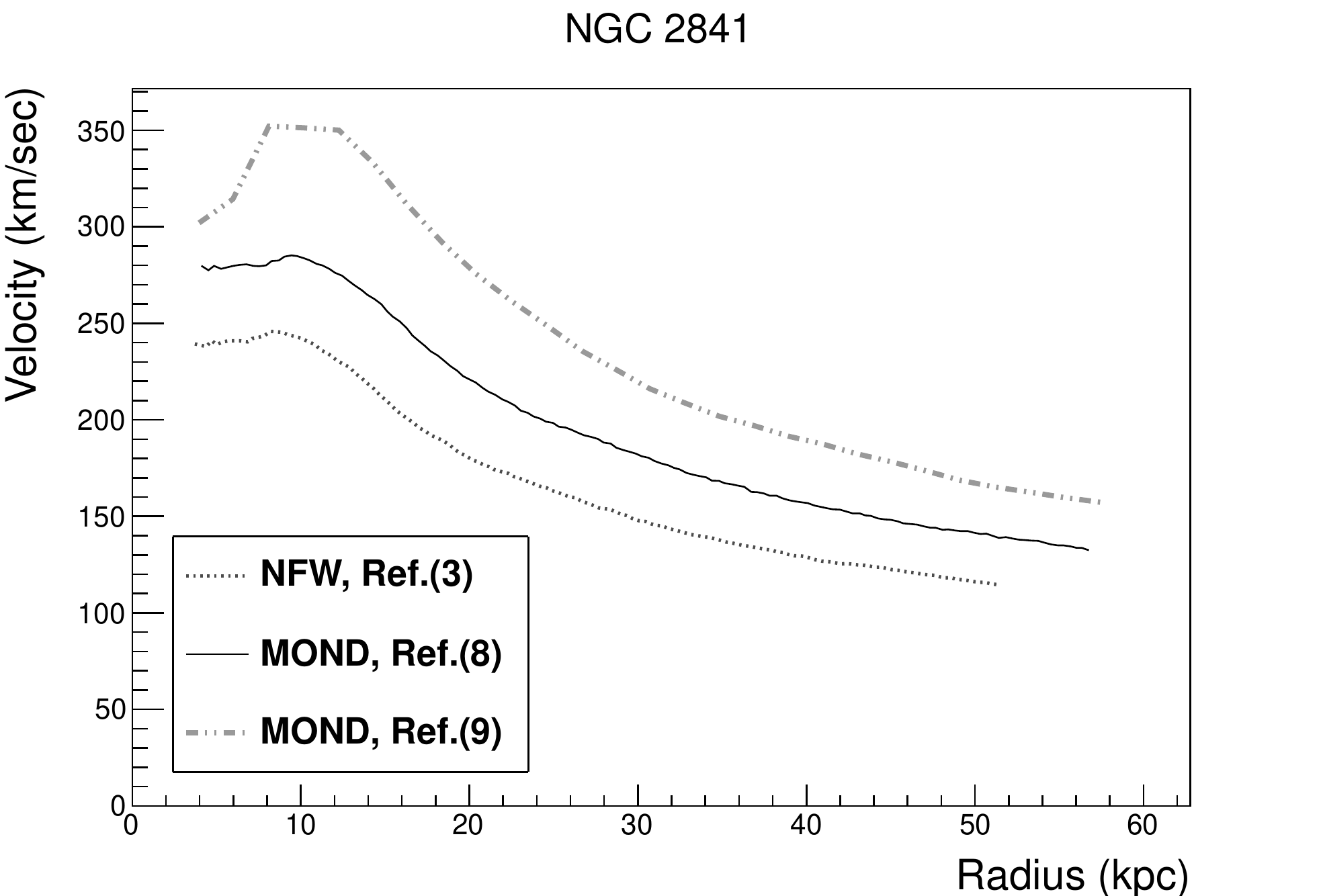}}
 \subfigure[]{\includegraphics[width=0.3\textwidth]{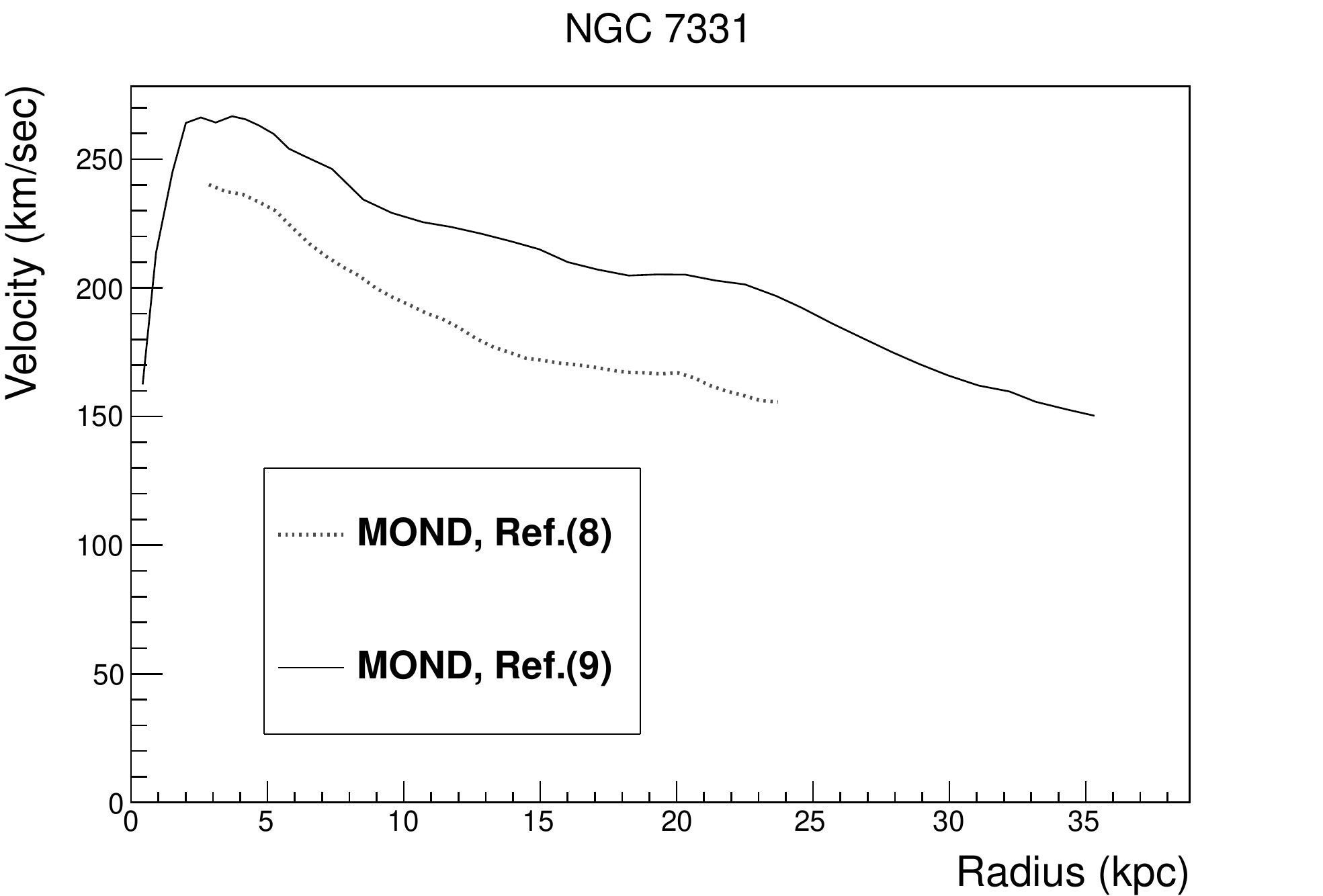}}
 \subfigure[  ]{\includegraphics[width=0.3\textwidth]{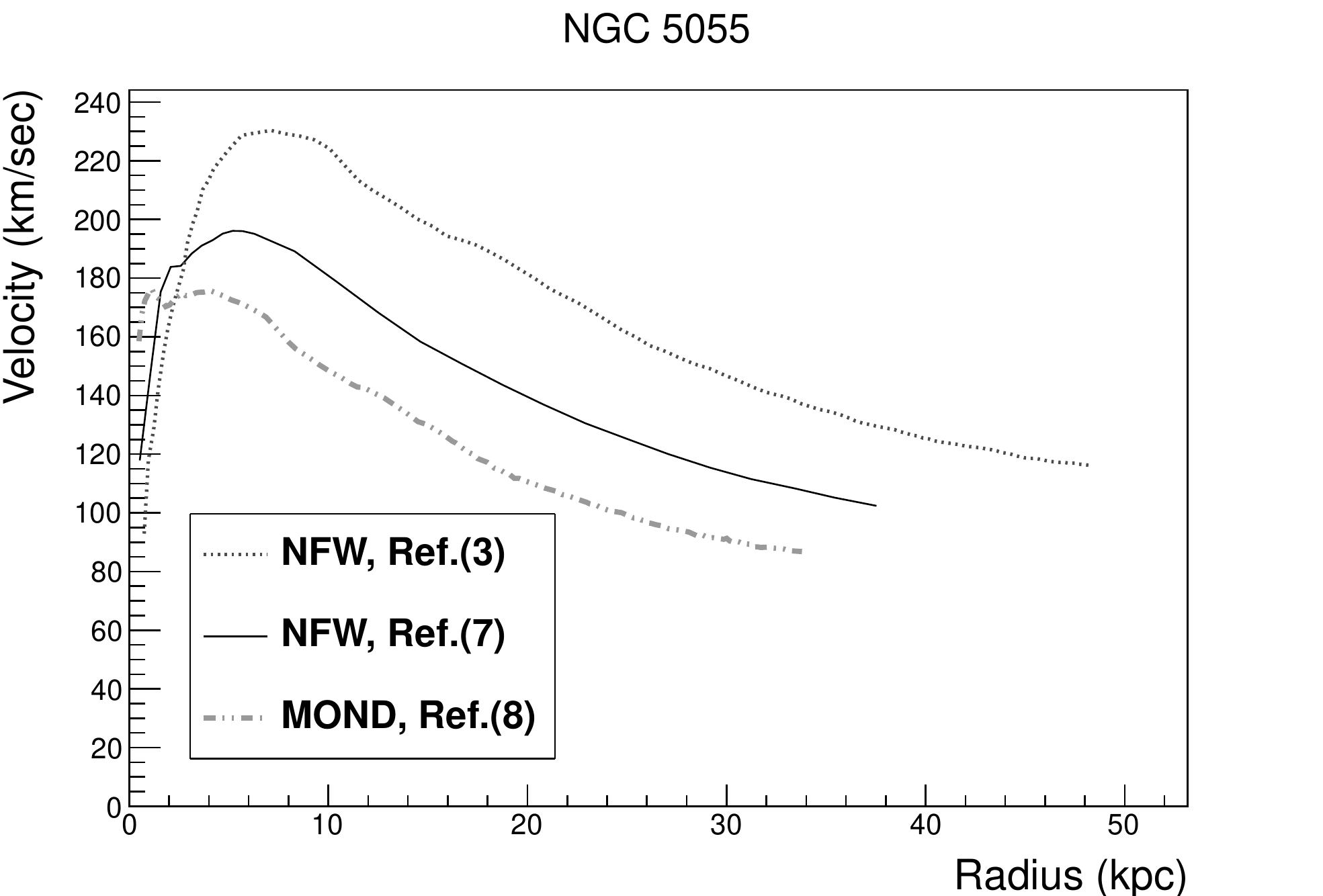}}\\
 
\subfigure[ ]{\includegraphics[width=0.3\textwidth]{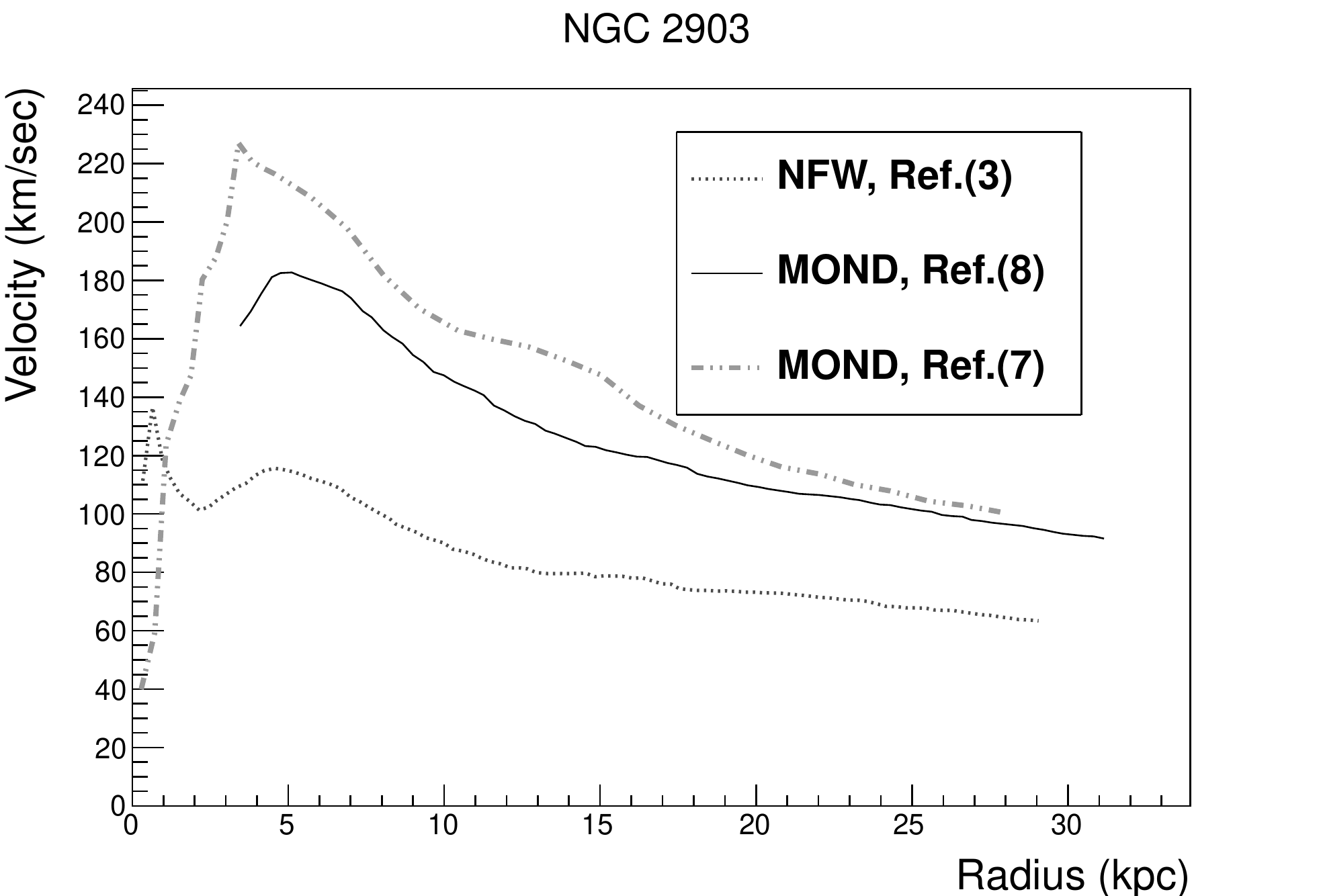}}
 \subfigure[  ]{\includegraphics[width=0.3\textwidth]{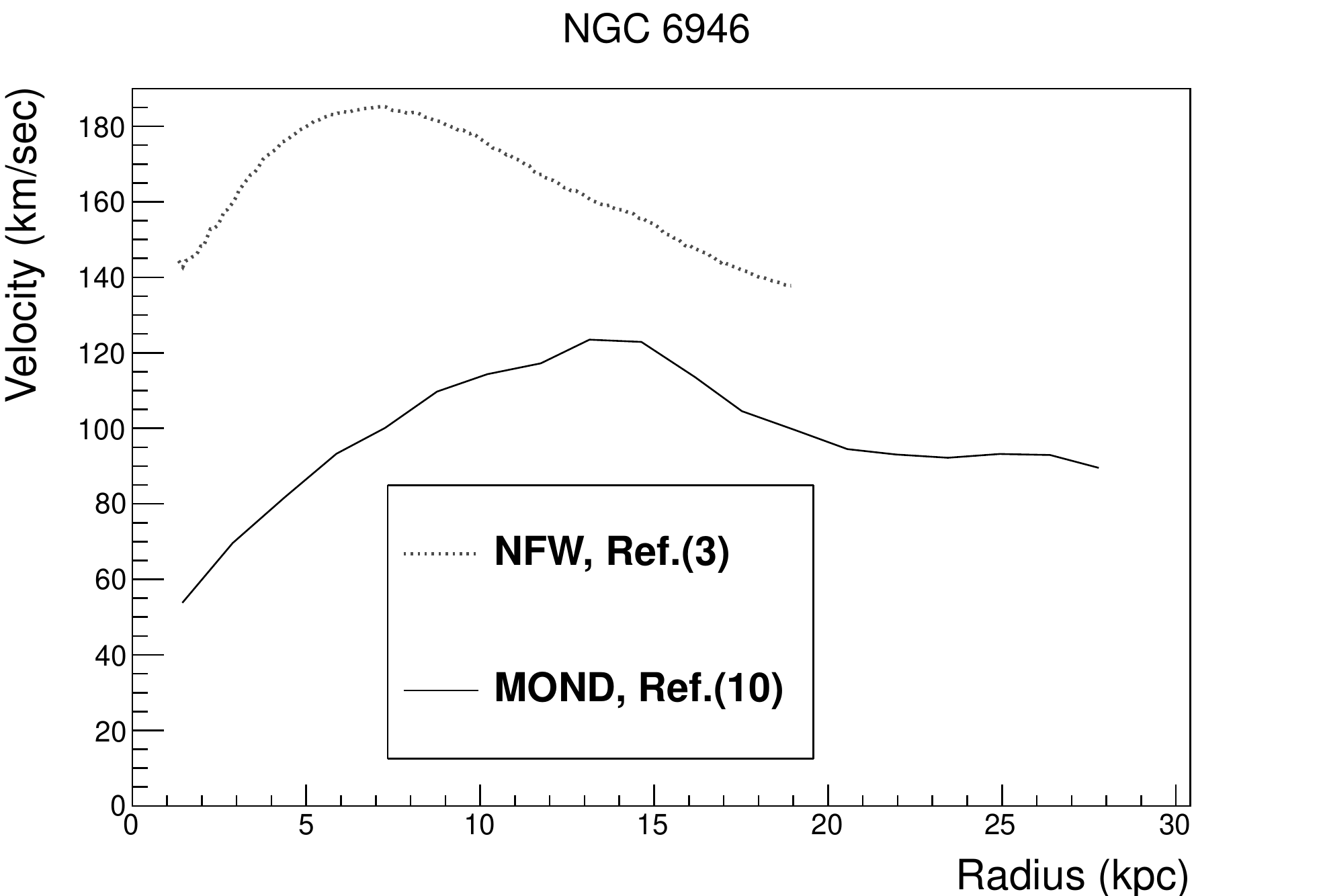}}
 \subfigure[ ]{\includegraphics[width=0.3\textwidth]{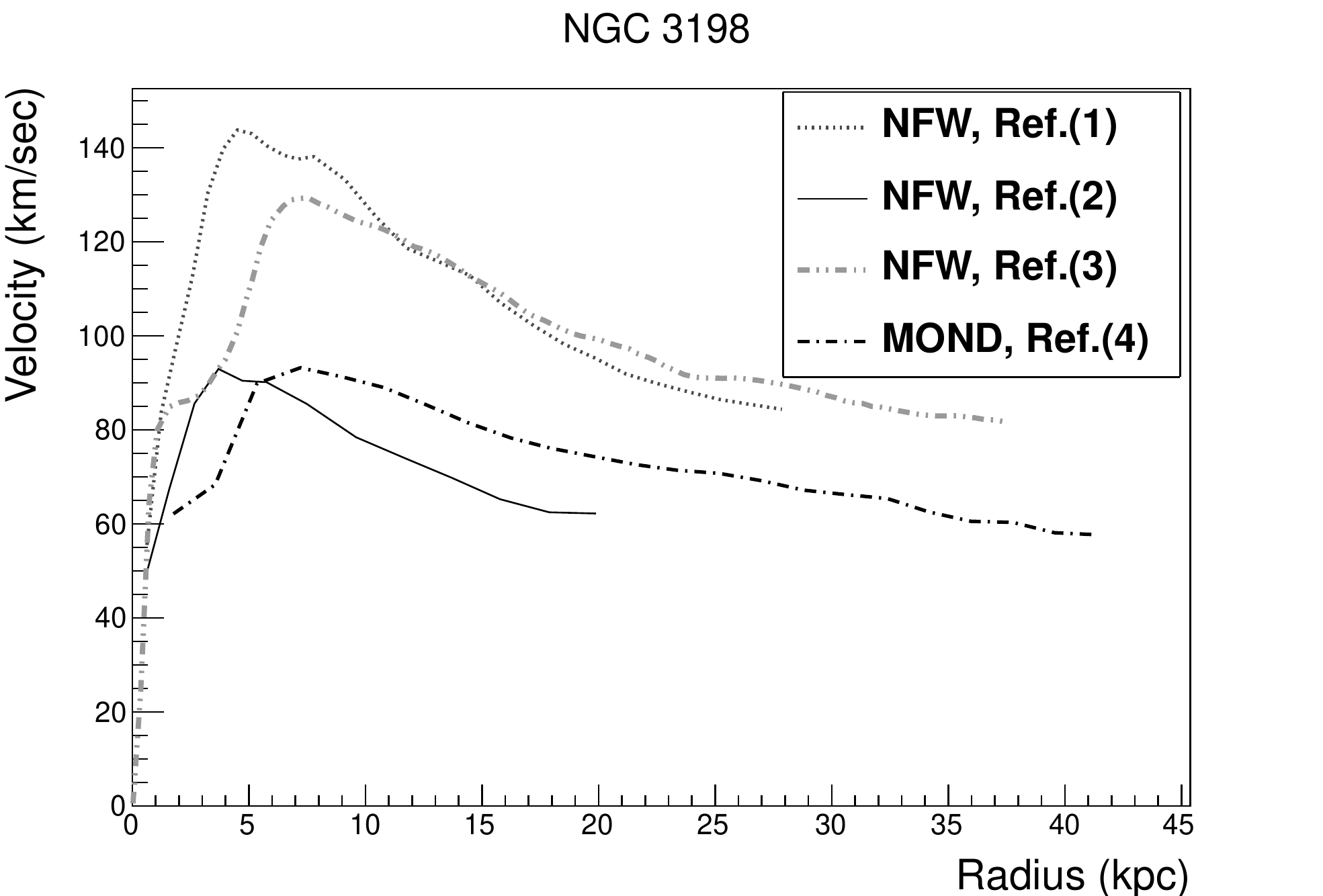}}\\
 
 \subfigure[ ]{\includegraphics[width=0.3\textwidth]{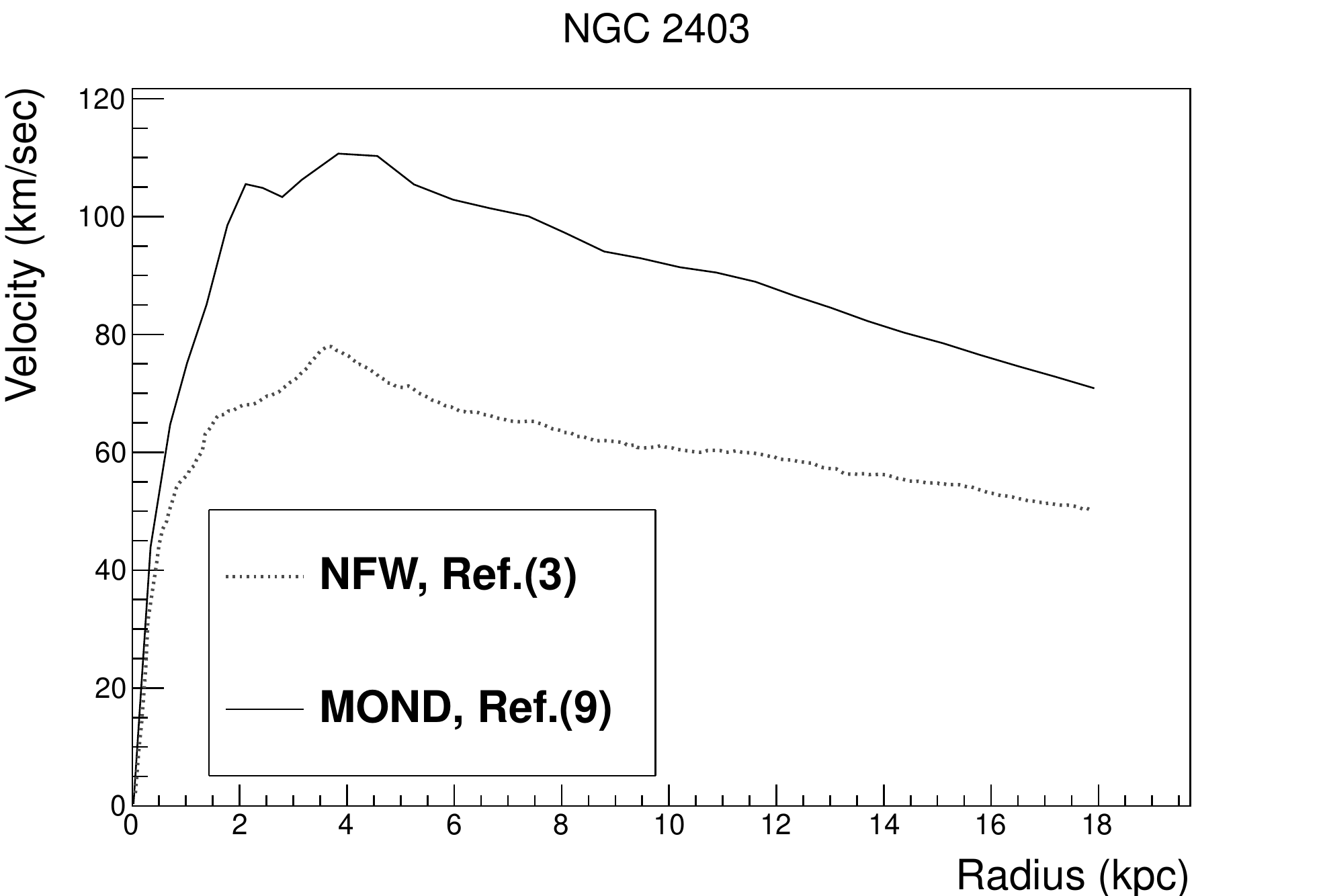}}
 \subfigure[]{\includegraphics[width=0.3\textwidth]{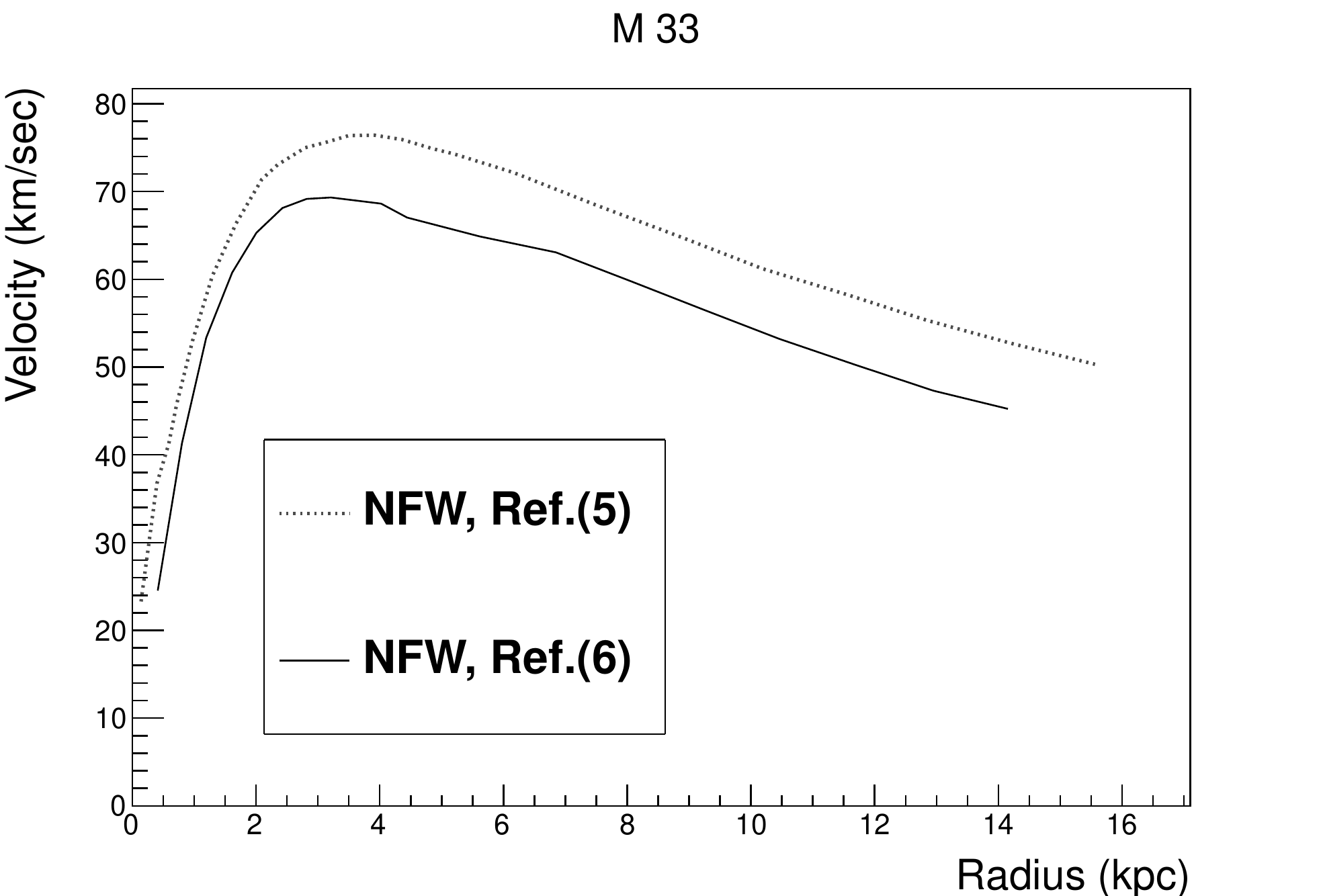}}
 \subfigure[  ]{\includegraphics[width=0.3\textwidth]{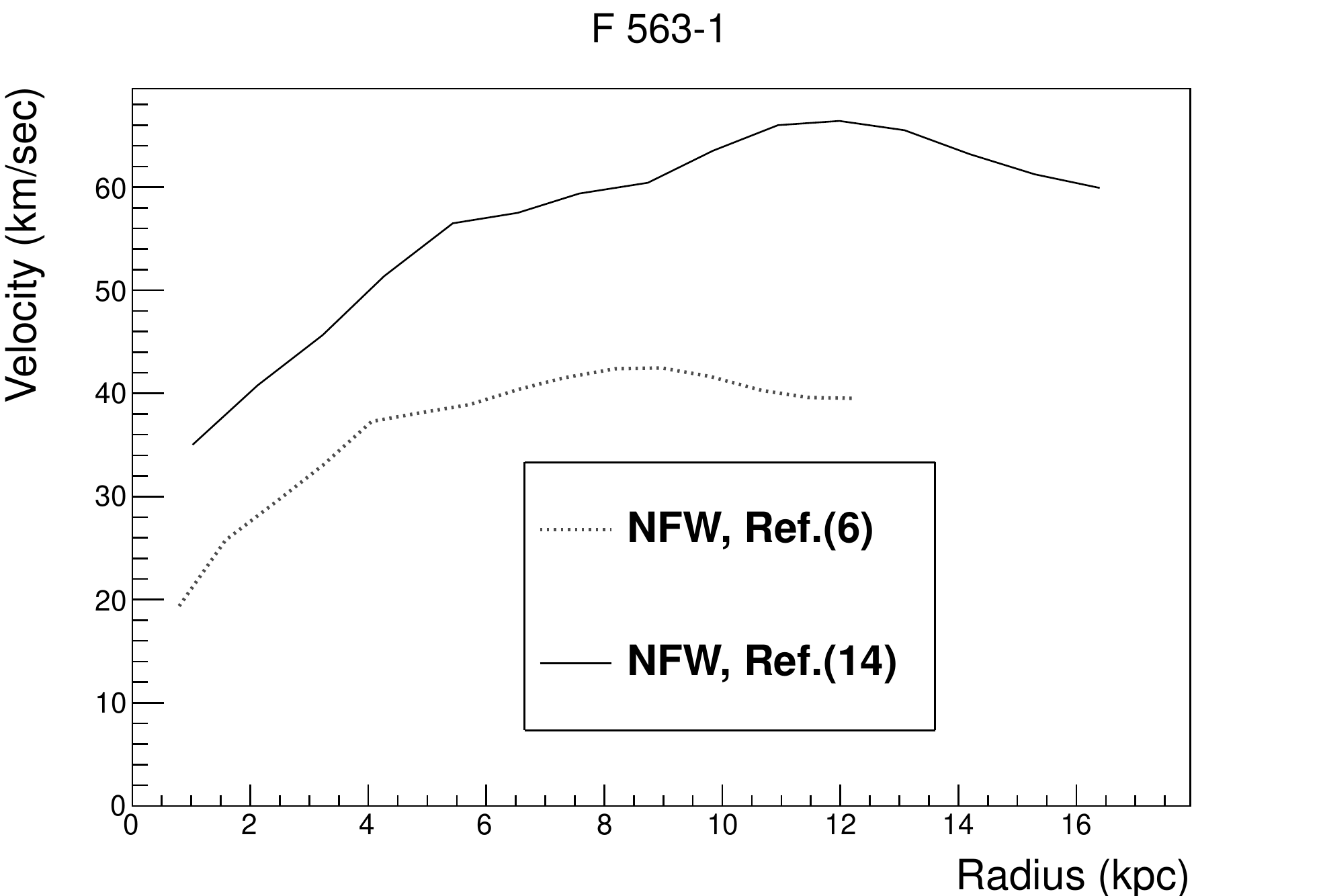}}\\
 
 \subfigure[  ]{\includegraphics[width=0.3\textwidth]{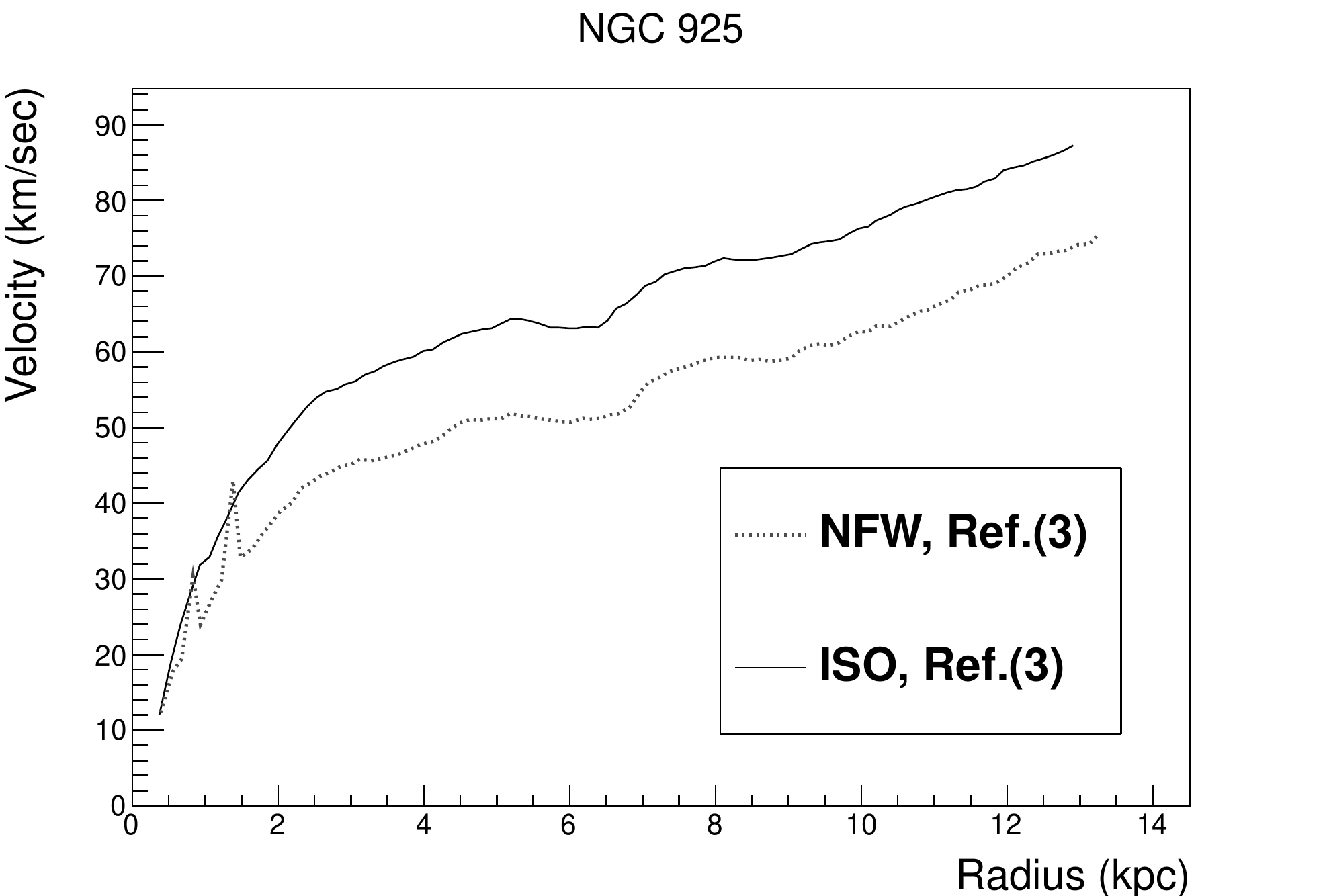}}
 \subfigure[  ]{\includegraphics[width=0.3\textwidth]{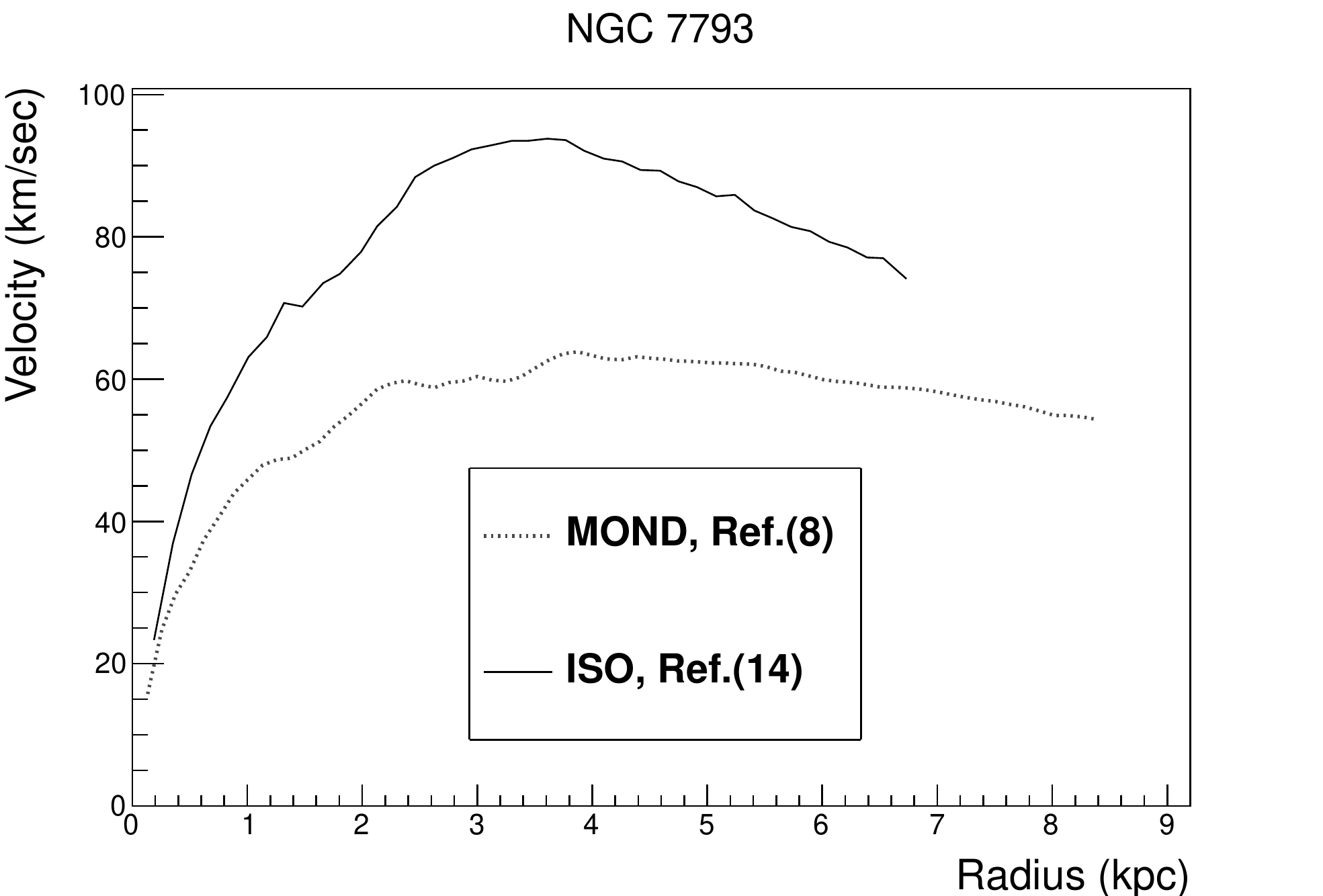}}
  \caption{Original reported rotation curve velocities due to the posited luminous mass  for spiral galaxies in this sample; the originating model  context  is  indicated in each  figure   legend.  Variations in the originating luminous mass profiles reported in the literature illustrate the under-constrained nature of population synthesis modeling, both in magnitude and geometry.    Originating  model   M/L ratios,   wavelength bands  and  references are   in Table~\ref{sumRESULTS}.}
 \label{fig:massmodels18}
\end{figure*}

Luminous matter modeling in spiral galaxies is an under-constrained
 field of study ~\citep{Conroy}.    Dark matter models and MOND are commonly used as constraints on luminous matter modeling.  Fig.~\ref{fig:massmodels18} compares original reported luminous profiles, fit with dark matter and MOND models, from    fourteen ($14$) of the galaxies in our sample.    As can be seen in  the figure,  neither
class of model   demonstrates consistent  predictions of the  luminous mass  in  a given galaxy. 
 
Estimates   of the total luminous galaxy mass,  $M_l$, commonly include 
 the masses of gas and dust,  $M_{g}$, stellar bulge,  $M_{b} $ and disk,  $M_{d} $,
   \begin{equation}
 M_{l}=M_{g}+M_{d} +M_{b}.
  \label{eq:lumMasses}
 \end{equation}
Measurements of the  individual component  come from a variety of observations for the
 surface brightness profiles at many wavelengths. The dominant gas
 mass is contributed by atomic hydrogen,    H\,{\sevensize\bf I}, observed at $21$-cm and by
 molecular gas, observed through CO rotational lines in the
 mm-regime. Generally, it is assumed that the $21$-cm line is optically
 thin in which case it can be translated directly into a column
 density of    H\,{\sevensize\bf I}. Estimates of the effects of optical depth (e.g. ~\citet{Braun}) imply that the actual    H\,{\sevensize\bf I} column density could be higher
 by up to $30\%$, but this likely is dependent on the inclination of the
 galaxy. In the case of CO, various elaborate calibration methods have
 led to a conversion factor of CO intensity to a molecular hydrogen
 column density. This has an associated uncertainty of up to a factor of two. 
  Generally, molecular gas surface densities are higher in the inner
 disks of galaxies and  H\,{\sevensize\bf I} column densities in the outer disks of
 galaxies. In terms of ionized gas, the only substantial mass is
 contributed by a diffuse ionized medium. The column density of this
 ionized medium is about $30\%$ of that of the    H\,{\sevensize\bf I} column in the solar
 neighborhood. Because it is only possible to obtain column density
 estimates in the Milky Way for this gas (from pulsar dispersion
 measurements), it is usually ignored in rotation curve
 modeling. Other ionized gas phases in    H\,{\sevensize\bf II} regions and hot X-ray
 emitting plasmas contribute little mass. The derived gas column
 densities are generally corrected for the presence of helium, by
 a factor of approximately 1.4. Lastly, the dust-to-gas ratio in the
 interstellar gas is only about $1\%$ so dust masses may be ignored in
 rotation curve modeling. ~(R.\,A.\,M.\, Walterbos, personal communication, 2014)

 Associating a mass surface density with the observed light profiles
 of the stellar components involves a suite of modeling assumptions
 used to reproduce the observed light intensity in various bands. The
 conversion of light into stellar mass   generally depends on the
 age(s) of the stellar populations present, their metallicities,
 corrections for obscuration by dust and the inclination of the
 galaxy.  The stellar populations present are a consequence of the
 star formation history, the initial mass function, and the
 evolutionary history of the galaxy. The most important of these
 factors are included in a population synthesis model (PSM) that is
 matched to the observed light distribution in various colors to yield
 a mass-to-light ratio (M/L).   Conversion of light into masses by the M/L is the largest uncertainty in luminous matter modeling~\citep{Toky} as evidenced by the variations in Fig.~\ref{fig:massmodels18}.   
 
 After the appropriate Poisson's law forces are calculated for   each galaxy component in
  Eq.~(\ref{eq:lumMasses}),   the
  orbital velocities are added in quadrature  by the assumption of spherical symmetry:
  \begin{equation} 
 v^2_{l} =  v^2_{g} + ( \rmn{M/L} )_d v^2_{d} +( \rmn{M/L} )_b v^2_{b}. 
  \label{eq:quadrature}
 \end{equation}  
 The  orbital rotation velocity associated with  the gas mass is  $v^2_{g}$, with  the stellar bulge  mass is   $(\rmn{M/L} )_b  v^2_{b} $  and  with the stellar disk  mass is  $(\rmn{M/L})_d v^2_{d}$.

     \section[]{Lorentz kinematics and the LCM derivation }
\label{sec:DERIVE} 
The LCM construction is based upon a careful identification of the relationship between the local flat frames
where physical measurements are made and   the underlying curved manifolds.   It is known from 
the differential geometry of   General Relativity,  that sufficiently precise measurements in the local flat frames can completely specify the underlying curved manifold's metric.    

In the flat rotation-curve problem,  we assume  spectra are such   sufficiently precise measurements, and that our local observations of the visible extent of spiral galaxies    gives enough information 
 to determine the underlying global curvatures  of the emitter and receiver galaxies from their baryonic masses.
 
  \subsection{Curvatures from luminous mass}
  \label{sec:curvatures}
  
   Curvatures in Schwarzschild metric space-times are indicated by     gravitational redshift  effects on photon frequencies:
  \begin{equation}
 \frac{\omega_o }{ \omega ( r) }=\left(\frac{1}{\sqrt{-g_{tt}} }\right)_r,
\label{eq:Clone}
\end{equation}
where $\omega_o$ is  the   characteristic photon frequency as   defined in section~\ref{sec:stationTotal}, $\omega (r)$ is  the  shifted frequency\footnote{ as 
  received by a stationary observer  at asymptotic infinity}.  The   time coefficient of the Schwarzschild metric is $g_{tt}$:
 \begin{equation}
 g_{tt}(r)=-\left(1- 2\frac{G M}{c^2r}\right), 
 \label{eq:timeportion}
 \end{equation}
where $G$ is  Newton's constant of gravity,   $M$ is  the   enclosed mass at some  radial distance $r$  from the center of 
 the mass distribution,  and
 $c$ is  the vacuum light speed.
 
 In the   weak field limit ~\citep{Hartle}
 the   metric coefficient becomes: 
 \begin{equation}
g_{tt}(r) \approx -1 + 2\frac{\Phi(r)}{c^2}.
\label{eq:weakfield}
\end{equation}
where   $\Phi$ is the Newtonian  scalar   gravitational potential. \\

Spiral galaxies are  treated as weak fields due to  the diffuse nature of the luminous mass distributions, such that dark matter theories   are  based on Newtonian kinematics  and Special Relativistic interpretations of spectra.  To extend  these flat space-time concepts to the slightly curved frames  of spiral galaxies, we   will make the following  two   caveats.
\subsubsection{Assumption $1$  }   
 The exterior Schwarzschild metric  is  a vacuum solution to Einstein's equations,
  intended for use outside   the central mass that is generating the curvature.   By invoking the  simplifying assumption of spherical symmetry generally used in galaxy analysis (see section \ref{sec:stationTotal}),   Gauss's law for  spherical mass distributions~\citep{Fowles} becomes a valid approximation.  Therefore,  at each  radius $r$,
   all   mass elements  external to   $r$ cancel by symmetry, such that the   exterior Schwarzschild  metric   exactly satisfies the Einstein equations as a vacuum solution.     Note that  any theory using this approximation necessarily loses information at small radii  where tidal forces dominate  or in the presence of severe symmetry breaking features.   Since the flat rotation-curve problem is one of large radii,  we find this to be an acceptable approximation to first order.    Each   galaxy manifold can be viewed   as foliations of Schwarzschild solutions with  increasing  radii. 
   
   In practice this assumption results in summing   the  stellar and gas contributions to the gravitational potential  $\Phi$ in Eq.~\ref{eq:weakfield}  under the assumption of spherical symmetry, though the contributions are individually  calculated in the correct   geometric Poisson equation.  This approach is  consistent with the standard treatment of gravitational  potentials in  rotation curve  calculations~\citep{Xue,Klypin} because,  to 
good approximation,  deviations from spherical symmetry are higher order corrections to the gravitational  potential~\citep{Binney,Chatterjee}.      
\subsubsection{Assumption $2$ }  
The  gravitational potential,  $\Phi(r)$, which   parametrizes the curvatures of interest in    Eq.~\ref{eq:weakfield}, is defined as an integral over   the Newtonian force  $F(r)$:
\begin{equation}
\Phi(r)+\Phi_o=-\int \frac{F(r) }{m} dr,
\label{eq:potentialgeneral}
\end{equation}  
where   $F(r) /m$ is the force per unit mass for  each individual  luminous mass component,  and  $\Phi_o$  is the integration constant of interest.

  This   integration constant is  generally set such that the potential $\Phi(r)\to 0$ as $r\to \infty$.  However, when considering     two arbitrary galaxies\footnote{the  emitter galaxy and the receiver galaxy (Milky Way)}, connected by a single photon,  it is a violation of energy conservation to   set   the respective integration constants to different values.   We select a single universal value for the integration constant,  taken to be zero,  though  a more physical     choice  may be found in   future dark energy  research.   Physically, this means that at large $r$    the gravitational potentials     go  to small  but non-zero  values.   
 
 \subsection{The convolution function} 

 The LCM mapping term,   $ v_{lcm}^2$,  is composed
  of three terms: 
\begin{equation}
v_{lcm}^2=   \kappa  v_{1}  v_{2},
\label{eq:convolutionFunc}
\end{equation}
where $\kappa$   is the curvature ratio,  and  $v_1$ and $v_2$ are    successive Lorentz transformations.  

We then     normalize the mapping,  $ v_{lcm}^2$,  by its     value  as $r\to \infty$, $(v^2_{lcm})_\tau$,    such that the relative curvature term in Eq.~\ref{eq:zonteLCM}   is:
    \begin{equation}
  \tilde{v}^2_{lcm}=\frac{v^2_{lcm}}{(v^2_{lcm})_\tau}.
  \label{eq:domain}
  \end{equation}
 
\subsubsection{The LCM curvature ratio: $\kappa$}  
 \label{kappa}
\begin{figure}
\includegraphics[scale=0.25]{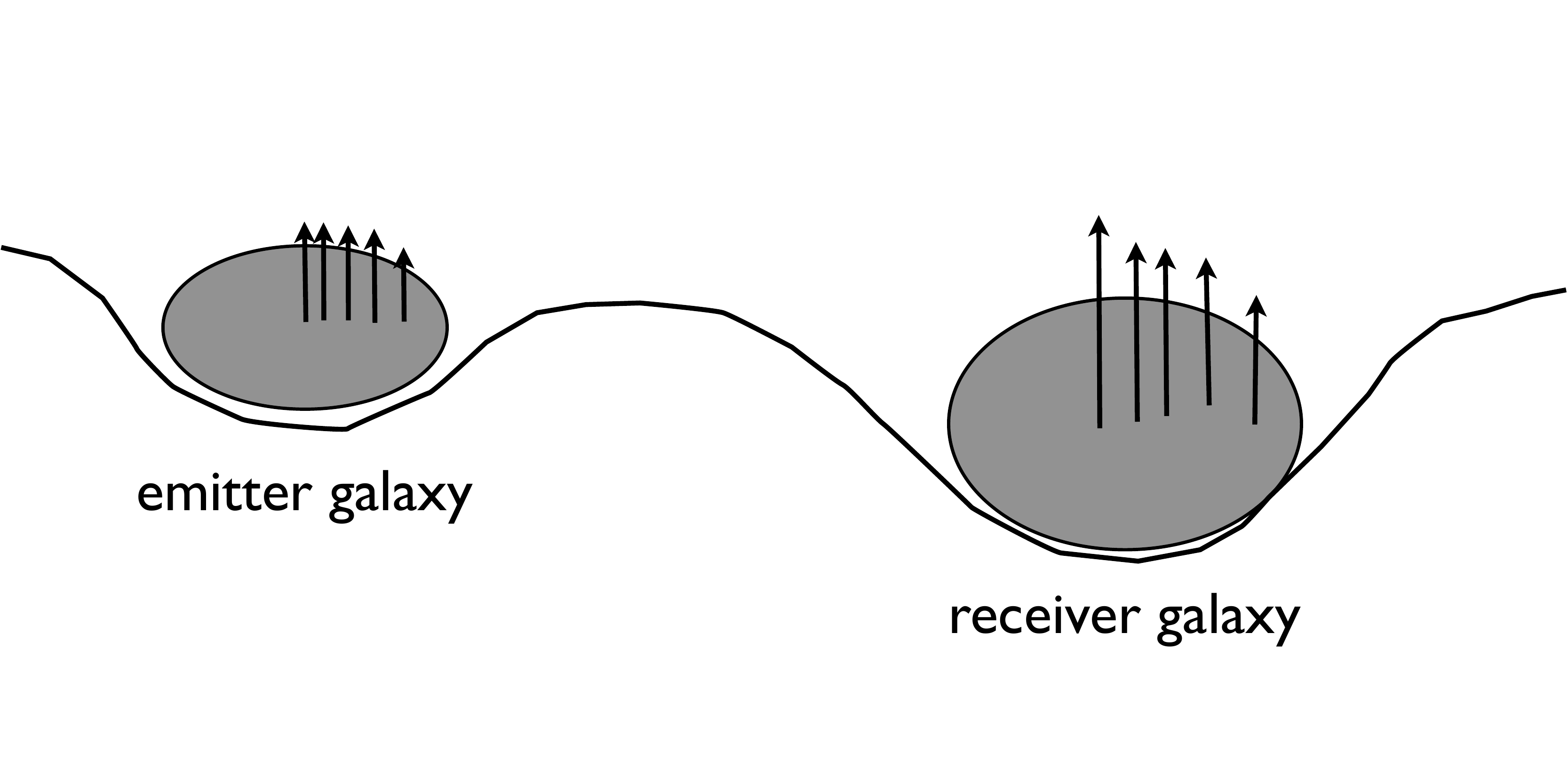} 
\caption{A cartoon of   emitter   and receiver galaxies.  The    vertical arrows represent the unit times at each point in the respective galaxy.   The relative length of the    unit time is  indicated by the length of the    arrow;  a longer unit time indicates that time runs slower. Unit times within the enclosed mass density are longer than those outside, due to the mass dependence of the time coefficient of the space-time metric.     $\kappa(r)$ is proportional to the ratio of the arrow lengths at  radius $r$, because the observed frequencies are measured with respect to the receiver's coordinate time. \label{kappaClocks}}
\end{figure}

 In traditional Lorentz transformations all   frames are symmetric with respect to the coordinate time  $t=\cosh \xi$  for  the rapidity angle, $\xi$.   Since   Lorentz transformation are used to interpret spectra  with  respect to the coordinate time of the Milky Way,  we need a rule to   re-express the coordinate time of the emitting galaxy in terms of that of the Milky Way.
   This is done with the curvature ratio $\kappa$, which is based upon the idea of  coordinate light speeds  $\tilde{c} $.   Fig.~\ref{kappaClocks} shows how  time is affected by the underlying curvature of space.
   
 Coordinate light speeds  are  a physical indicator   of curvature; the degree to which
 $\tilde{c}<c $ indicate the increase in path length due to curvature \citep{Narayan}.  As viewed by an  external observer at asymptotic infinity, who can not \emph{see}  manifest curvature but   has knowledge of the line-of-sight travel distance of the light, curvature is indicated by the difference between the vacuum light speed and the measured coordinate light speed, 
 $c-\tilde{c}(r)$.
 
  Coordinate  light speeds arise in the solution to the  wave equation for light~\citep{Cisn}.  In the Schwarzschild case,    the    
   effective index of refraction, $n(r)$, relates the two light speeds:
     \begin{equation}
 n(r) \tilde{c }=\left(\frac{1}{\sqrt{-g_{tt}} }\right)_r  \tilde{c}=c,
 \label{eq:index}
\end{equation}
by the gravitational redshift (Eq.~\ref{eq:Clone}).

  The  ratio  $\kappa$   scales the coordinate time of  the emitter galaxy relative to that of the  Milky Way in terms of        deviations  from flatness as a function of radius:
\begin{equation}
\kappa(r)=\frac{c-\tilde{c}_{gal}(r)}{c-\tilde{c}_{mw}(r)},
\label{eq:kappa}
\end{equation}
 where  $\tilde{c}_{gal} (r)$  and  $\tilde{c}_{mw} (r)$ are the  respective   coordinate light speeds of the emitter   and receiver galaxies.  The value of the ratio at the limit of the data,  $\kappa_\tau$,  approximates the value when $r\to \infty$ when all the luminous mass is enclosed.
 
\subsubsection{Equivalent Lorentz Doppler-shift formula}
 
 The most general  form of the  Lorentz transformation is the   exponential mapping:
 \begin{equation}
 \Lambda= e^{\chi}=\sum_{n=0}^{\infty} \frac{\chi^n}{n!}
 \end{equation}
 where   $\chi=-\xi S$   is the  product of the rapidity angle $\xi$ and  the generator of the rotation $S$. The rapidity 
 angle defines the relationship between two frames in the hyperbolic space-time of Special Relativity.
 
   The Doppler-shift formula in Eq.~\ref{eq:dataLorentz}  comes from such a Lorentz transformation.   
    In hyperbolic form,  the Lorentz transformation matrix for   a boost in the $x$ direction is
\begin{align}
\Lambda_x&=
  \left( \begin{array}{cccc}
\cosh \xi &-\sinh \xi & 0&0 \\
  -\sinh \xi & \cosh \xi  & 0&0\\
  0&0& 1&0\\
  0&0& 0&1 
 \end{array} \right) \\
 &= exp \left[- \xi \left(\begin{array}{cccc}
0&1 & 0&0 \\
1 & 0  & 0&0\\
  0&0& 0&0\\
  0&0& 0&0
 \end{array} \right)\right] .
    \end{align} 
This form  yields the Lorentz Doppler-shift formula by  rotating   a   photon's 4-vector $(\omega,  k_i)$     through $\xi$, where $k_i=2\pi/\lambda_i$,   $\lambda$ is the wavelength, and   the indices of  the   spatial basis are $i=1,2,3$. The hyperbolic form of the Lorentz
Doppler-shift formula for line of sight photons is:
 \begin{equation}
\frac{v}{c}= \tanh \xi= \frac{e^\xi - e^{-\xi}}{e^\xi + e^{-\xi}}. 
\label{eq:LorentzDefine}
\end{equation}
We can extend  this  formalism to the  mapping of the small relative curvatures between a pair of emitter 
and receiver galaxies.  
The resulting   parameter  $v$ from    Eq.~\ref{eq:LorentzDefine} will   relate the emitter to  receiver galaxy,  and should  not to be confused with  a physical speed. 

To generalize the Doppler formula as given in  Eq.~\ref{eq:LorentzDefine},   we identify the   mapping  factor $e^\xi$    with   the ratio of the received to emitted frequencies,  $\omega_s/\omega_o$. In Special Relativity the two frames are symmetric, so it is meaningless to pin frequencies to specific frames.  
However,   transitioning to gently curved frames  it becomes necessary to   consistently identify the mapping factor. 

  In what follows, we   always identify the numerator with the   receiver's reference frame
 and the denominator with the emitter's reference frame:
  \begin{equation}
  e^\xi(r)=\omega_{\rm{receiver}}(r)/\omega_{\rm{emitter}}(r).
  \label{eq:array}
  \end{equation} 
 
\subsubsection{ Mapping $v_1$ } 
 The first LCM term, $v_1$,  looks at the gravitational redshift frequencies  $\omega(r)$ (Eq.~\ref{eq:Clone}) and rephrases them in terms of an equivalent Doppler-shift formula via Lorentz transformations.  This term is a 2-frame map of the  emitting and receiving galaxies, created  by
their  respective  gravitational   potentials from  luminous mass.

By the  convention in Eq.~\ref{eq:array}, the curved 2-frame
   mapping  factor  is: 
 \begin{equation}
e^{\xi_{c}}(r)=\frac{\omega_{mw}(r)}{\omega_{gal}(r)},  
\label{eq:specific}
\end{equation}
for  the   redshift frequencies  of the emitter galaxy, $\omega_{gal}(r)$,    and  the receiver galaxy, $\omega_{mw}(r)$. 
 
Consistent with the form of Eq.~\ref{eq:LorentzDefine}, the first LCM mapping  is then  
\begin{equation}
\frac{v_{1}}{c}=\frac{ e^{\xi_{c}} - e^{-\xi_{c}}}{e^{\xi_{c}}+ e^{-\xi_{c}}}.
\label{eq:prime1}
\end{equation}
All quantities are     functions of radius except for the vacuum light speed.   Fig.~\ref{v_1graphic}    gives a visual representation of    such a mapping. 
\subsubsection{Mapping  $v_2$ } 
\begin{figure}
\includegraphics[scale=0.25]{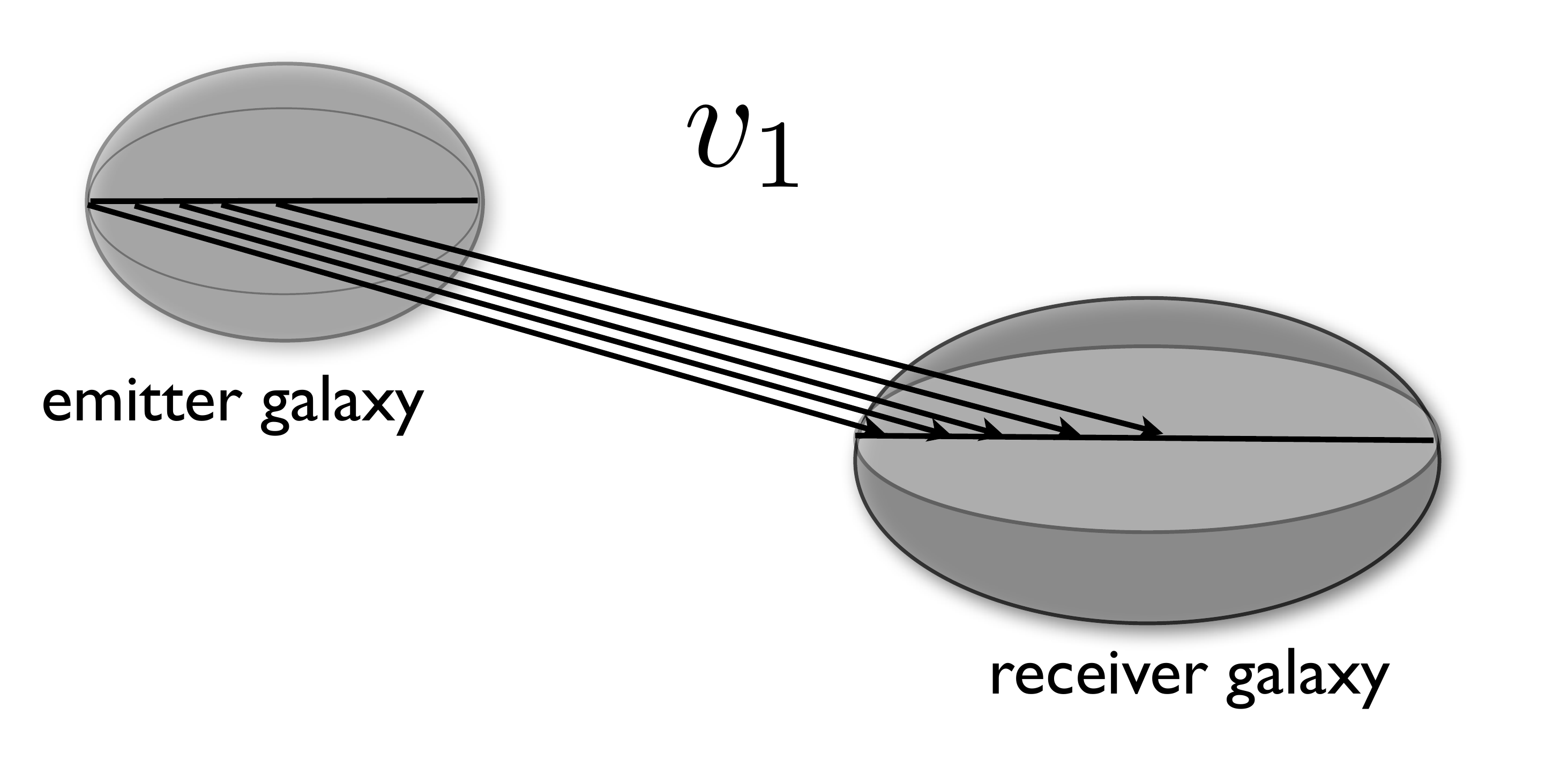} 
\caption{The respective small curvatures of two galaxies are mapped from emitting galaxy to receiving galaxy (Milky Way), as a function of radius.    This is done with $v_1$ using Lorentz-group kinematics.\label{v_1graphic}}
\includegraphics[scale=0.25]{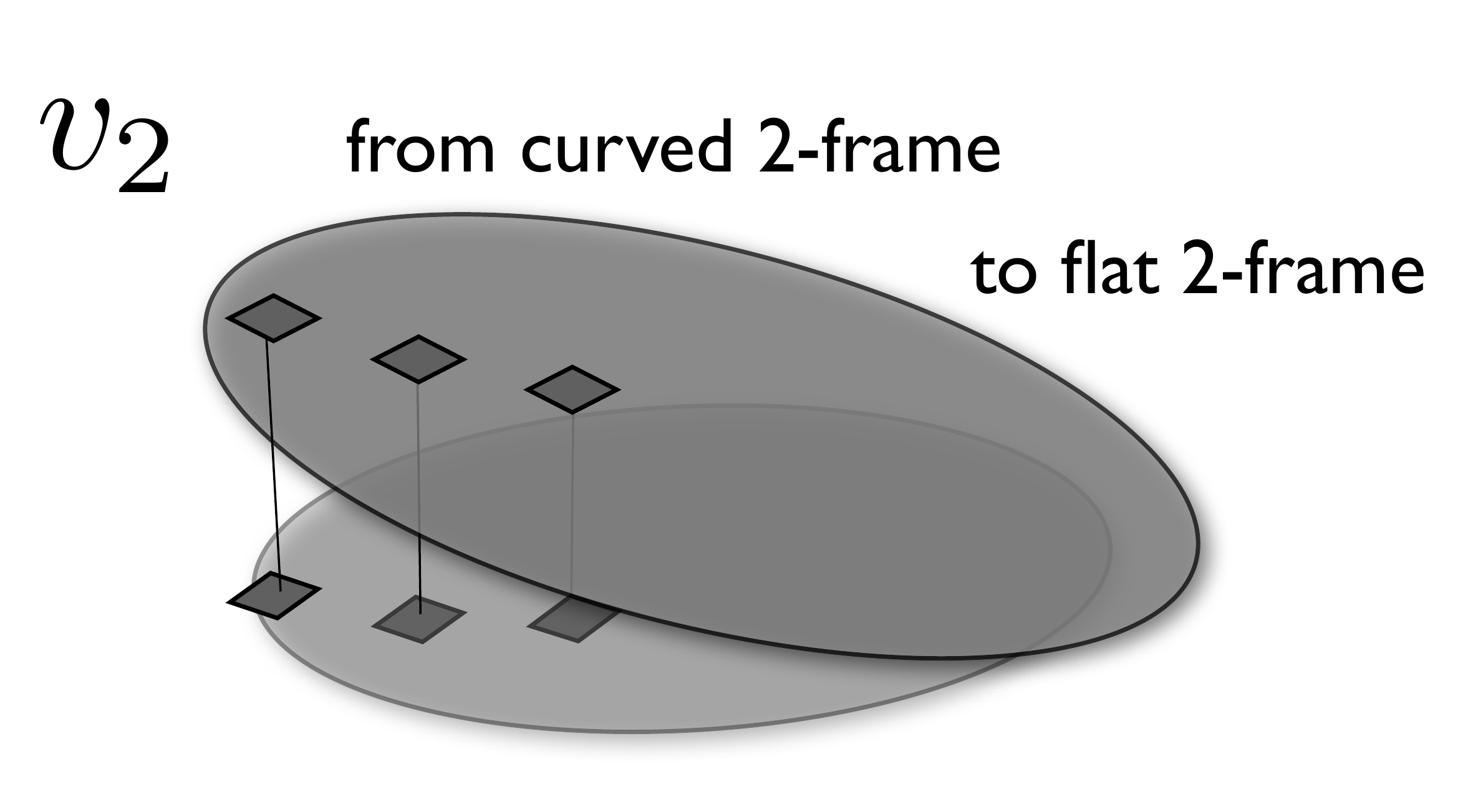} 
\caption{The  final step in adjusting the received frequency shifts to the frame of the observer is mapping the curved 2-frame   onto the flat 2-frame where observations are made.   This conversion is done with $v_2$ as a function of radius,  using Lorentz group kinematics.    \label{v_2graphic}}
\end{figure}

 The second LCM term, $v_2$, looks at the requisite transformation from the curved 2-frame  (indicated by  Eq.~\ref{eq:specific}) to the flat 2-frame where physical measurements 
 are made. Fig.~\ref{v_2graphic} gives a visual of    such a mapping.  The term  $v_2$ is also phrased in terms of an equivalent Doppler-shift  via Lorentz transformations.     That observations are always made in  flat frames is demonstrated  by the constancy of the local speed of 
light as measured by all observers.   

We define    the  flat frames by those   frequencies which wouldhave been measured if  the    Keplerian rotation curve velocities  $v_{l}(r)$ from the luminous mass in Eq.~\ref{eq:quadrature} were observed.    Keplerian velocities are calculated in a purely  Newtonian context   and  so describe  our best understanding  of the absence of all curvature.  The  shifted frequencies  $\omega_{l}(r)$ are defined by:
     \begin{equation}
 \frac{v_{l}(r)}{c}=
 \frac{
 \frac{\omega_{l}(r) }{\omega_{o}}
-  \frac{\omega_{o} }{\omega_{l}(r)}}{ 
  \frac{\omega_{l}(r)}{\omega_{o}}
+  \frac{\omega_{o} }{\omega_{l}(r)}} .
 \label{eq:MPflat}
\end{equation} 
for  the characteristic  frequency $\omega_o$ defined in Eq.~\ref{eq:dataLorentz}.   
Consistent with  in Eq.~\ref{eq:array}, the   flat 2-frame mapping  factor is:
 \begin{equation}
  e^{\xi_{f}}(r) =   \frac{\omega_{l} (r)}{\omega_{o}}.
    \label{eq:flatflat}
     \end{equation}
     
The $v_2$  mapping involves four frames instead of two, since the two curved frames and the two flat frames have already been mapped onto each other.   To generalize the mapping factor  used in the standard Lorentz boost formula (Eq.~\ref{eq:array})  for the  $v_2$ mapping factor, we write the ratio of the receiver 2-frame (flat) to emitter 2-frame (curved) as a square:   
\begin{equation}
 (e^{ \xi_2} )^2= \frac{e^{\xi_{f}}}{ e^{\xi_{c}}  }
 \label{eq:FCFtwo}
\end{equation} 
where  $e^{\xi_{f}}$   and $ e^{\xi_{c}}$  are functions of $r$. 

\begin{figure}
\includegraphics[scale=0.25]{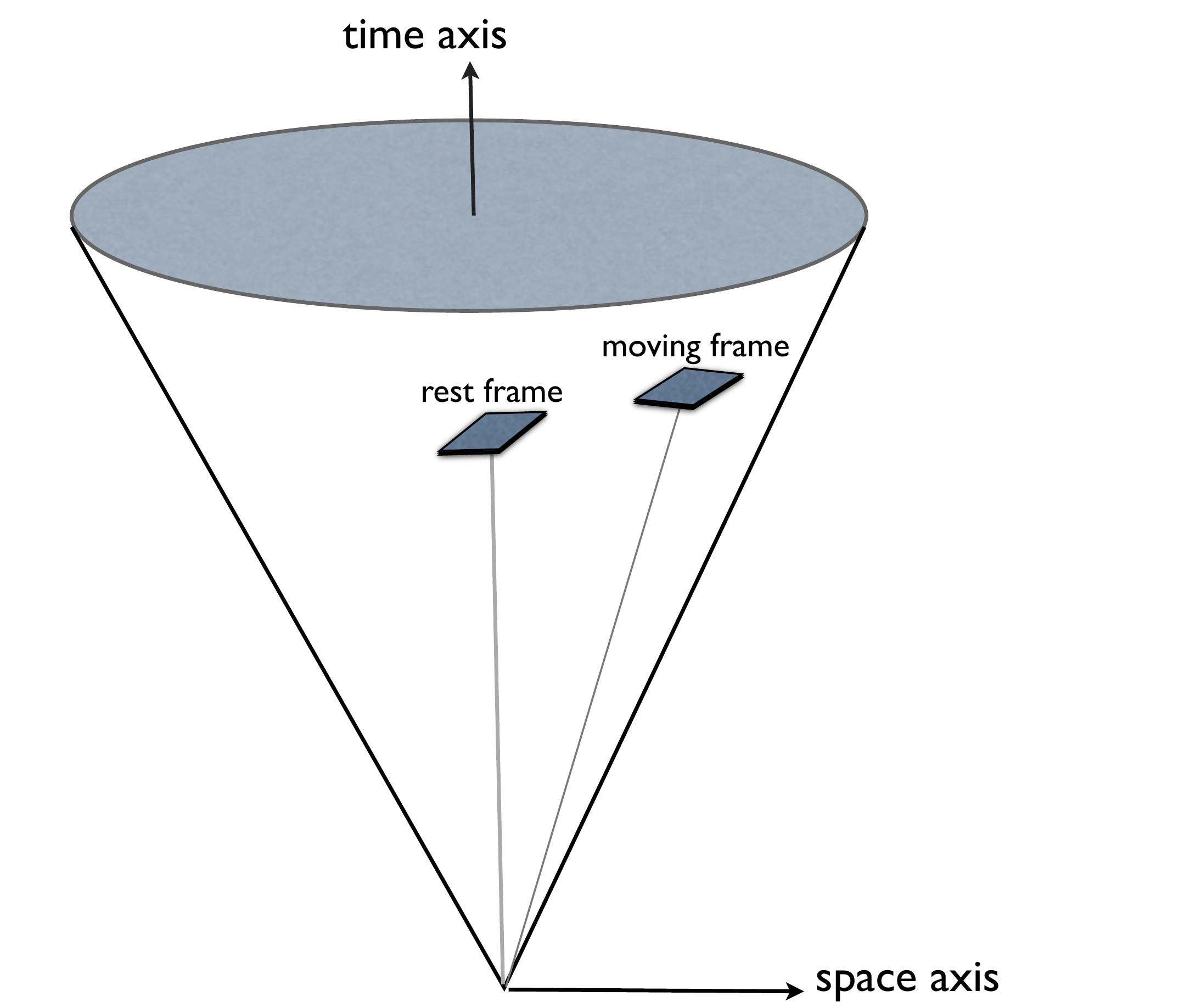} 
\caption{The identification of the rest frame and the moving frame in the light cone. \label{lightconegraphic}}
 \end{figure}

Before we   write  the final form of the  4-frame map, we   consider the idea of relative motion invoked in the Lorentz transformation.  In   Special Relativity,   Lorentz boosts are always defined as   positive rotations  away from the rest frame,  which is    
  aligned with the vertical  time axis of the light cone (see  Fig.~\ref{lightconegraphic}). 
  Since all frames are symmetric in Special Relativity, specific   identification of which frame is the  rest frame is   meaningless.   
However in the LCM we have included small deviations from flatness, and so we identify    the moving frame  with  the 
 curved 2-frame mapping factor $e^{\xi_{c}}$ and  the rest frame with the   flat 2-frame mapping factor  $e^{\xi_{f}}$. This identification is made based upon 
    the respective clocks (i.e. the coordinate times in each map), which  reflect  our concepts of  the moving and rest frames. 
 
Once this identification is made, we note that we are transforming  from the moving frame 
   to the rest frame --- effectively a  
   reverse boost.  We characterize   the reverse boost    by the   reciprocal    of the  Lorentz transformation    in Eq.~\ref{eq:LorentzDefine}: 
 \begin{equation}
\frac{v_2 }{c}= \coth \xi_2= \frac{e^{\xi_2 } +e^{-\xi_2 }}{e^{\xi_2 } - e^{-\xi_2 }}. 
\label{eq:hyperbolicreference}
\end{equation}  
Re-written for algebraic convenience,   to accommodate a 4-frame map, the final 
LCM mapping is:
 \begin{equation}
\frac{v_2 }{c}=  \frac{e^{ 2\xi_2  }+1}{e^{ 2\xi_2 } - 1}, 
\label{eq:hyperbolico}
\end{equation}  
 where   all quantities are     functions of radius except for the vacuum speed of light and the characteristic frequency $\omega_o$.   \\
 

  \section[]{ LCM Sample and  Results}
\label{sec:RESULTS}
In this section we   describe 
the four Milky Way luminous mass distributions  (section~\ref{MW}), the rotation curve fitting protocol  (section~\ref{fitting}),  the  sample  of emitter galaxies (section~\ref{sample}), 
 and   the  results of the LCM fits (section~\ref{results}). 
 
\subsection{Milky Way luminous mass models}
\label{MW}  
 
As previously describe, in the LCM  we map each emitter galaxy onto  the receiver galaxy (i.e. the Milky Way) to derive the relative curvature term $v_{lcm}$.   The luminous matter profile of the Milky Way is notoriously difficult to determine because we   are observing the  system from the inside (e.g.  interpreting  H\,{\sevensize\bf I}   outside the solar radius is particularly difficult~\citep{Car}).     We compare each of the  emitter galaxies to four $(4)$ different Milky Way   luminous mass models to accommodate for the variations in the Milky Way  luminous mass modeling.     

\begin{figure}
\includegraphics[scale=0.45]{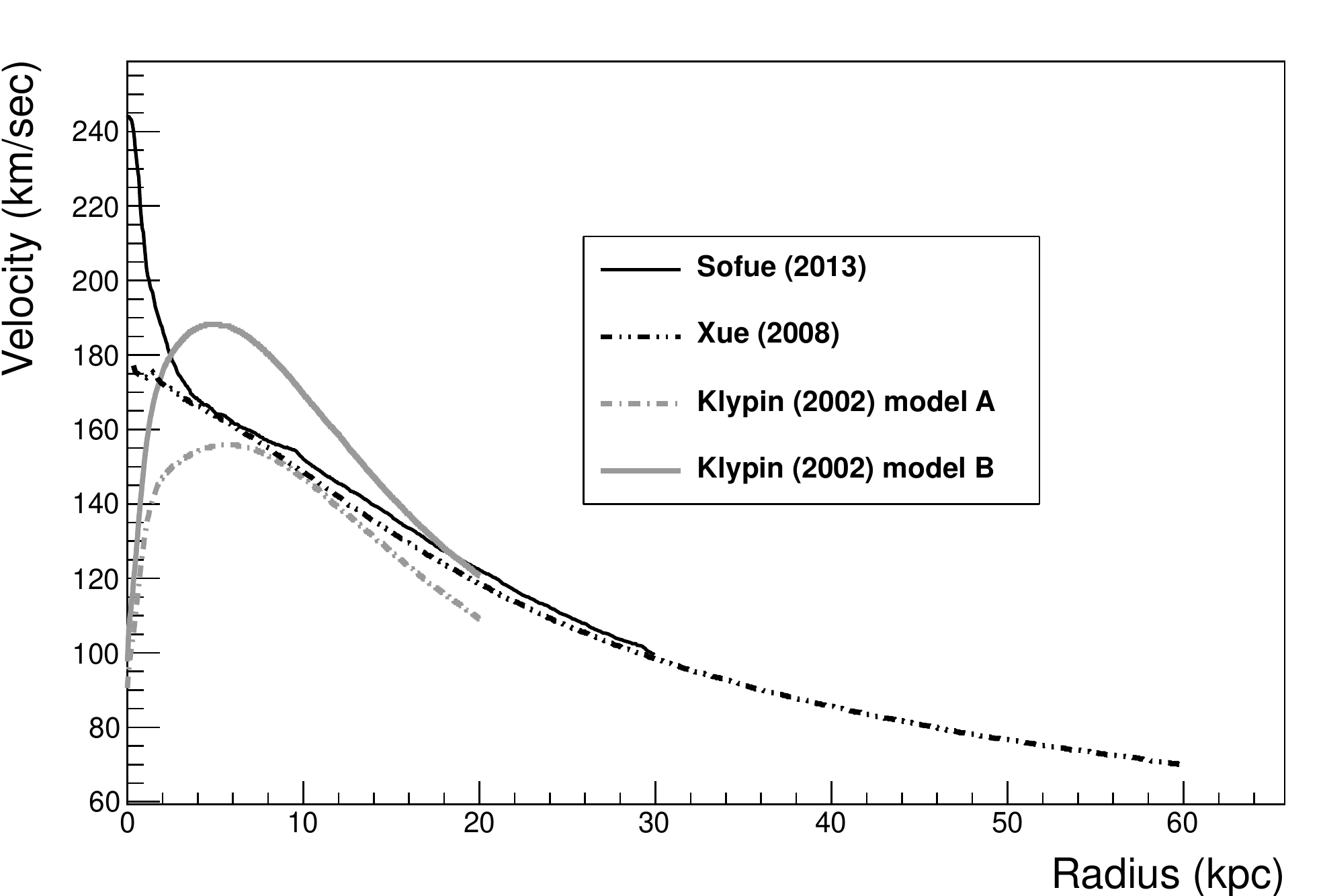} 
\caption{The Milky Way orbital velocities due to the luminous mass  profiles reported in each of the four references used in this work.\label{MWlum}}
\end{figure}

The four   models   differ in key features: inner rise rate, asymptotic velocity at large $r$, and coverage.  Mass components and total coverage   are described in  Table~\ref{tab:MWlum} and the total luminous profiles of each Milky Way model are plotted in Fig.~\ref{MWlum},  in terms of the resulting orbital velocities for a test particle. 

The Milky Ways from ~\citet{Sofue} and ~\citet{Xue}     differ in the inner  rise rates of the orbital velocities,  and   the  Keplerian fall-offs  at large $r$ differ only in magnitude.  The Sofue model has the largest    luminous mass of all four Milky Ways.   The two Milky Ways from ~\citet{Klypin} (models A and B) differ in that model B includes angular momentum sharing  for in-falling baryons to the disk, and so    the Keplerian rotation  curve from model B is globally higher than  model A.
  
\begin{table}
 \centering
 \begin{minipage}{140mm}
  \caption{Milky Way Luminous Mass Models \label{tab:MWlum}}
  \begin{tabular}{@{}llcc@{}}
  \hline
 Galaxy     	 &coverage &$M_{bulge} $&$M_{disk} $\\ 
 \hline
 Sofue 	 & 30 kpc 	& $1.8$	&$6.8$ \\
 Xue 		&60 kpc	&$1.5$	&$ 5.0$\\
 Klypin,  A	&15 kpc	& $0.8$	&$4.0$ \\
 Klypin,  B  &15 kpc	&$1.0$	&$5.0$\\
\hline 
\end{tabular}\\ 
 Masses in   units  of $M_{\sun} \times 10^{10}$
 \end{minipage}
\end{table}
\subsection{LCM rotation curve fitting protocol}
 \label{fitting}
In this paper we fit    the predicted LCM rotation curve velocity $v_{rot}$ (Eq.\ref{eq:zonteLCM}) to   the reported rotation curve data $v_{obs}$ (Eq.~\ref{eq:dataLorentz}).  The LCM prediction  and fit    are calculated using the MINUIT minimization software as implemented in the ROOT data-analysis package~\citep{ROOT}. 

The fit procedure is as follows:
\begin{enumerate}
 \item The   luminous mass components reported in each reference are   digitized using the software package Graph Click \citep{Graphclick};
 \item   The associated    Newtonian gravitational potential $\Phi$ is calculated for each component and the components are  summed (Eq~\ref{eq:potentialgeneral});
  \item We parametrize  the Schwarzschild metric as a function of radius and calculate the  convolution function,  $v_{lcm}$, for the selected Milky Way mass model, (Eq.~\ref{eq:domain})
  \item The  minimization procedure  explores the parameter space to find an optimal fit of $v_{rot}$  to    $v_{obs}$; the parameter space includes   the stellar disk and bulge M/L constrained (separately) to be between  $0.00$ and $5.00$,  reported gas fraction constrained  between $0.75$ and $1.5$,  and $\tilde{a}$ free;
   \item The  best-fit  values for: M/L (bulge and disk), gas scalings,    $\tilde{a}$,   and the associated  $\chi^2$ values for the fit are   reported in Tables~\ref{sumRESULTS} and \ref{sumRESULTSbigger}.
\end{enumerate}

The initial luminous mass profiles used in the LCM fits come from four different fitting models:  NFW, ISO, MOND and Maximum Light (MaxLight).  In all of these model fits   M/L was free,   with the notable exception of those galaxies sampled from Ref.~8 \citep{Blok}.    
   Comparisons between the  original fitting models'   reported M/L and the  $\chi^2$ values to  those of   the LCM fits  are reported   in Table~\ref{sumRESULTS}.  The NFW, ISO, MOND or  MaxLight  $\chi^2$ values reported here  have been recalculated using the same MINUIT software used for the LCM fits so that the  $\chi^2$ values can be compared appropriately. 
   
   The ranges used to  constrain  M/L  are motivated by  significant   variations for a given galaxy's luminous mass profile,  even when fit reports for the M/L are done with the same model, see Fig.~\ref{fig:massmodels18}.   Reported gas profiles      endure less  variation than stellar      M/L, but      still vary significantly for a  single system~\citep{Frat,Blok}. The  M/L reported in the literature, which are used as the LCM starting mass estimates,   are reported in Table~\ref{sumRESULTS}.  

\subsubsection{Error estimates and  reduced $\chi^2_r$ values}
All figures reported here indicate the uncertainties   as reported in the literature.
There is currently no  standard practice  as to how to quantify the uncertainties associated with   rotation curve data  ~\citep{Gent,Blok,JNav,San96},  such that resulting $\chi^2_r$ values can not be interpreted in a uniform way.  Often the  uncertainty comes from a fit of the  tilted ring model to the  H\,{\sevensize\bf I} velocity field.  In other analyses uncertainty estimates come from    differences between the approaching- and receding-side velocity fields  ~\citep{Blok,Gent,Toky} .  

 \subsection{The sample}
 \label{sample}
 The LCM sample reported in this paper  represents twenty-three $(23)$    galaxies,   in forty-two $(42)$ different  data sets.  The galaxies were randomly selected, with the only selection criteria  being some effort to represent both galaxies smaller and larger than the Milky Way and to compare different model-based assumptions of the luminous mass profiles. We report the   different  galaxies   in roughly decreasing order of size.   In  Table~\ref{sumRESULTS}   average results (over the   four $(4)$ different  Milky Way luminous mass models) are reported as well as the original    model contexts, M/L values,  associated wavelength bands,  and  $\chi^2$ values.   The  $\chi^2$ values reported here for the originating models (NFW, MOND, ISO or MaxLight, from which we use  the reported stellar and gas profiles as starting estimates for LCM   luminous profiles) have been calculated using the same points and in the same fitting code as the associated LCM fits.

In sections~\ref{Blok}-\ref{Bottema} we describe  some  features  and basic assumptions from some of the  references for the emitter galaxies.   

  \begin{table*}
 \centering
 \begin{minipage}{140mm}
  \caption{Summary  of results and M/L for originating   model \&  LCM   \label{sumRESULTS}}
  \begin{tabular}{@{}lllcccccccc@{}}
  \hline
   Galaxy     	  &Ref.~&  Band M/L 	&Model 	& \multicolumn{3}{c}{\underline{ Model Fit Results}}	& & \multicolumn{3}{c}{\underline{LCM Fit Results  (average)}}  \\
	     \hline
   		 &  	 & 			& 		& $\rmn{M/L}_{disk}$ &$\rmn{M/L}_{bulge}$	& $\chi^2_r$			& &$\rmn{M/L}_{disk}$&$\rmn{M/L}_{bulge}$&$\chi^2_r$ \\ 
 \hline
 F 563-1	&2	&I			&NFW	&2.3		&--		&0.05& 	&12.92	&--		&0.16 \\
F 563-1	&13	&B			&  NFW	&2.6		&--		&0.14& 	&4.40	&--		&0.22 \\
M 31 		& 12 	&B 			&ISO		&7.5		&3.5	 	&0.36 & 	&13.0  	&2.17		&0.30  \\
M 31 		& 15 	&R 			&NFW 	&0.93	&3.0	 	&13.45& 	&1.26	&2.10	&10.6  \\
M 33		& 5	 &	K 		& NFW	&0.70 	 & -- 		&2.46&  	 &0.46	&-- 		& 0.84 \\
M 33		&6  	& 	B		&MIdm$^{a}$&0.80 	& -- 		&0.16&  	& 0.70	&-- 		&0.18 \\
NGC 891* 	&11 	 & 3.6$\,\umu$m&MOND   	& 0.5		& 2.0		&5.30  & 	 &0.92	&  1.40 	&1.06  \\
NGC 891	& 11 	 &3.6$\,\umu$m &MaxLight&0.9 	&1.64 	&1.10  & 	 & 1.03  	&1.49	&1.15 \\ 
NGC 925*	&3	&3.6$\,\umu$m &NFW	&0.65	&--		&2.94& 	&0.36	&--		&6.45 \\
NGC 925*	&3	&3.6$\,\umu$m &ISO	&0.18	&--		&2.40& 	&0.003	&--		&3.81 \\
NGC 2403	 & 3	& 3.6$\,\umu$m& NFW	&0.41 	& -- 		&4.56& 	&0.41 	&-- 		& 0.88 \\
NGC 2403	 &9	 & 	 B		& MOND	&1.60 	& -- 		&3.67 & 	&0.99 	&-- 		& 2.30 \\
NGC 2841 &3	 & 3.6$\,\umu$m&  NFW	&0.74 	& 0.84	&0.45 & 	& 0.99 	&1.29 	&0.44 \\
NGC 2841*& 8	 &3.6$\,\umu$m&MOND  	&0.89 	&1.04	&0.87& 	& 0.84	&1.09	 & 0.43\\
NGC 2841 &9	&	B 		& MOND	&8.3 		& 0.83	 &34.7& 	& 5.10 	&2.91 	& 1.94 \\  
NGC 2903 	&3	 & 3.6$\,\umu$m&NFW   	&0.61	&1.3		&0.55& 	 & 2.07	&0.65	&0.75\\
NGC 2903 	&8	 &3.6$\,\umu$m&MOND 	&1.71		& -- 		& 0.61& 	 & 2.03 	&-- 		&0.76\\
NGC 2903 	&10	 &B	   		&MOND   	&3.60 	& -- 		&10.71& 	 & 2.94	&-- 		&2.10\\
NGC 3198 & 1 	 &r			&NFW	& 3.8		&--		&  1.34 & 	&3.26 	& --		 &1.65 \\
 NGC 3198 & 2 	&r			&NFW	& 1.4 		&-- 		& 0.67  &   	 &2.87 	& -- 		&1.38  \\
 NGC 3198 & 3 	&3.6$\,\umu$m&NFW	&0.80	& --  		&5.40  &   	 &0.64 	& -- 		&4.61   \\
NGC 3198	& 4	&3.6$\,\umu$m&MOND	& 0.33	&  --		&3.50  & 	& 0.61 	& --	 	&3.39  \\
NGC 3521 & 8	 &3.6$\,\umu$m &MOND	&0.71 	&-- 		&0.97 & 	 & 0.72   	&--		&0.80  \\
NGC 3726	& 10	&B 			&MOND	&1.00	&--		&3.57& 	&1.15		&--		&3.12 \\
NGC 3953	& 10	&B 			&MOND	&2.7		&-- 		&1.35& 	&1.72		&--		&0.62 \\
NGC 3992	& 10	&B 			&MOND	&4.93	&-- 		&0.50& 	&4.04 	&--		&0.37 \\
NGC 4088	& 10	&B 			&MOND	&1.16		&-- 		&1.70& 	&1.64 	&--		&1.39 \\
NGC 4138	& 10	&B 			&MOND	&3.5		&-- 		&2.12& 	&3.21	&--		&2.34 \\
NGC 5055	 & 3	&3.6$\,\umu$m &  NFW	&0.79	&0.11 	&17.23&  	&0.49	&0.44	&2.85 \\
NGC 5055	 &7	&B			&  NFW	&3.20 	& -- 		& 3.27 &  	& 2.51 	&-- 		&2.55 \\
NGC 5055	&  8	&3.6$\,\umu$m	 & MOND	&0.55 	&0.56 	&1.12& 	& 0.56 	&0.36 	&1.16\\
NGC 5533 	& 10 	&B 			&MOND	&0.6		&7.2	 	&1.57 & 	&3.00 	&6.67	&0.73  \\
NGC 5907	& 10	&B 			&MOND	&1.6 		&6.8 	 	&0.44& 	&1.41		&4.29	&0.24 \\
NGC 6946	& 3	 &  3.6$\,\umu$m&NFW	&0.64  	& 1.00	& 4.34 & 	& 0.35  	&0.66 	& 1.42  \\
NGC 6946	&  10	 & 	 	B	&  MOND	& 0.5		&   -- 		&   3.03& 	&0.70  	&--		& 0.46  \\ 
NGC 7331	&8	 & 3.6$\,\umu$m&  MOND	&0.4 		&1.22	& 0.45& 	& 0.40 	&0.95	&0.27 \\
NGC 7331*& 9	 & 	 B		& MOND	&2.0 		&  1.8  	&9.32 & 	& 1.41 	&1.29	&4.47 \\
NGC 7793	&8	&3.6$\,\umu$m	&MOND	&0.28	&--		&4.61& 	&0.13	&--		&7.99 \\
NGC 7793*&14	&B			&ISO		&2.6		&--		&1.08& 	&2.55	&--		&0.59 \\
NGC 7814 	&11 	 & 3.6$\,\umu$m&ISO   	& 0.68  	&0.71		& 0.25& 	 &3.17  	&0.54   	&0.44 \\ 
NGC 7814* & 11	 & 3.6$\,\umu$m &MaxLight&9.25  	&0.64 	&6.11  & 	 & 3.81  	&0.56 	& 0.50 \\ 
UGC 6973	& 10	&B 			&MOND	&2.7		&-- 		&23.5 & 	&1.49  	&--		&0.07 \\
\hline 
\end{tabular}
 {\bf Notes:}\\
     Values in this table  are   averaged over   the   four emitter-receiver (Milky Way) galaxy  pairings. Galaxies for which one LCM fit-pair failed is indicated by (*).   MIdm$^{a}$ is a model independent dark matter fit.
 Mass-to-light ratios   in units of $M_{\sun} / L_{\sun}$.   Reduced $\chi^2$ per degree of freedom indicated by  $ \chi^2_r$. \\
{\bf References:} 
 1.~\citet{Bege}, 2.~\citet{JNav}, 3.~\citet{Blok} , 4.~\citet{Maria}, 
5.~\citet{Cor03}, 6.~\cite{CoSa00},   7.~\citet{Batt},   8.~\citet{Gent},   9.~\citet{Bot},   10.~\citet{SanMcGa},
  11.~\citet{Frat},   12.~\citet{Car},   13.~\citet{giraud2000universal},   14.~\citet{Dicaire}, 15.~\citet{Klypin}. \\
    \end{minipage}
\end{table*}

\subsubsection{Ref.~ 3 parameters    ~\citet{Blok}, NFW  }
\label{Blok}

From  Ref.~3 ~\citep{Blok}  we sample seven galaxies: NGC 3198, NGC 5055, NGC 2403, NGC 2841, NGC 6946, NGC 2903 and NGC 925. These galaxies are   reported in the context of 
NFW fits with fixed M/L,  consistent with   the PSM of ~\citet{BelldYong} with a
 Salpeter ~\citep{Salpeter}  initial mass function. For NGC 925 we  sample both the NFW and the  ISO fits, both with M/L fixed.
 
  The Ref.~3 rotation curve data comes from the 
 H\,{\sevensize\bf I}\, THINGS survey ~\citep{Walter}.  The THINGS data is generally considered to be the highest resolution rotation curve data  available. The Ref.~3 sample is chosen  based on favorable inclination angles and regular rotations.   Rotation curve derivations are based on    tilted-ring rotation curve analysis, fit with third-order Gauss-Hermite   polynomials.     Gas estimates are based on  a   thin gas disk  geometry from   the   H\,{\sevensize\bf I}\,  flux maps reported by  ~\citet{Walter}, scaled by $1.4$ to account for helium and metals.   Molecular gas is included as a slight increase in the M/L  of the stellar disk.   Stellar M/L   are reported in   the $3.6\,\umu$m  band to avoid the uncertainties prevalent in  B-band from  hot dust and young stars.    Ref.~3 decomposes the stellar light  into a bulge and disk with a double exponential disk model, avoiding the canonical $R^{1/4}$ bulge distribution.     
 
\subsubsection{ Ref.~8 parameters   ~\citet{Gent}, MOND}
\label{TaM}

 From Ref.~8~\citep{Gent}  we sample  six galaxies: NGC 2903,   NGC 2841, NGC 5055, NGC  7331, NGC 7793 \&  NGC 3521.  These galaxies are reported in the context of 
MOND fits and all  galaxy rotation curves  come from Ref.~3~\citep{Blok}.  Ref.~8  uses  the Ref.~3 geometry and gas assumptions (above),  with    stellar  M/L of the disk and  bulge    free,  distances    constrained,  and   the    $\umu$ ``simple'' interpolating function.  They report  M/L    in the $3.6\,\umu$m   band. 

Of those galaxies sampled here, they report that all are consistent with the PSMs of ~\citet{BelldYong} with the exception of  NGC 2903 and NGC 7331.      As evidenced in Fig.~\ref{fig:massmodels18},  while NGC 2841 velocities associated with the luminous profiles reported in Ref.~3 and Ref.~8 differ by approximately $30\,  \rmn{km\,  s^{-1}}$  and those for  NGC 5055 by about $60\,  \rmn{km\,  s^{-1}}$, both sets of luminous models are reported as consistent with the same PSMs, demonstrating the degree of freedom within current PSMs. 

\subsubsection{Ref.~10 parameters   ~\citet{SanMcGa}, MOND}
\label{SanMc02}

 From Ref.~10~\citep{SanMcGa}  we sample  ten galaxies: NGC 5533, NGC 4138, NGC 5907,  NGC 3992, NGC  2903, NGC 3953,  UGC 6973, NGC 4088,  NGC 3726,  and  NGC 6946. These galaxies are
  reported in the context of  MOND fits,   at Ursa Major distances. In Ref.~10 the distance to the Ursa Major system is   reported as $15.5$ Mpc, but Cepheid based recalibration indicates $18.6$ Mpc~\citep{Tully}. 
  They model the luminous mass with   thin   stellar and gaseous disks  and a spheroidal stellar bulge, for M/L free.    Most systems are treated with coplanar circular motions, but, when appropriate, complexities for   bars and warping of the gas layer are   modeled.     M/L  is reported in the  B-band.  Neutral hydrogen is scaled by $1.4$ to account for primordial helium.   

\subsubsection{Ref.~9 parameters    ~\citet{Bot}, MOND}
\label{Bottema}

 From ~\citet{Bot} we sample  three galaxies: NGC 2841, NGC 7331, and NGC 2403.  These galaxies are reported in the context of  MOND fits at Cepheid distances. The only geometry assumption they report is a thing stellar disk.      M/L    is free and  reported in the  B-band. Ref.~9    gas is reported from a surface density distribution equal to the observed  H\,{\sevensize\bf I}   scaled by a factor of $1.3$ to account for primordial helium.    The contribution of the gas to the rotation curve is fixed and does not depend on the distance to the galaxy.  

 \subsection{Results}
\label{results}
In this section we make notes on individual galaxies and groups of galaxies with interesting features that   constrain    PSM and imply an interpretation of the  LCM free parameter $\tilde{a}$. 

In Table~\ref{sumRESULTS} we   report the LCM average results for M/L and reduced $\chi^2_r$  values from the LCM fits plus a    comparison  to     fits using the originating models (NFW, ISO, MOND, or  MaxLight).  
 Individual emitter galaxy fit results for  each Milky Way luminous mass model are reported in  Table~\ref{sumRESULTSbigger}.    LCM fits for seven ($7$) out of   $165$  emitter-receiver galaxy pairings failed; these failures  indicate problems with the  initial  luminous mass profiles.
     
Figs.~\ref{fig:resultssmall}-\ref{galaxiesSmallest} show   LCM fits  using  the ~\citet{Xue} Milky Way.   We selected the Xue   Milky Way for the figures because it has the largest   radial coverage.  In the one case where the LCM fit with the Xue Milky Way failed, we   instead show   the  ~\citet{Sofue} Milky Way pairing, as indicated in the figure caption.

\begin{figure}
\includegraphics[scale=0.45]{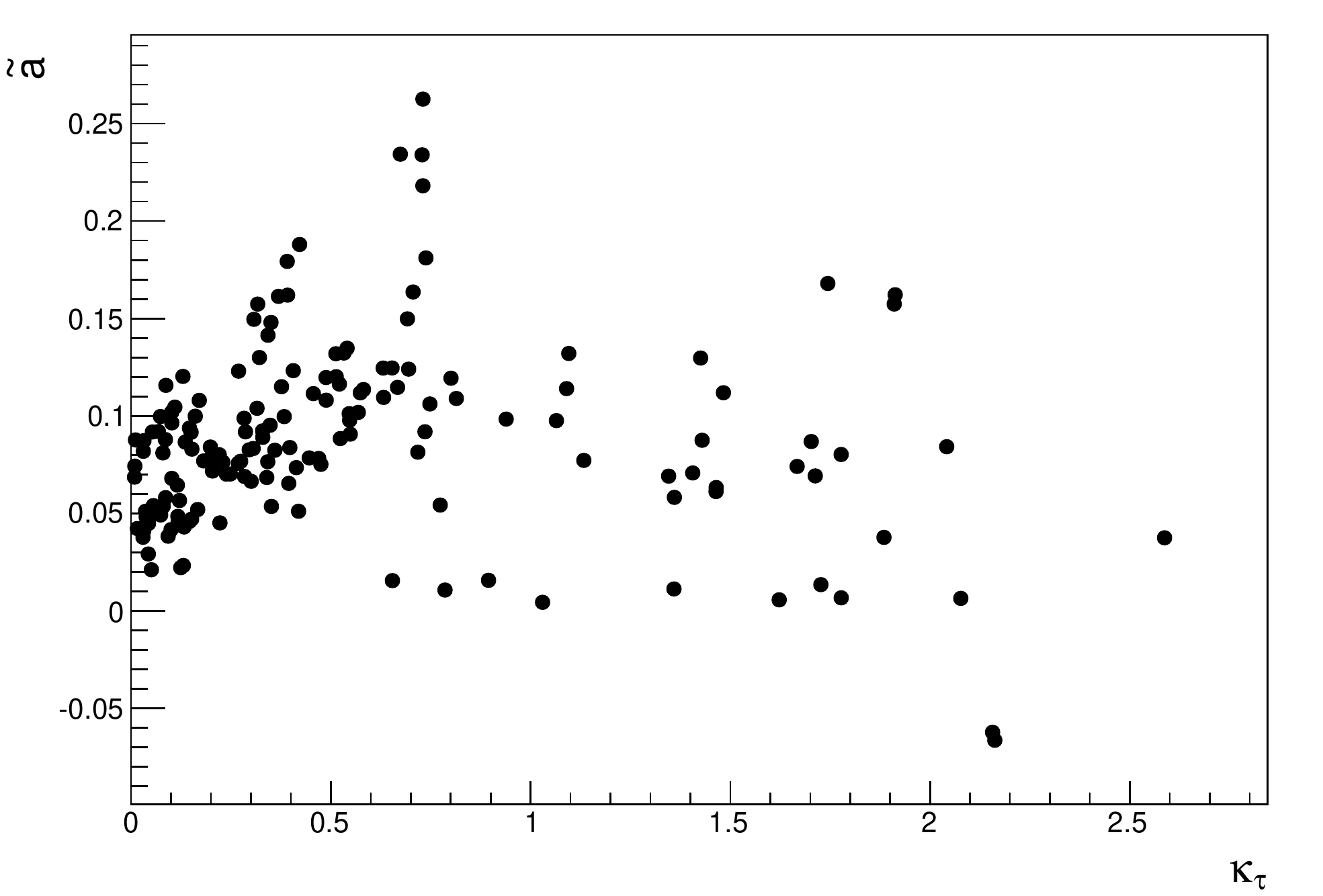} 
\caption{Free parameter $\tilde{a}$ versus $\kappa_\tau$. Each dot represents one 
emitter/receiver galaxy pair.    Mean value for the sample  presented here $\tilde{a}=0.09\pm0.05$.   \label{fig:alphaDistrib1}} 
\end{figure}

     The distribution of values for $\tilde{a}$, 
      has a mean  value of  $\tilde{a}=0.09\pm0.05$ and does not exhibit an obvious dependence on the relative curvature ratio, $\kappa_\tau$, as indicated in Fig.~\ref{fig:alphaDistrib1}.   In section~\ref{n891n7814} we present a  conjecture regarding a  possible physical correlation between $\tilde{a}$ and   density gradients in the luminous matter.

  \begin{table*}
 \centering
 \begin{minipage}{160mm}
  \caption{LCM emitter-galaxy pairs Results \label{sumRESULTSbigger}}
  \begin{tabular}{@{}llcccccc@{}}
  \hline
emitter&receiver Milky Way   &$\kappa_\tau$&$\tilde{a}$	&gas  scaling	& $ \rmn{M/L}_{disk}$&$ \rmn{M/L}_{bulge}$&$\chi^2_{R}$   \\
  \hline
  F 563-1  Ref.~2	&Sofue		&0.04	&0.03	&1.5		&9.59 	&--	&0.15 \\
			&Klypin A		&0.07	&0.06	&1.5		&9.37 	&--	&0.16\\
			&Klypin B		&0.08	&0.05	&1.5		&9.45 	&--	&0.16 \\
			&Xue		&0.10	&0.07	&1.5		&9.33	 &--	&0.16   \\
 F 563-1  Ref.~13	&Sofue		&0.05  	&0.02	 &1.5		&4.48	&--	&0.21\\
			&Klypin A		&.10 		&0.04 	&1.5		&4.36	&--	&0.22\\
			&Klypin B		&0.09  	&0.04	 &1.5		&4.40	&--	&0.22\\
			&Xue		&0.11		 &0.05	 &1.5		&4.34	&--	&0.22 \\
M 31  Ref.~12 		&Sofue		&0.89	&0.02	& 0.75	& 8.98	&1.88 	&0.30 \\
				&Klypin A		&2.16	&-0.07	&0.75	&17.90	&1.94	&0.28 \\
				&Klypin B 		&2.16	&-0.06	&0.75	&17.67	&2.22	&0.28  \\
				&Xue		&2.59	&0.04	&0.75	&7.41		&2.63   	&0.33 \\
M 31  Ref.~15		&Sofue		&0.94	&0.10	&--		&0.55 	&3.18		&9.52 \\
				&Klypin A		&1.88 &0.04	&--		&2.02		&2.44 	&13.8  \\
				&Klypin B 		&1.10	&0.13 &--		&1.61		&1.39 	&9.48   \\
				&Xue			&1.43	&0.13	&--		&0.85		&1.38 	  &9.53\\
M 33 Ref.~5		&Sofue		&0.04	&0.04	&	1.5	&0.47	&--	&0.78\\
			&Klypin A		&0.09	&0.09	&	1.5	&0.45  	&--	&0.87 \\
			&Klypin B		&0.08	&0.08	&	1.5	&0.46  	&--	&0.87\\
			&Xue		&0.10	&0.10	&	1.5	&0.46  	&--	&0.83\\
M 33  Ref.~6	&Sofue		&0.04	&0.05	&0. 89	&0.72  	&--	&0.19\\
			&Klypin A		&0.07	&0.10	&1.00	&0.70	&--	&0.18\\
			&Klypin B		&0.07	&0.09	&1.02	&0.70 	&--	&0.18\\
			&Xue		&0.09	&0.12	&1.05		& 0.69	&--	&0.18\\
NGC 891 Ref.~11  \footnote{Citation's  M/L assumptions from MOND fit}
 				&Sofue		& 0.65	&0.02	&0.75	&0.86	&1.24	&0.95 \\
				&Klypin A* 	&--		&	--	&	--	& --		&--		&--   \\
				&Klypin B		&1.36	&0.01	&0.75	&0.93	&1.40	&1.06\\
				&Xue		&1.73		&0.01	&0.75	&0.92	&1.39	&1.06 \\
 NGC 891 Ref.~11 \footnote{Citation's  M/L assumptions from  MaxLight fit}
				&Sofue		& 0.79	&0.01		&0.75 	&0.98	&1.37		&1.08\\
				&Klypin A		&1.78		&0.01		&0.75	&1.04	&1.55		&1.16\\
				&Klypin B		&1.62	&0.01		&0.75	&1.05		&1.53		& 1.17\\
				&Xue		&2.08	&0.01		&0.75	&1.04	&1.53		&1.17 \\
NGC 925 Ref.~3\footnote{original citation  M/L assumptions: NFW}	&Sofue*		
							&--		&--		&--		&--		&--	&--\\
			&Klypin A		&0.03	& 0.04	& 1.5		& 0.36 	&--	&6.44 \\
			&Klypin B		&0.03	&0.04	&1.5		&0.37	&--	&6.43\\
			&Xue		&0.04	&0.05	&1.5		& 0.36	&--	&6.48	 \\
NGC 925 Ref.~3\footnote{original citation  M/L assumptions: ISO}	&Sofue*		&--		
									&--		&--		&--		&--	&--\\
  			&Klypin A		&0.01	&0.07 	& 1.5		&0.003 	&--	&3.80 \\
			&Klypin B		&0.01	&0.07	&1.5		&0.003	&--	&3.80\\
			&Xue		&0.01 	&0.09	&1.5		&0.003	&--	&3.83	 \\
NGC 2403 Ref.~3&Sofue		&0.06	&0.05	& 1.5		&0. 42	&--	&0.87\\
			&Klypin A		&0.11		&0.10	&1.5		&0.40	&--	&0.88 \\
			&Klypin B		&0.10	&0.10	&1.5		& 0.41	&--	&0.88\\
			&Xue		&0.13	&0.12	&1.5		& 0.41	&--	&0.88\\
 NGC 2403  Ref.~9&Sofue		& 0.07	&0.05	&	1.5	& 1.01	&--	&2.24\\
			&Klypin A		&0.15		&0.09	&	1.5	& 0.98	&--	&2.30\\
			&Klypin B		&0.14	&0.09	&	1.5	&0. 99	&--	&2.30\\
			&Xue		&0.17		&0.11		&	1.5	& 0.98	&--	&2.35 \\
\hline
\end{tabular}\\
  References as in Table~\ref{sumRESULTS}.  Mass-to-light ratios   in units of $M_{\sun} / L_{\sun}$.  
Milky Way: ~\citet{Sofue}, ~\citet{Klypin},   \& ~\citet{Xue}.  The quantity $\kappa_\tau$ is defined in section~\ref{kappa}.
(*) indicates an   LCM fit which failed.  
 \end{minipage}
\end{table*}

\begin{table*}
\addtocounter{table}{-1} 
 \centering
 \begin{minipage}{160mm}
  \caption{Continued: LCM emitter-receiver galaxy pairs}
  \begin{tabular}{@{}llcccccc@{}}
  \hline
 emitter 		&receiver Milky Way&$\kappa_\tau$&$\tilde{a}$	&gas  scaling	& $ \rmn{M/L}_{disk}$&$\rmn{M/L}_{bulge}$&$\chi^2_{R}$   \\
  \hline 
  NGC 2841 Ref.~3	 	&Sofue		&1.13		&0.08		&0.75	& 1.40	& 2.50	&0.48\\
				&Klypin A		&0.73	&0.26		&1.5		& 0.63	& 0.60	&0.19 \\
				&Klypin B		&0.73	&0.23		&1.5		& 0.67	& 0.69	&0.16\\
				&Xue		&1.91		&0.16		&1.5		& 1.24	& 1.36	&0.96\\
NGC 2841 Ref.~8 	 	&Sofue		& 0.51	&0.13		&1.50		&0.76	&1.12		&0.10\\
				&Klypin A*		&--		&--			&--		&--		&--		&--	\\
				&Klypin B		&0.67	&0.23		&1.50		&0.55	&0.95	&0.10\\
				&Xue		&1.74		&0.17			&1.50		&1.20	&1.21		&1.10\\
NGC 2841 Ref.~ 9		&Sofue		&0.58	&0.11			&1.50		& 3.73	&2.56	&0.28  \\
				&Klypin A		&1.46	&0.06		&0.75	& 5.46	&3.13	&1.28  \\
				&Klypin B		&1.36	&0.06		&0.75	& 5.53	&3.17		&1.36\\
				&Xue		&1.91		&0.16		&1.50		& 5.70	&2.80	&4.82\\
   NGC 2903 Ref.~3 &Sofue	&0.28	&0.07	&	1.5	&2.27	&0		&1.14\\
			&Klypin A	&0.49	&0.12	&	1.5	&2.00	&0		&0.72 \\
			&Klypin B	&0.46	&0.11		&	1.5	&2.04	&0		&0.74\\
			&Xue	&1.06	&0.10	&	0.75	&1.97		&2.57	&0.38\\ 
NGC 2903 Ref.~8&Sofue	& 0.24	&0.07	&	1.5	&2.17		&--		&1.02\\
			&Klypin A	&0.41	&0.12	&	1.5	&1.94	&--		&0.53\\
			&Klypin B	&0.38	&0.12	&	1.5	&1.96	&--		&0.54\\
			&Xue	&0.53	&0.13	&	1.5	&2.07	&--		&0.95\\
NGC 2903 Ref.~10&Sofue	&0.30	&0.07	&	1.5	&3.16	&--		&2.99   \\
			&Klypin A	&0.52	&0.12	&	1.5	& 2.78	&--		&1.00   \\
			&Klypin B	&0.49	&0.11		&	1.5	 & 2.85	&--		&1.04\\
			&Xue	&0.65	&0.12	&	1.5	 &2.95 	&--		&3.39 \\			
 NGC 3198 Ref.~1&Sofue		&0.15		&0.05	&1.5		&3.34	&--&1.45\\
			&Klypin A		&0.27	&0.08	&1.5		&3.21	&--&1.83 \\
			&Klypin B		&0.25	&0.07	&1.5		&3.24	&--&1.84 \\
			&Xue		&0.33	&0.09	&1.5		&3.24	&--&1.47\\
 NGC 3198  Ref.~2&Sofue		& 0.15	&0.05	&1.5		&2.91	&--&1.27 \\
			&Klypin A		&0.30	&0.08	&1.5		&2.84	&--&1.42\\
			&Klypin B		&0.27	&0.08	&1.5		&2.87	&--&1.43\\
			&Xue		&0.35	&0.10	&1.5		&2.84	&--&1.39 \\
 NGC 3198 Ref.~3&Sofue		&0.12	&0.05	&1.5		&0.73	&--&3.94\\
			&Klypin A		&0.16	&0.10	&1.5		&0.53	&--&5.44 \\
			&Klypin B		&0.15		&0.09	&1.5		&0.55	&--&5.35 \\
			&Xue		&0.29	&0.09	&1.5		&0.75	&--&3.69 \\
 NGC 3198  Ref.~4&Sofue		& 0.13	&0.04	&1.5		&0.65	&--&2.94 \\
			&Klypin A		&0.22	&0.08	&1.5		&0.57	&--&4.36\\
			&Klypin B		&0.20	&0.07	&1.5		&0.58	&--&4.31\\
			&Xue		&0.32	&0.10	&1.5		&0.64	&--&1.95 \\
  NGC 3521 Ref.~8 & Sofue	&0.42	&0.05 	&1.50		&0.79	&--		&0.68\\
			&Klypin A	&0.74	&0.09 	&1.50 	&0.69	&--		&0.85\\
			&Klypin B	&0.72	&0.08 	& 1.50	&0.74	&--		&0.86\\
			&Xue	&0.81 	&0.11		&1.50		&0.65	&--		&0.83\\ 
NGC 3726 Ref.~10&Sofue	&0.12	&0.06	&0.75	&1.15		&--&2.40 \\
			&Klypin A	&0.23	&0.08	&1.5		& 1.17 	&--&3.86\\
			&Klypin B	&0.20	&0.08	&0.75	& 1.17	&--&3.46\\
			&Xue	&0.27	&0.12	&1.50		& 1.10	&--&2.76\\
NGC 3953 Ref.~10&Sofue	& 0.20	&0.08	&0.75	& 1.86	&--		&0.63   \\
			&Klypin A	&0.37	&0.16	&0.75	& 1.68	&--		&0.61  \\
			&Klypin B	&0.35	&0.15		&0.75	& 1.73	&--		&0.62 \\
			&Xue	&0.42	&0.19	&0.75	& 1.63	&--		&0.61 \\
NGC 3992 Ref.~10&Sofue	& 0.36	&	0.08	&	1.00	& 4.52	&	--	&0.29 \\
			&Klypin A	&0.54	&	0.13	&	0.75	& 3.86	&	--	&0.45\\
			&Klypin B	&0.51		&	0.12	&	0.75	& 3.99	&	--	&0.48\\
			&Xue	&0.71		&	0.16	&	1.50	& 3.79	&	--	&0.27\\
\hline
\end{tabular}\\
  References as in Table~\ref{sumRESULTS}.  Mass-to-light ratios   in units of $M_{\sun} / L_{\sun}$.   Milky Way: ~\citet{Sofue}, ~\citet{Klypin},   \& ~\citet{Xue}.  The quantity $\kappa_\tau$ is defined in section~\ref{kappa}.   
(*) indicates an  LCM fit which failed. 
 \end{minipage}
\end{table*}

\begin{table*}
\addtocounter{table}{-1} 
 \centering
 \begin{minipage}{160mm}
  \caption{Continued:LCM emitter-receiver galaxy pairs}
  \begin{tabular}{@{}llcccccc@{}}
  \hline
emitter&receiver Milky Way   &$\kappa_\tau$&$\tilde{a}$	&gas  scaling	& $ \rmn{M/L}_{disk}$&$ \rmn{M/L}_{bulge}$&$\chi^2_{R}$   \\
\hline
NGC 4088 Ref.~10&Sofue	&0.17		&0.05	&0.75 	&1.15		&--		&1.38 \\
			&Klypin A	&0.33	&0.09	&1.11		&1.10		&--		&1.39\\
			&Klypin B	&0.31	&0.08	&1.06	&1.12		&--		&1.40\\
			&Xue	&0.38	&0.10	&1.20	& 1.09	&--		&1.40\\
 NGC 4138 Ref.~10&Sofue	&0.22	&0.05	&1.50		&3.24	&--		&2.40 \\
			&Klypin A	&0.45	&0.08	&1.50		&3.20	&--		&2.26 \\
			&Klypin B	&0.41	&0.07 	& 1.50	&3.23	&--		&2.34\\
			&Xue	&0.52	&0.09 	& 1.50	&3.19	&--		&2.31\\
NGC 5055 Ref.~3	&Sofue	& 0.34	&0.07	&0.90  	& 0.39	&4.83	&1.46\\
			&Klypin A	&1.03 	&0.004	&1.50		&0.66	&6.50	&5.21 \\
			&Klypin B	&0.55	&0.10	&0.75	&0.35 	&4.40	& 1.87\\
			&Xue* 	&-- 		&	--	&--		&--		&--		&--\\
 NGC 5055 Ref.~7 &Sofue	&0.35	&0.05	&1.5		&2.77 	&--		&2.93\\
			&Klypin A	&0.57	&0.10	&1.5		&2.31	&--		&1.76\\
			&Klypin B	&0.55	&0.09	&1.5		&2.43 	&--		&1.68\\
			&Xue	&0.75	&0.11		&1.5		&2.51 	&--		&3.83\\
NGC 5055 Ref.~8	&Sofue	& 0.39	&0.07	&	1.5	&0.62	& 0.42 	&1.26 \\
			&Klypin A	&0.63	&0.12	&	1.5	&0.50	& 0.33 	&0.88 \\
			&Klypin B	&0.63	&0.11		&	1.5	&0.54	& 0.37 	&0.96\\
			&Xue	&0.80	&0.12	&	1.5	&0.58	& 0.31 	&1.53\\
  NGC 5533 Ref.~10	  &Xue		&2.04	&0.08	&1.50		&3.0		&6.67  	&0.73 \\
  
NGC 5907 Ref.~10&Sofue	& 0.34	&0.08	&0.75	&1.54		&4.72 	&0.20 \\
			&Klypin A	&0.57	&0.11		&0.75	& 1.34 	&4.27 	&0.30\\
			&Klypin B	&0.55	&0.10	&0.75	&1.39 	&4.43 	&0.29\\
			&Xue	&0.69	&0.15		&0.95	&1.37 	&3.75 	&0.19\\
			
 NGC 6946 Ref.~3&Sofue	&0.18	&0.08	&1.5		& 0.38	&0.70	&1.37\\
			&Klypin A	&0.34	&0.14	&1.5		& 0.34	& 0.65	&1.44 \\
			&Klypin B	&0.32	&0.13	&1.50		& 0.35	& 0.67	&1.42 \\
			&Xue	&0.39	&0.16	&1.50		& 0.33	& 0.64	&1.46 \\
 NGC 6946  Ref.~10&Sofue	&0.12	&0.06	&0.75	& 0.71	&--		&0.46 \\
			&Klypin A	&0.22	&0.08	&0.75	& 0.71	&--		&0.36\\
			&Klypin B	&0.20	&0.07	&0.75	& 0.72	&--		&0.38\\
			&Xue	&0.28	&0.10	&1.5		& 0.67	&--		&0.62 \\
NGC 7331 Ref.~8 	&Sofue	&0.40	&0.08	&0.75	&0.44	& 1.04	&0.34\\
			&Klypin A	&0.69	&0.12	&1.07		&0.41	&0.88	&0.17 \\
			&Klypin B	&0.67	&0.11		&0.75	&0.44	&0.88	&0.17 \\
			&Xue	&0.74	&0.18	&0.75	&0.31	&1.01		&0.40 \\
  NGC 7331 Ref.~9  & Sofue	&0.48	&0.08	&0.75	&1.90	& 0.25	&6.63 \\
			 &Klypin A*
					&--		&--		&--	 	&--		&--		&--\\
			 &Klypin B	&1.09 	&0.11		&0.75	&0.21	&1.64	&1.50 \\
			 &Xue	&1.48	&0.11		&0.75	&2.13	&1.32	&5.28 \\
NGC 7793  Ref.~8	&Sofue		&0.02	&0.04	&0.75	& 0.12	&--	&7.93\\
			&Klypin A		&0.03	&0.09	&0.75	& 0.12	&--	&8.12 \\
			&Klypin B		&0.03	&0.08	&0.75	& 0.12 	&--	&8.11 \\
			&Xue		&0.05	&0.09	&1.50		& 0.15	&--	&7.81 \\
NGC 7793  Ref.~14&Sofue*		&--		&--		&--		&--		&--	&--\\
			&Klypin A		& 0.13	&0.02	&1.5		&2.55 	&--	&0.56 \\
			&Klypin B		&0.12	&0.02	&1.5		&2.55 	&--	&0.58 \\
			&Xue		&0.16	&0.03	&1.5		&2.55 	&--	&0.60 \\			
NGC 7814 Ref.~11\footnote{Citation's  M/L assumptions from ISO fit  } 
  				&Sofue		&0.77	&0.05	&0.75	& 3.32	&0.55	&0.23 \\
				&Klypin A		&1.67		&0.07	&0.75 	& 2.81	& 0.60	&0.41 \\
				&Klypin B 		&1.41		&0.07	&1.5		& 3.4		& 0.53	&0.82 \\
				&Xue		&1.70		&0.09	&0.75	&3.17		&0.51		&0.32\\
NGC 7814 Ref.~11 \footnote{Citation's  M/L assumptions from  MaxLight fit  }
 				&Sofue*		&--		&--		&	--	&	--	&--		&--\\
				&Klypin A		&1.71		&0.07	&0.75	&3.28	&0.61 	&0.45\\
				&Klypin B		&1.46	&0.06	&1.50		&4.50	&0.54	&0.70\\
				&Xue		&1.78		&0.08	&0.75	&3.63 	&0.53 	&0.36\\
UGC 6973 Ref.~10&Sofue	&0.15		&0.08 	&0.75 	&1.51		&--		&  0.07 \\
			&Klypin A	&0.32	&0.16	&0.75 	&1.47		&--		&0.06\\
			&Klypin B	&0.31	&0.15		&0.91	&1.50		&--		&0.07 \\
			&Xue	&0.39	&0.18	&1.13		&1.48	&--		&0.07\\
 \hline
\end{tabular}\\
 References as in Table~\ref{sumRESULTS}.  Mass-to-light ratios   in units of $M_{\sun} / L_{\sun}$.   
Milky Way: ~\citet{Sofue}, ~\citet{Klypin},   \& ~\citet{Xue}.  The quantity $\kappa_\tau$ is defined in section~\ref{kappa}.
(*) indicates an   LCM fit which failed. 
\end{minipage}
\end{table*}

 \subsubsection{M 31}
  
M 31 is an early-type spiral galaxy of  large angular size, in our Local group.  
The proximity of   M 31  to the Milky Way  presents   difficulties in disentangling the gas kinematics of the two galaxies.  It  remains an open question if M 31 or the Milky Way is the  largest member of the Local Group.  There is general agreement on the M 31 rotation curve  between $10$ and $30$ kpc ~\citep{Sof81}.  The bulge of M 31 is thought to be an old stellar population which is almost twice as massive  and more compact than that of the Milky Way ~\citep{Klypin}. 

We compare   rotation curves  as  reported in    Ref.~12 ~\citep{Car}  and Ref.~15~\citep{Klypin}. 
 Both Ref.~12   and Ref.~15  luminous profiles are from    dark matter fits. The  luminous matter profile in Ref.~12 is   from    an ISO  fit and that in Ref.~15 is from    an NFW fit with concentration taken from N-body simulation  (consistent with $\Lambda\rmn{CDM}$).  
   Ref.~12 disentangles M 31 and Milky Way gas  velocities by using   only the approaching side kinematics of the rotation curve data    reported in     ~\citet{1985ApJ...299...59C}.  
    Both references use the  stellar    parameters   as reported  in ~\citep{walt87,walt88}. 
 
Comparison of the  Ref.~12 and Ref.~15 M/L results    reflect the under-constrained nature of PSM, reflected in  the two reported luminous profiles shown in Fig.~\ref{fig:massmodels18}.
  The Ref.~12  ISO fit returns an  anomalously high  disk $ \rmn{M/L}_B =7.5$ and a    bulge    consistent with ~\citet{walt87,walt88} of $3< M/L_B <4$.   Most fit results for M 31 return a M/L of the  disk   twice that of the bulge ~\citep{1989Kent}, however Ref.~ 15 disputes this trend with their  varying concentration  treatment, which  returns  M/L  respectively  of  $0.93$ and  $3.0$ for the disk and bulge,   consistent with PSM of ~\citet{BelldYong}.  
Ref.~15  attributes the difference   to   dark matter treatments which use    constant density dark  halos and do not account for  absorption.   ~\citet{Widrow} also notes extinction effects from dust in the disk.  
 While the LCM fits for M 31 in Fig.~\ref{fig:resultssmall} do not obviously favor one luminous model over the other, the stability of individual pairing results reported in Table~\ref{sumRESULTSbigger} indicate the 
 Ref.~15 luminous profile is more physically acceptable.     As will be discussed in  section~\ref{n891n7814}, the M 31  $\tilde{a}$ values are consistent with a conjecture relating $\tilde{a}$ to   density gradients in the luminous matter.  Based on that conjecture,    the Ref.~15  luminous profile is again more physically representative   an old, heavy stellar bulge. 
  
\subsubsection{ NGC 5533} 

 NGC 5533 is an early-type galaxy at an estimated distance of $54$ Mpc. NGC 5533 has  significant side-to-side asymmetries and a warp. 
 
  We sample  NGC 5533 as reported in  Ref.~10~\citep{SanMcGa} with an associated luminous profile from a MOND fit.  
  The Ref.~10   rotation curve data is from  ~\citet{San96}.  
  The luminous profile geometry reported by Ref.~10  is calculated using  a double exponential model similar to those used in all Ref.~3 and Ref.~8 galaxies.  
  The Ref.~10  gas is from    H\,{\sevensize\bf I}  surface densities, reported as   $ 3.0 \times 10^{10} M_{\sun}$.   Ref.~10 notes that a large fraction of total light is in the central bulge and that  higher bulge M/L are consistent with expectations for older, less actively star forming spheroidal sub-systems. 

Unfortunately,   only one Milky Way luminous mass model in our sample has   sufficient radial coverage to adequately fit this galaxy  ~\citep{Xue}.   However,  NGC 5533  is retained
because there are so   few galaxies   larger than the Milky Way. We return M/L  of the bulge a little lower than reported in Ref.~10, and maximal gas scalings of $1.5$.  The Ref.~10 bulge M/L is about ten times that of the disk. Our LCM fits reduce this proportion to     about a factor of two ($2$).    Fit results are shown in  Fig.~\ref{fig:resultssmall}.   As will be discussed in the conjecture in  section~\ref{n891n7814}, the NGC 5533  $\tilde{a}$ values (in Table~\ref{sumRESULTSbigger}) indicate a stellar bulge for NGC 5533 which is less dense than that of M 31 as reported in Ref.~15. 

 \subsubsection{NGC 891 \& NGC 7814 \label{n891n7814}}
 
  The comparisons between the  NGC 891 and NGC 7814 as reported in Ref.~11
 ~\citep{Frat}  are probably the most important in our sample.   NGC 891 and NGC 7814 are edge-on spiral galaxies at respective  distances of $9.5$ and $14.6$ Mpc.   The  Ref.~11 rotation curves are from  \citet{Oosterloo:2007se},  with a velocity resolution of $4.12\,  \rmn{km\,  s^{-1}}$,  derived from  concentric rings and Gaussian  fits to the velocity field.   M/L is reported   in the $3.6\,\umu$m  band, and    the geometries of an  exponential disk and spherical bulge are used.  Because these two galaxies have similar total enclosed luminous mass at the limit of the data as indicated by 
  $\kappa_\tau$ (see section~\ref{kappa}), 
 very similar rotation curves beyond  $3-4~\rmn{kpc}$, and similar inclinations, the direct comparison of their 
   free parameter $\tilde{a}$ values  in Table~\ref{sumRESULTSbigger} can be tied to morphology.  
 NGC 891 is disk-dominated and NGC 7814 is bulge-dominated.  
 
 The comparison of the LCM results for NGC 891 and NGC 7814 are    the  most direct indication of a physical interpretation for    $\tilde{a}$.  The  lower density gradient in the luminous matter associated with the disk-dominated galaxy NGC 891  is  associated  with low  $\tilde{a}$ values as compared to those for the  higher density gradient  bulge-dominated NGC 7814.  We see this trend throughout the sample,  with one possible exception; therefore we make the following conjecture:    $\tilde{a}$ increases with   increasing density gradients in the luminous matter.  

We note that it  remains possible that  $\tilde{a}$ instead  implies a relation between  the dark matter halo and the luminous matter profile, because NGC 891 is commonly referred to as having a minimal dark matter halo,  consistent with    vanishingly small  $\tilde{a}$ values.      
 
Ref.~11 reports   rotation curves fits for  NGC 7814 and NGC 891 in  three different model contexts: Maximum light (MaxLight),  ISO and MOND.  For  NGC 7814,  we sample the  ISO  and   MaxLight luminous   profiles.  For  NGC 891,  we sample the  MOND and  MaxLight luminous   profiles.   Ref.~11  MOND fits are based on the  $\umu$-simple interpolating function as described in ~\citet{Gent}.   The LCM results for both galaxies  demonstrate  convergence of the reported  M/L from very different starting estimates. The  NGC 7814   MaxLight fit result M/L   of the  bulge was  unrealistically high,   taken to indicate  the necessity of a dark matter halo.   The LCM lowers that    M/L  without a   halo.  In   Fig.~\ref{fig:resultssmall}  we   report only one of the  NGC 891  LCM fits (from MOND starting estimates of M/L)    in the interest of space,   given that the two  LCM luminous profiles and fits were essentially the same, and       for   NGC 7814 LCM results from the starting  assumptions of M/L from     ISO   are   indicated by (A) and those from MaxLight   are indicated by  (B).

\subsubsection{NGC 2841}

 NGC 2841 is an early-type, large  spiral galaxy larger in radial extent than the Milky Way.  The controversial distance estimates to  this galaxy are     the   Cepheid distance of $14.1$ Mpc versus   that indicated by   MOND fits of $23$ Mpc.  The larger MOND distance is supported by the lowest luminosity   supernova type IA event ever observed~\citep{Gent,Bot}, which would be of standard luminosity at the MOND distance.   The Cepheid based distance  is considered the most precise distance measure, though some unresolved issues persist~\citep{HubbleCepheid}. 

  We sample   three different references for NGC 2841:   an NFW fit   from Ref.~3~\citep{Blok} and   MOND fits from Ref.~8~\citep{Gent}   and Ref.~9~\citep{Bot}.   The fits in Ref.~ 3 and 9 are at the 
 Cepheid distance, and that from  Ref.~8 is  at the constrained Cepheid distance of $15.60$ Mpc.  Ref.~3 and Ref.~8 rotation curve derivations and modeling assumptions  are   described in sections~\ref{Blok} and \ref{TaM}. The Ref.~3 and Ref.~8 rotation curves   agree   with that   in  Ref.~9 within the  error. The Ref.~9 rotation curves is     from    ~\citet{BegeTH}.   Ref.~3    notes    excess surface brightness in the central radii of NGC 2841, not accounted for by their exponential model.   
 
 The  major difference between these three  luminous profiles as reported is   the stellar mass decomposition.   In    Ref.~3 and Ref.~8 the  disk M/L is   slightly smaller than that of the bulge.   In Ref.~9,  the   disk M/L     is  ten times that of the bulge.  The   M/L reported in  Ref.~3    are fixed consistent with ~\citet{BelldYong} PSM.    As shown in Fig.~\ref{fig:resultssmall}, the LCM  returns three luminous profiles and associated rotation curve fits which are very similar in all features.   However, what distinguishes these three LCM fits are the resulting  $\tilde{a}$ values. 
The    luminous assumptions  in  Ref.~3 and Ref.~8 return the largest  $\tilde{a}$ values   in the sample, versus those from Ref.~9 which are very close to the mean value (Fig.~\ref{fig:alphaDistrib1}). The scaling of $\tilde{a}$ with steeper  density gradients in  the luminous matter (section~\ref{n891n7814})   indicates that  future directed photometric investigation  of NGC 2841 can   identify which model is most physically representative.
     
\subsubsection{NGC 7331}

NGC 7331 is an early-type spiral with strong spiral arm structure. 

 We sample two NGC 7331 rotation curves and associated MOND fits,  using   different distance estimates and associated luminous profiles.  Ref.~9~\citep{Bot}   uses the cepheid based distance ($14.7$ Mpc) and  
and  Ref.~8~\citep{Gent}   the  constrained   distance ($13.43$  Mpc).   The Ref.~8 rotation curve   is higher resolution, and shows a rise in the final radii which is not observed in Ref.~9.  The difference is attributed to  differing assumptions of the galaxy's inclination ~\citep{Blok} . 

The major difference between these two data sets are the luminous profiles; in  Ref.~9 the M/L of the    bulge and disk   are approximately equal, and  in Ref.~8 the M/L of the bulge  is about three ($3$) times that  of the disk. 
 PSM ~\citep{BelldYong}   for this galaxy indicate  a M/L of the  bulge which is bigger by a factor of approximately $1.5$~\citep{Blok}.  As can be seen in Fig.~\ref{fig:resultssmall}, the two LCM fits are roughly equivalent past $10$ kpc, though differing coverage makes the two data sets difficult to directly compare, though the LCM fits from Ref.~8 assumptions are more stable.  Consistent   with the conjecture in section~\ref{n891n7814}, the  larger $\tilde{a}$  values are associated with fits using the  larger bulge assumptions in   Ref.~8.    The total gas reported in  Ref.~9  is $1.4\times 10^{10}M_{\sun}$, with associated       rotation curve velocities approximately  $30\,  km\,   s^{-1}$ higher than   those for the gas reported in Ref.~8.   The majority of   LCM pairings  for both data sets  return minimal gas scalings for this galaxy.  A comparison to NGC 3521 is made in the next section. 
 
 \subsubsection{NGC 3521}
   
NGC 3521 was one of the first cases of a genuinely declining  H\,{\sevensize\bf I}   rotation curve,  although not a Keplerian descent. However,   we sample    data from Ref.~8~\citep{Gent} which  does not confirm the descending rotation curve. There is a small decrease in circular velocity at high radii, but not as  previously reported. 

  Ref.~8 reports  MOND fit results for   M/L free and  the constrained distance  of $7.50$\,  Mpc, from the 
  Hubble flow distance of $10.7\,  \pm\,  3.2$ Mpc.  See section~\ref{TaM} for a description of the 
associated assumptions for rotation curve derivations and luminous mass modeling.
 In  ~\citet{Blok}   the fixed PSM   of the disk M/L is $0.73$.   Ref.~8  returns a value of $0.71$.   The LCM fit return an average value of $0.72$.     LCM fits returns maximal gas scalings of $1.5$.   
 
\begin{figure*}
 \centering
 \subfigure[M 31,   Ref.~15]{\includegraphics[width=0.33\textwidth]{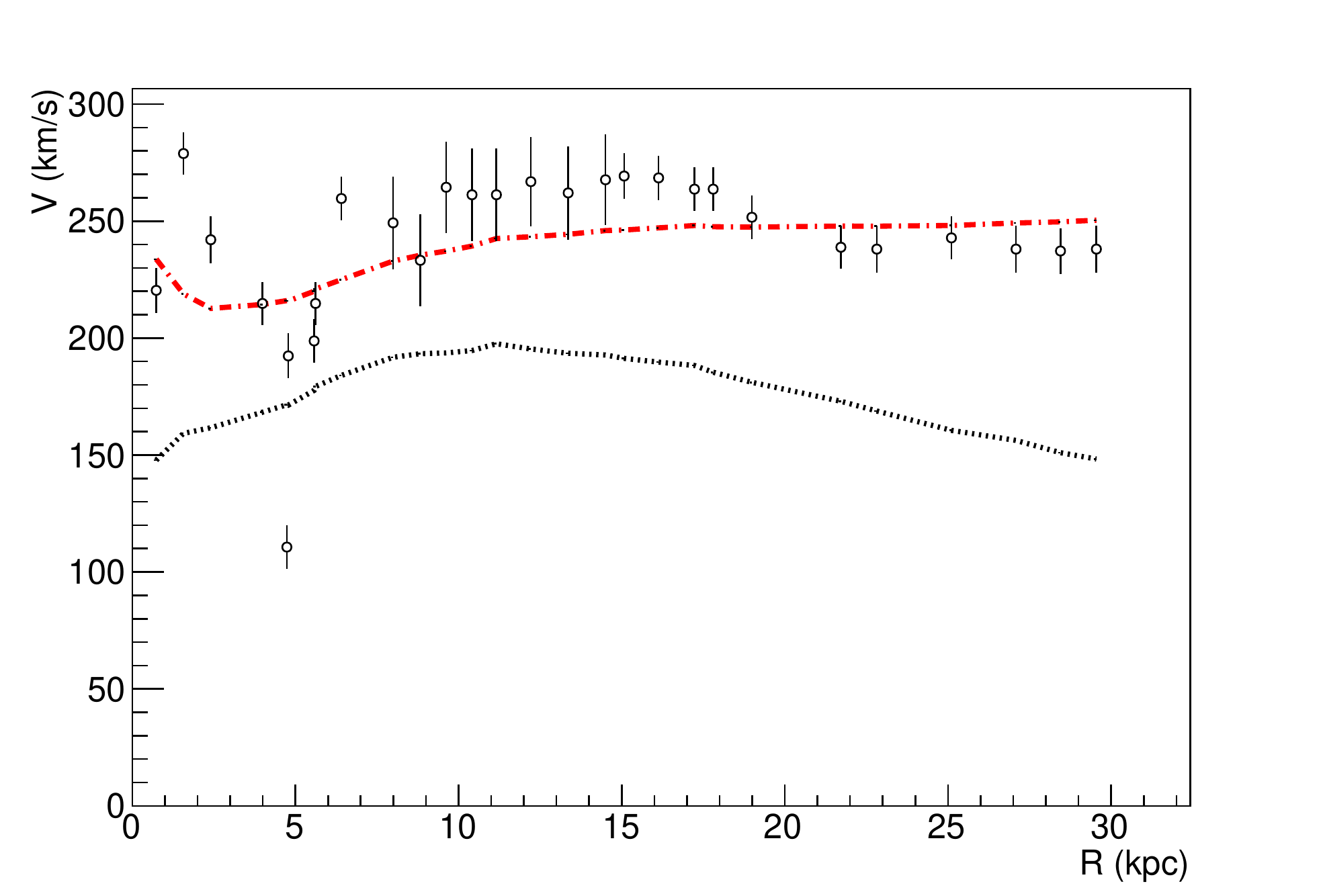}}
\subfigure[M 31,  Ref.~12]{\includegraphics[width=0.33\textwidth]{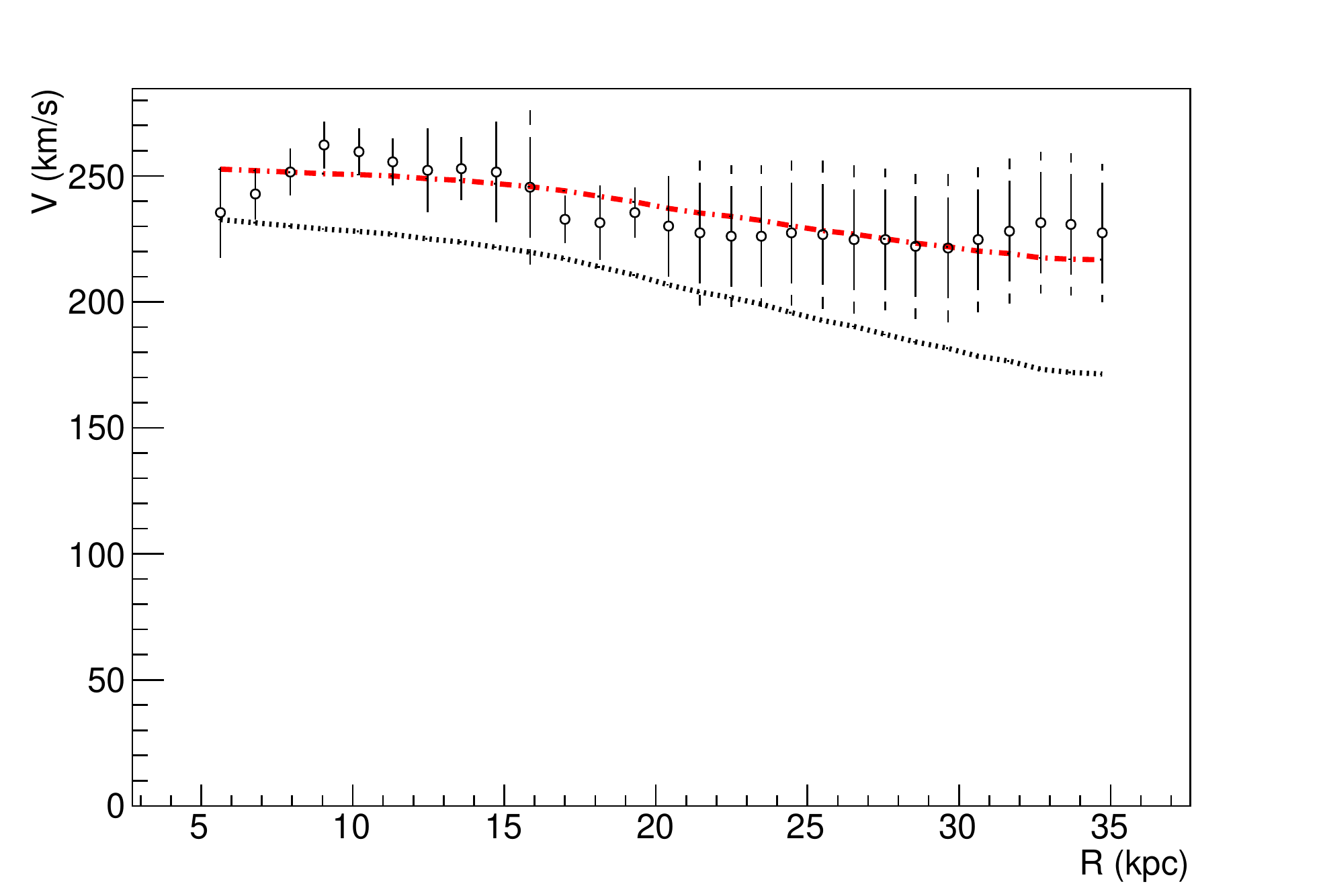}}
\subfigure[NGC 5533, Ref.~10]{\includegraphics[width=0.33\textwidth]{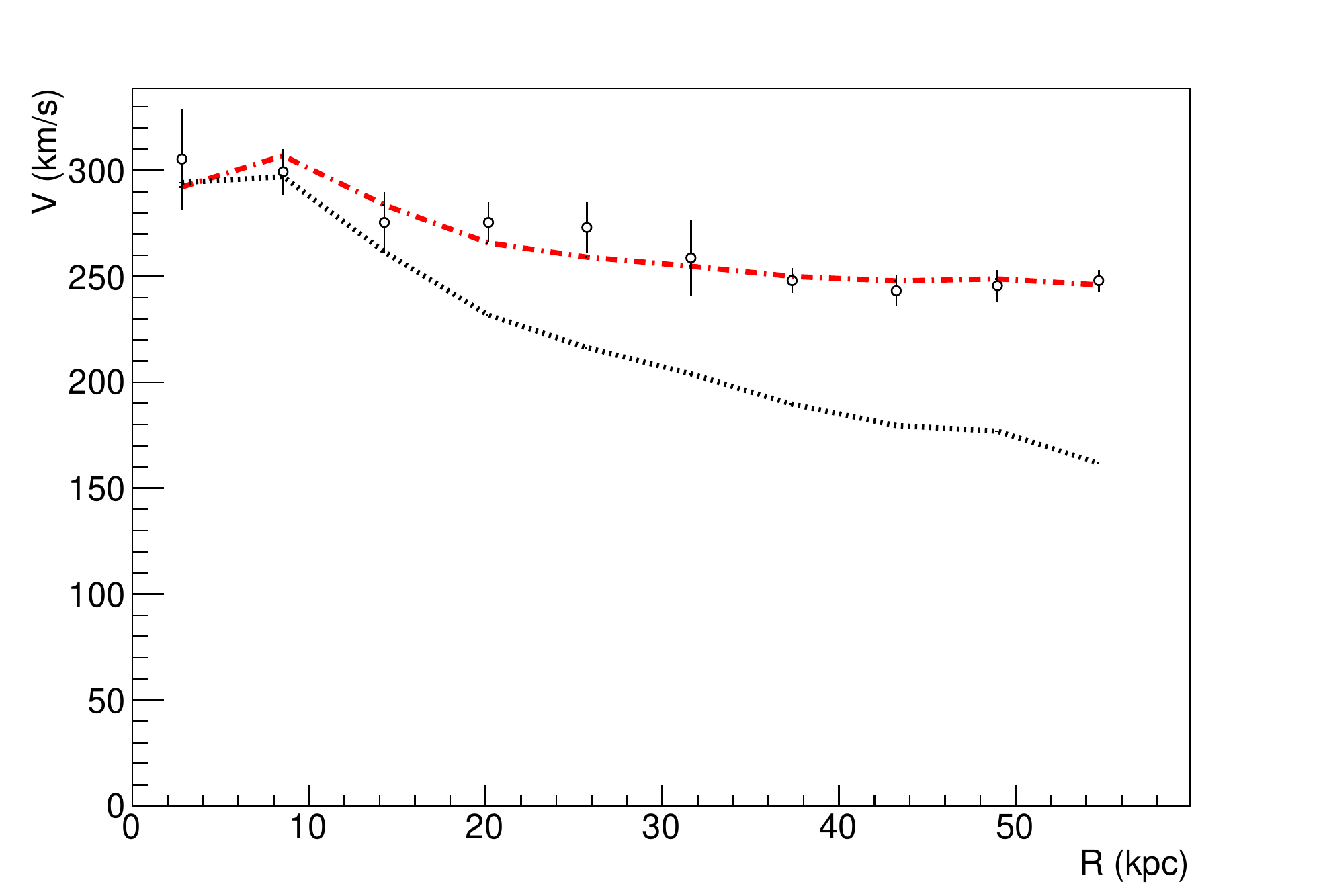}}\\

\subfigure[ NGC 7814,  Ref.~11 from  A ]{\includegraphics[width=0.33\textwidth]{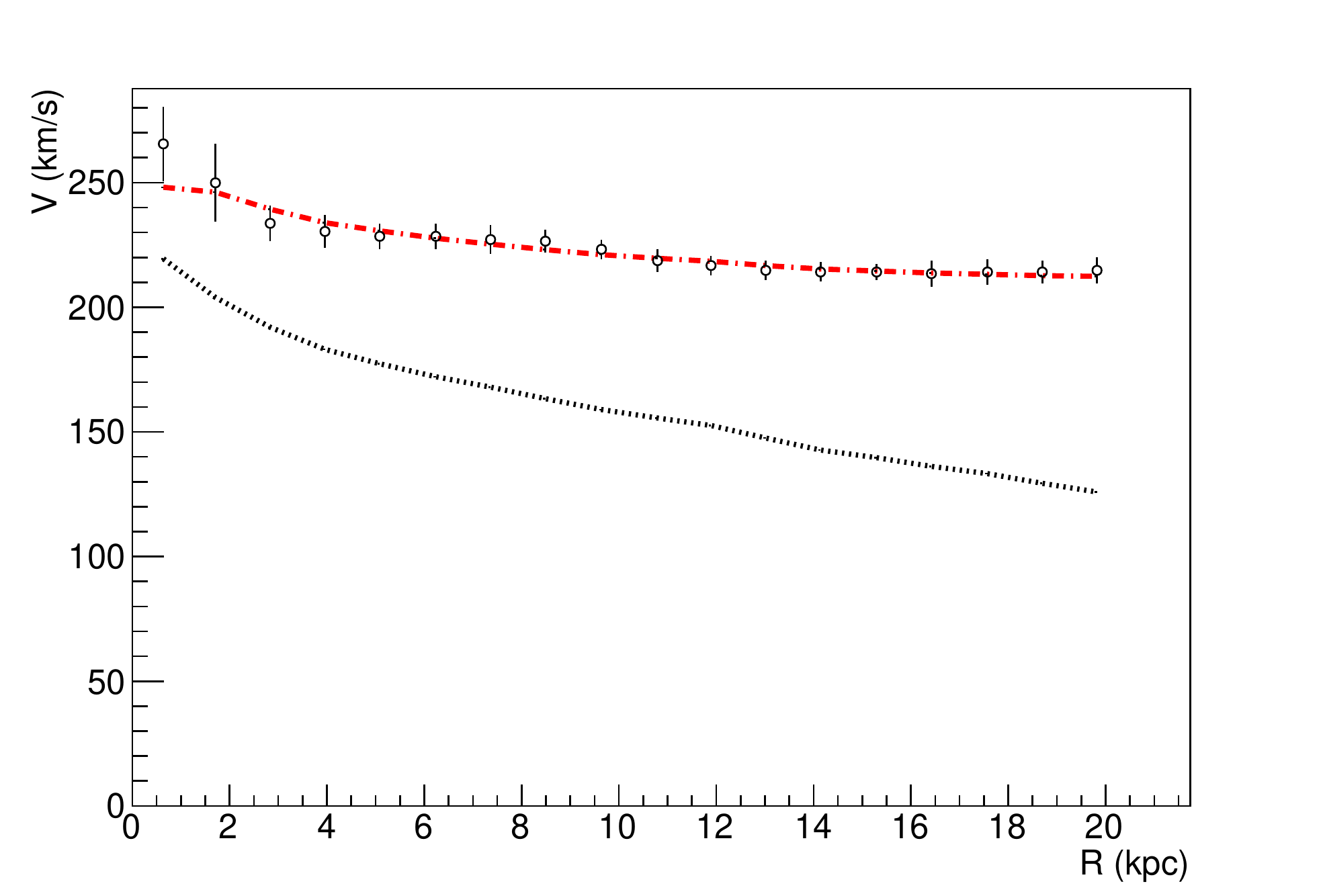}}
\subfigure[ NGC 7814,  Ref.~11 from  B ]{\includegraphics[width=0.33\textwidth]{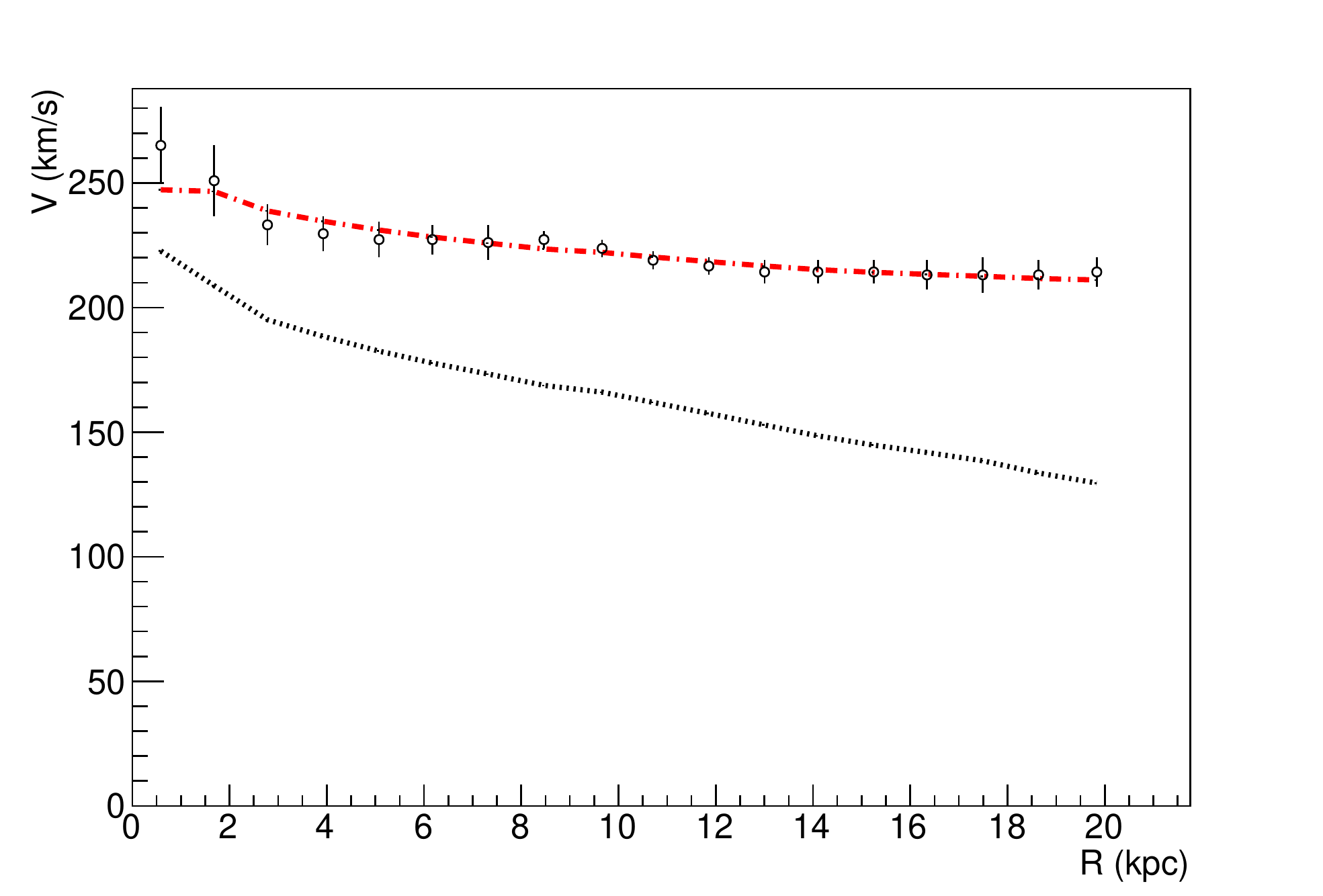}}
\subfigure[ NGC 891, Ref.~11  ]{\includegraphics[width=0.33\textwidth]{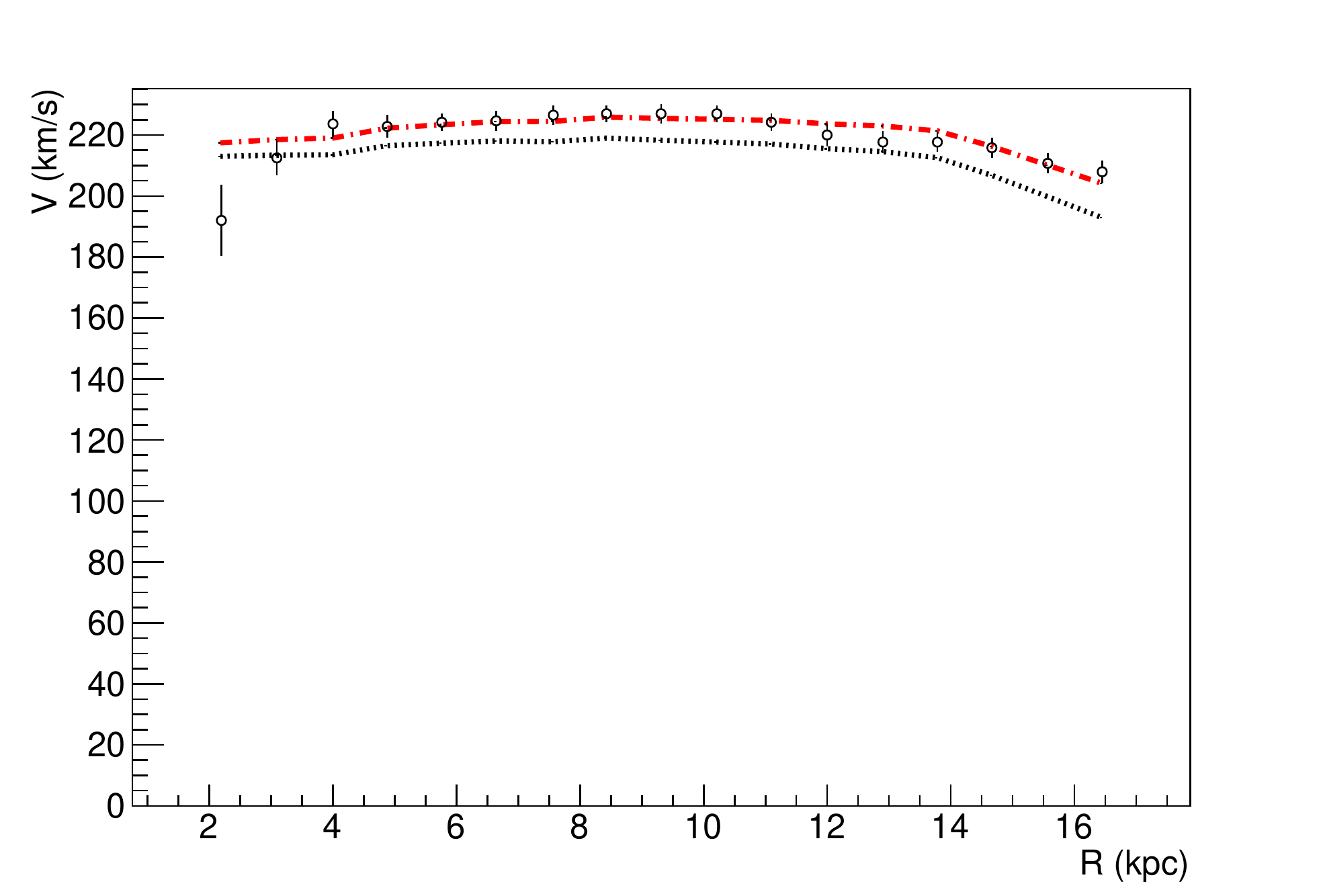}}\\

\subfigure[ NGC 2841,  Ref.~3 ]{\includegraphics[width=0.33\textwidth]{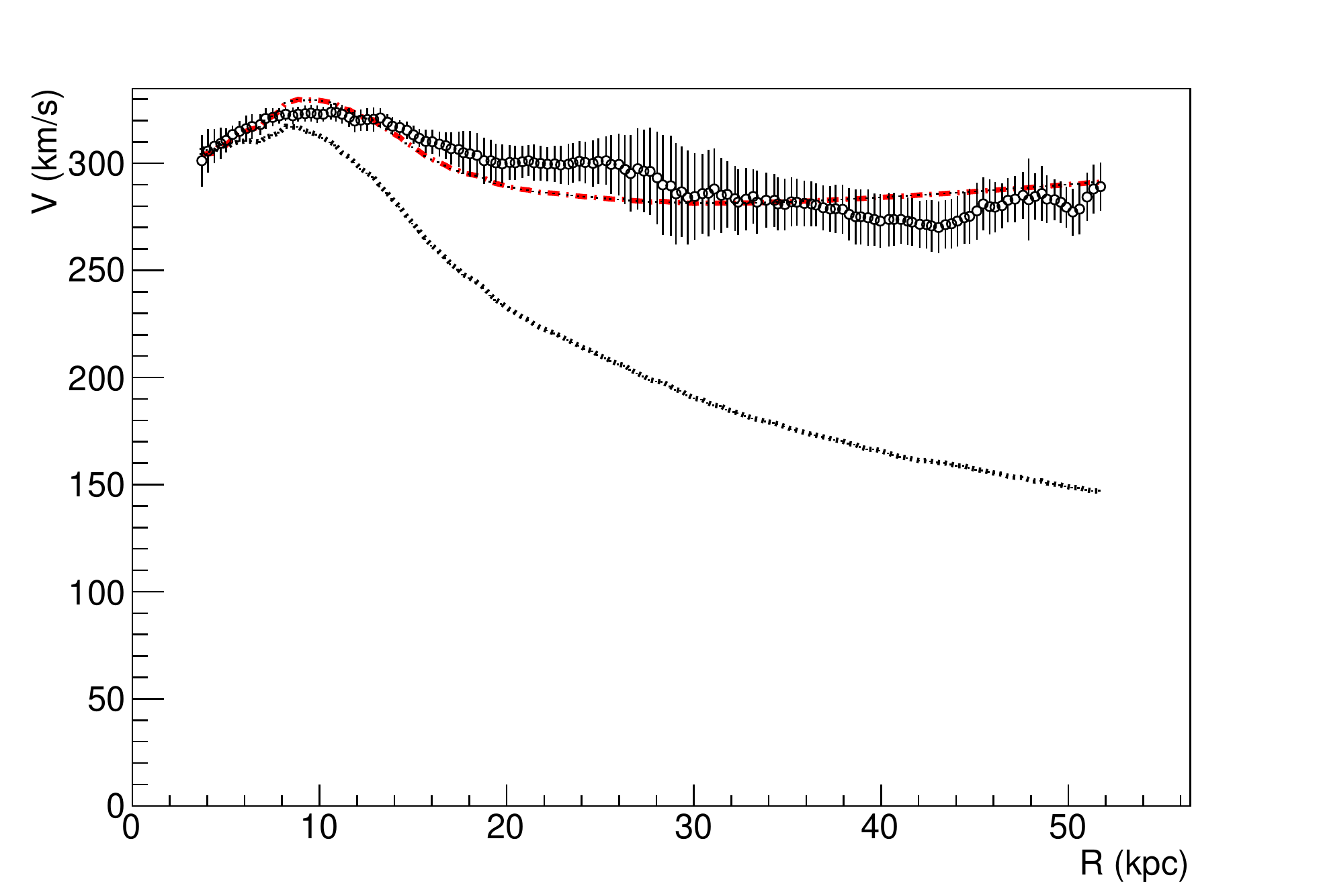}}
\subfigure[ NGC 2841, Ref.~8 ]{\includegraphics[width=0.33\textwidth]{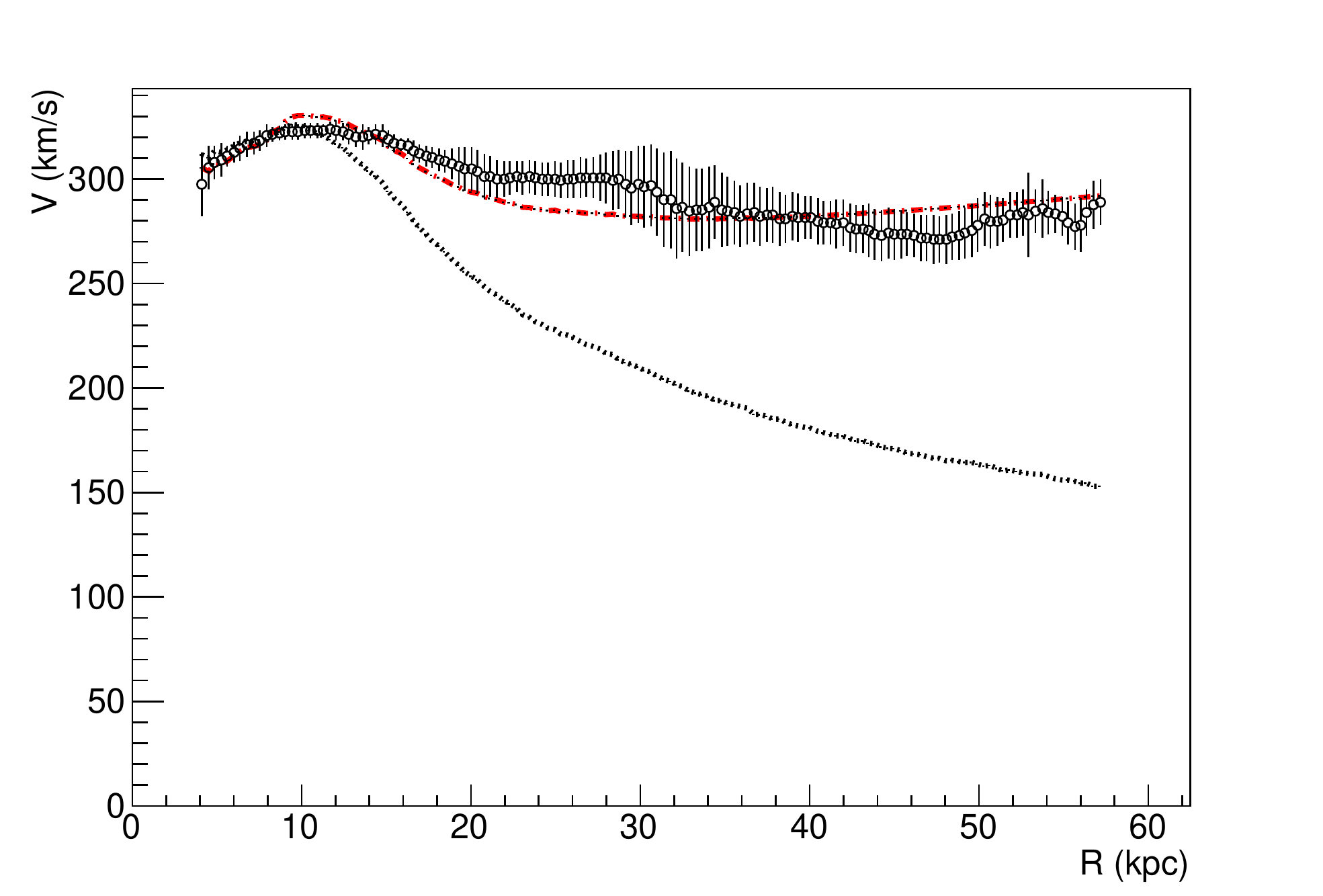}}
\subfigure[ NGC 2841,  Ref.~9 ]{\includegraphics[width=0.33\textwidth]{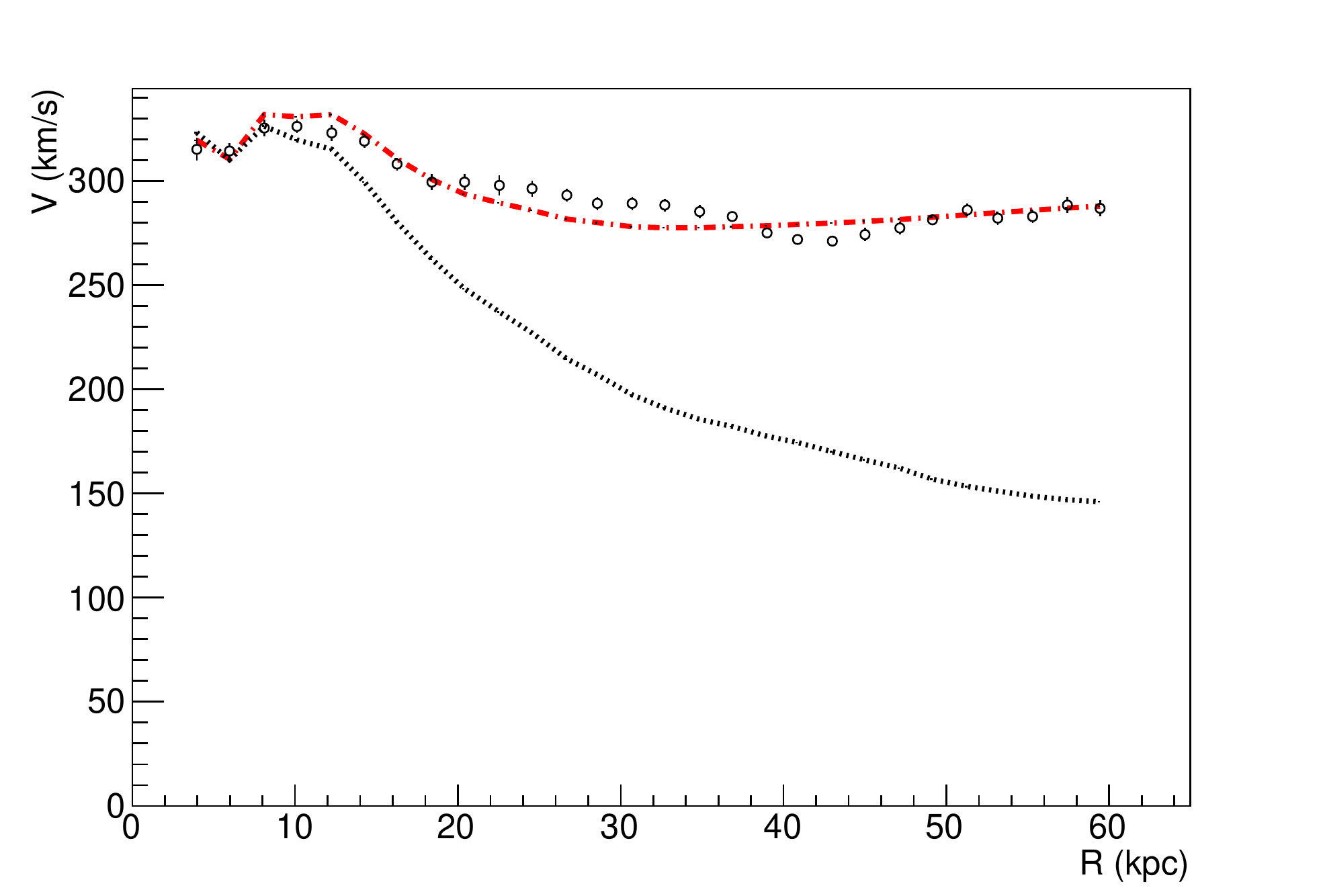}}\\

\subfigure[NGC 7331,~Ref.~8 ]{\includegraphics[width=0.33\textwidth]{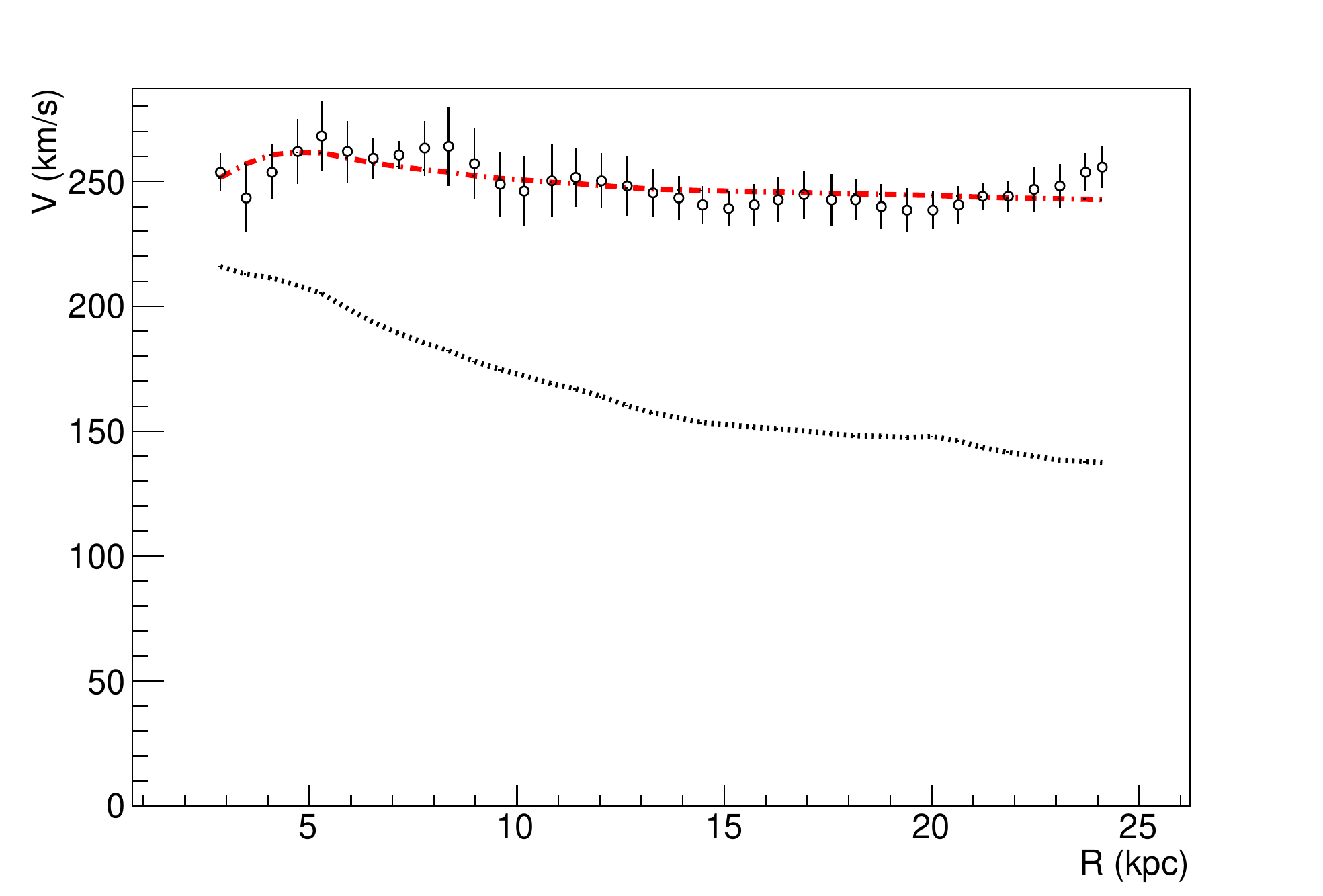}}
\subfigure[NGC 7331,~Ref.~9 ]{\includegraphics[width=0.33\textwidth]{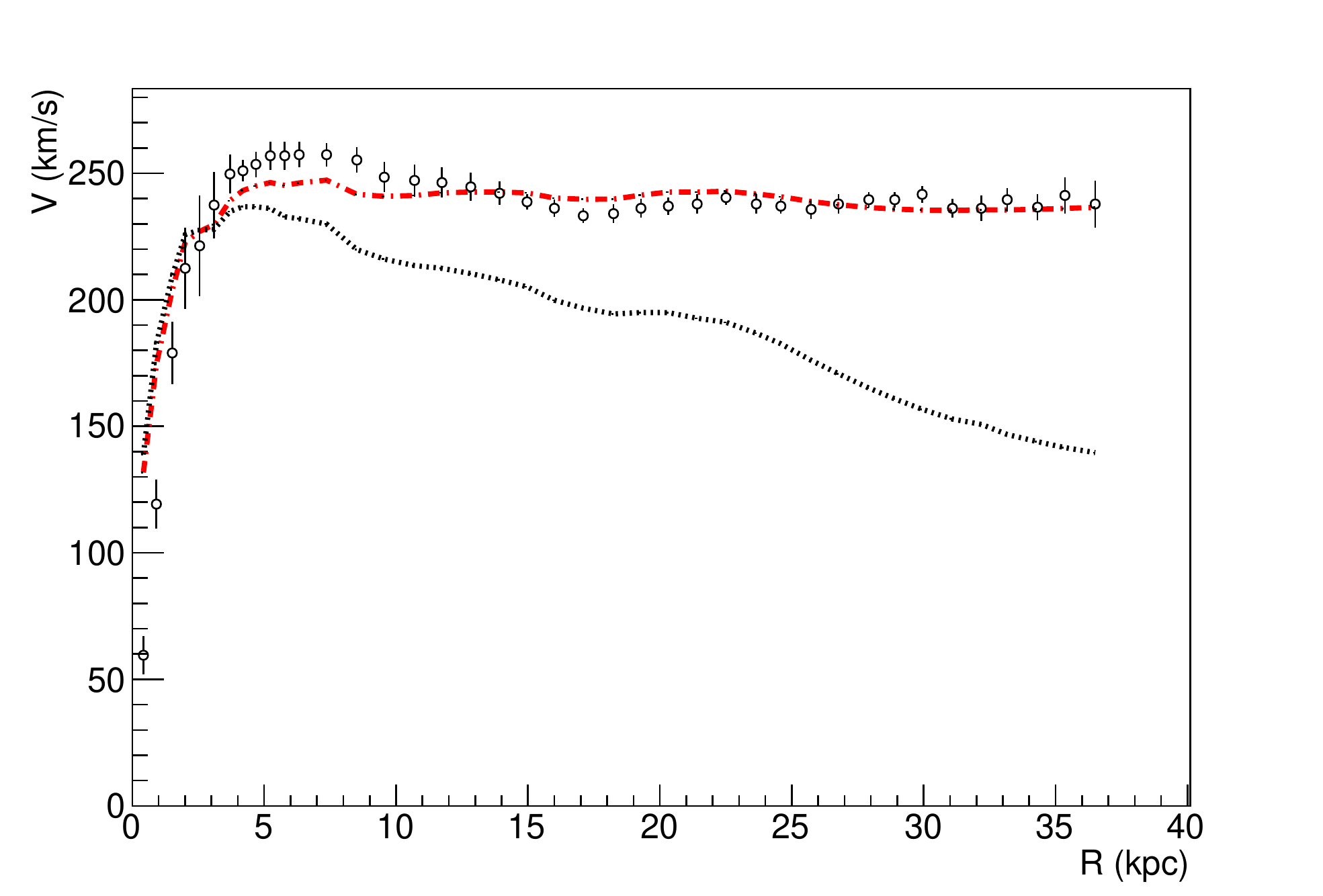}}
\subfigure[NGC 3521, Ref.~8  ]{\includegraphics[width=0.33\textwidth]{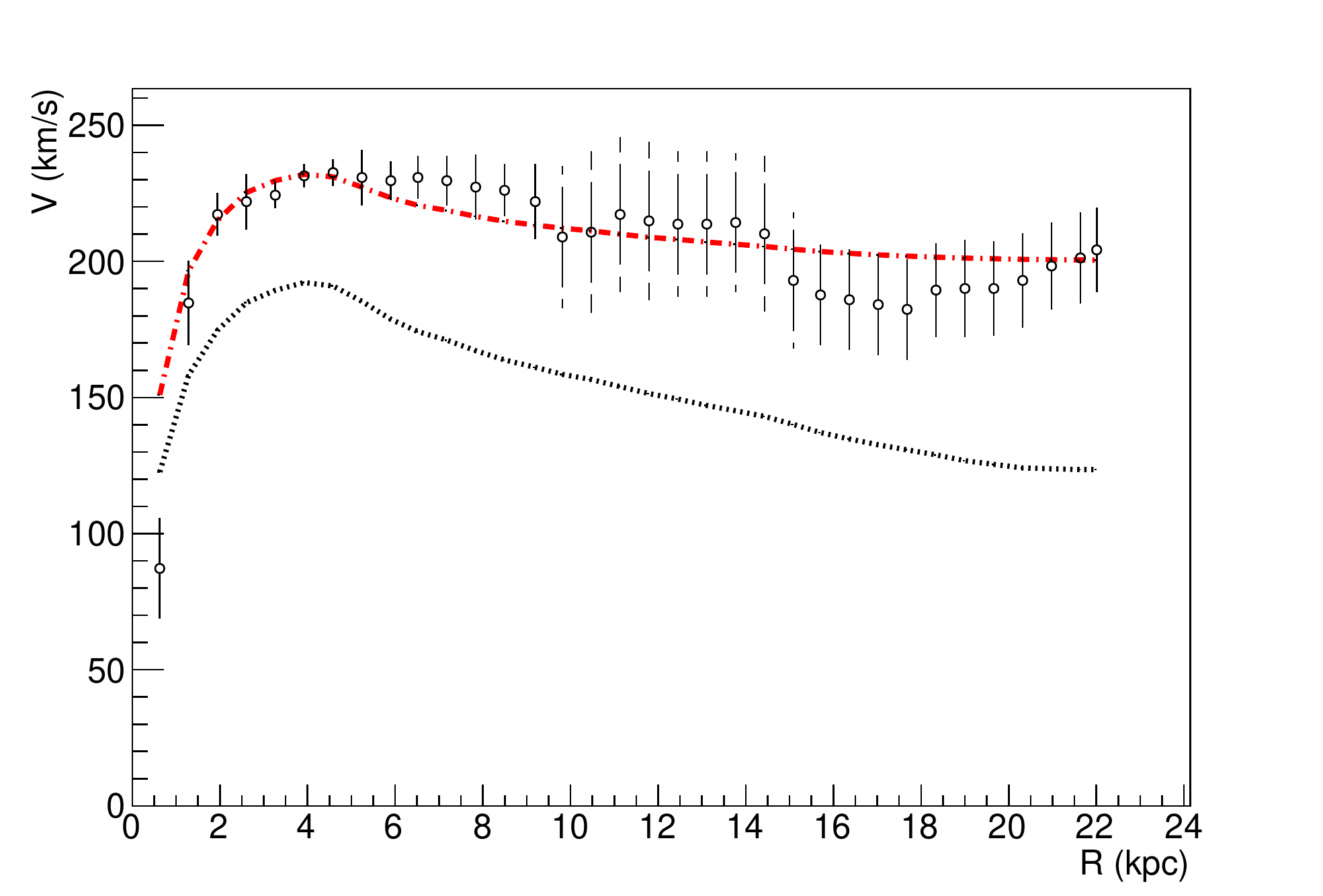}}
\caption{LCM rotation curve fits.   In all panels: black circles represent the observed  rotation velocities, thin bars represent the reported uncertainties, dotted curves show  the best-fit LCM  Newtonian contributions from the  luminous matter  and LCM resulting rotation curve   is the  red  dotted-dashed  line.   References  are as in Table~\ref{sumRESULTS}.}             
\label{fig:resultssmall}   
\end{figure*}

 An  interesting feature for this analysis  is the distribution of the luminous mass. While the total enclosed luminous mass at the farthest extent of the data  (indicated by   $\kappa_\tau$) for   NGC 3521 and NGC 7331 (as reported in Ref.~8) are approximately equal,  the average values of the free parameter $\tilde{a}$ for  NGC 3521 are  about $40\%$ smaller.  With no reported bulge component for NGC 3521 as compared to NGC 7331, this is consistent   with the conjecture in section~\ref{n891n7814}.  Fit results are shown in  Fig.~\ref{fig:resultssmall}.

\subsubsection{NGC 5055}
  
   NGC 5055  is a  rotationally dominated spiral galaxy    at a favorable inclination for kinematic study. It  has a 
nearby companion galaxy UGC 8313,  at a projected distance of $50$ kpc~\citep{Batt}.    

   We sample  three rotation curves:  Ref.~3~\citep{Blok}, Ref.~7~\citep{Batt}, and  Ref.~8~\citep{Gent}. 
     The rotation curves in Ref.~3 and Ref.~8,  as compared to  Ref.~7,  have a
   more pronounced rise rate   and higher velocities at large radii.  Ref.~3 and Ref.~8 derive their rotation curves as described in Sections ~\ref{Blok} and \ref{TaM}. 
 Ref.~7 derives their rotation curve velocity   using the conventional intensity-weighted mean  Gaussian profile fits. 
   Distance indicators for the three authors are: Ref.~3 uses the Hubble flow distance of $10.1$ Mpc,   Ref.~7  uses a Cepheid distance  of $7.2$ Mpc~\citep{Pierce}  and    Ref.~8  uses a constrained   fit distance of $7.07$ Mpc.   The gas mass reported in   Ref.~7.~\citep{Batt} is the total    H\,{\sevensize\bf I}  mass of $6.2 \times 10^9 M_ {\sun}$, scaled for helium abundances.  
   
   The luminous profile reported in Ref.~ 3 is    fixed by PSM, that in 
      Ref.~7 is  from a NFW  fits  and in Ref.~8  from a MOND fit.   All three authors  report   their resulting M/L are       consistent with PSM, though   in Fig.~\ref{fig:massmodels18} they appear geometrically different. In   Ref.~3,  the   M/L  of the bulge   is seven times smaller than that of the disk.  In   Ref.~8, the   M/L  of the    bulge and disk     are approximately equal.  In  Ref.~7  there is no  stellar bulge. 
      
   LCM fits from the  Ref.~3 luminous profile are   highly unstable, see  Table~\ref{sumRESULTSbigger}.    The LCM  fits using the  Ref.~7 and Ref.~8  luminous profiles (both at essentially the same estimated distance) are visually (see  Fig.~\ref{galaxiesSmaller}) difficult to compare given the higher resolution data from Ref.~8 and different extents of the coverage.  However, if we hope to address the question of whether or not mass models should include a bulge, it is necessary to   agree on a set of rotation curves, associated uncertainties,  and  luminous M/L in the same wavelength bands.     Consistent   with the conjecture in section~\ref{n891n7814}, the LCM fit results  from Ref.~8 assumptions (with   a bulge component)  have larger $\tilde{a}$ values than those from    Ref.~7 assumptions.  
 The LCM fits (from Ref.~8) do return  a bulge M/L  which is smaller  than that of the disk, by a factor of approximately $1.6$.

\subsubsection{NGC 4138, NGC 5907 and NGC 3992}

 The  rotation curves for  Ursa Major galaxies  NGC 4138, NGC 5907 and NGC 3992  as  reported in Ref.~10~\citep{SanMcGa}, are associated with luminous profiles  from  MOND fits.   The rotation curve data  is from     ~\citet{Ver98,San96}.  The Ref.~10 rotation curve derivation and luminous mass modeling assumptions   are described in section~\ref{SanMc02}.  These three galaxies are grouped   because they have similar enclosed mass at the limit of the data  (indicated by   $\kappa_\tau$).  

Comparison of these three galaxies,  and their associated $\tilde{a}$,  may challenge the conjecture in section~\ref{n891n7814}.  The    $\tilde{a}$ ranking, in increasing order is:  NGC 4138, NGC  5907, NGC  3992.  Correlating increasing $\tilde{a}$ with increasing density gradients in the luminous matter is challenged by the fact that in the Ref.~10 analysis,   NGC  5907  has a bulge, whereas   the  high surface brightness galaxy  NGC 3992 does not.  The NGC 5907  bulge M/L is proportionally smaller than that of the disk,  by  about  a factor of two, so it   remains possible that   NGC 3992 has a steeper density gradient.   ~\citet{Bott3992} reports that  NGC 3992   is  a highly non-exponential,      extreme Freeman type II galaxy.     The LCM fits for  NGC 4138 indicates that Ref.~10 over-estimates the luminous mass at small radii.  Fits for NGC 4138, NGC 5907 and NGC 3992 are shown in Fig.~\ref{galaxiesSmaller}. 

\subsubsection{NGC 2903}

NGC 2903 is a high surface brightness galaxy, for which we sample the reported rotation curves and associated luminous profiles from Ref.~3~\citep{Blok}, Ref.~8~\citep{Gent}   and  Ref.~10~\citep{SanMcGa}.    
Distance indicators for the three authors are:  Ref.~3 (NFW fit)   uses the   brightest stars distance of  $8.9\,\pm 2.2$ Mpc, Ref.~8 (MOND fit)   uses the constrained   brightest stars distance of   $ 9.56 $ Mpc, and Ref.~10  (MOND fit)  uses the Hubble flow distance  $6.4$ Mpc from ~\citet{BegeTH}.  The 
 Ref.~3   and  Ref.~8 rotation curve   agree  with that reported in ~\citet{BegeTH}, though Begeman's rotation curve rises more steeply in the inner radii.  Ref.~3 attributes this  difference to Begeman not using a full tilted-ring  model in the inner parts of the galaxy, therefore not accounting for the sudden changes in the P.A. and inclination in his analysis.  See sections~\ref{Blok}, \ref{TaM} and \ref{SanMc02}  for details of the total  light decomposition and rotation curve derivations for these three references.  Due to non-circular motions and the presence of  a bar, constraints on the inner radii of NGC 2903 remain open questions.  The gas profile used in  Ref.~ 10 is    from ~\citet{BegeTH}, of $2.4\,\pm 0.1 \times 10^{9}M_{\sun}$.  Out to a radius of $14.5$ kpc,  the  Ref.~3 and Ref.~8 gas profiles agree with    Ref.~10.  However at larger radii,  the gas profiles of Ref.~3 and 8 are far larger than that in   Ref.~10. 

\begin{figure*}     
 \centering 
  \subfigure[ NGC 5055, Ref.~3  vs. Sofue]{\includegraphics[width=0.33\textwidth]{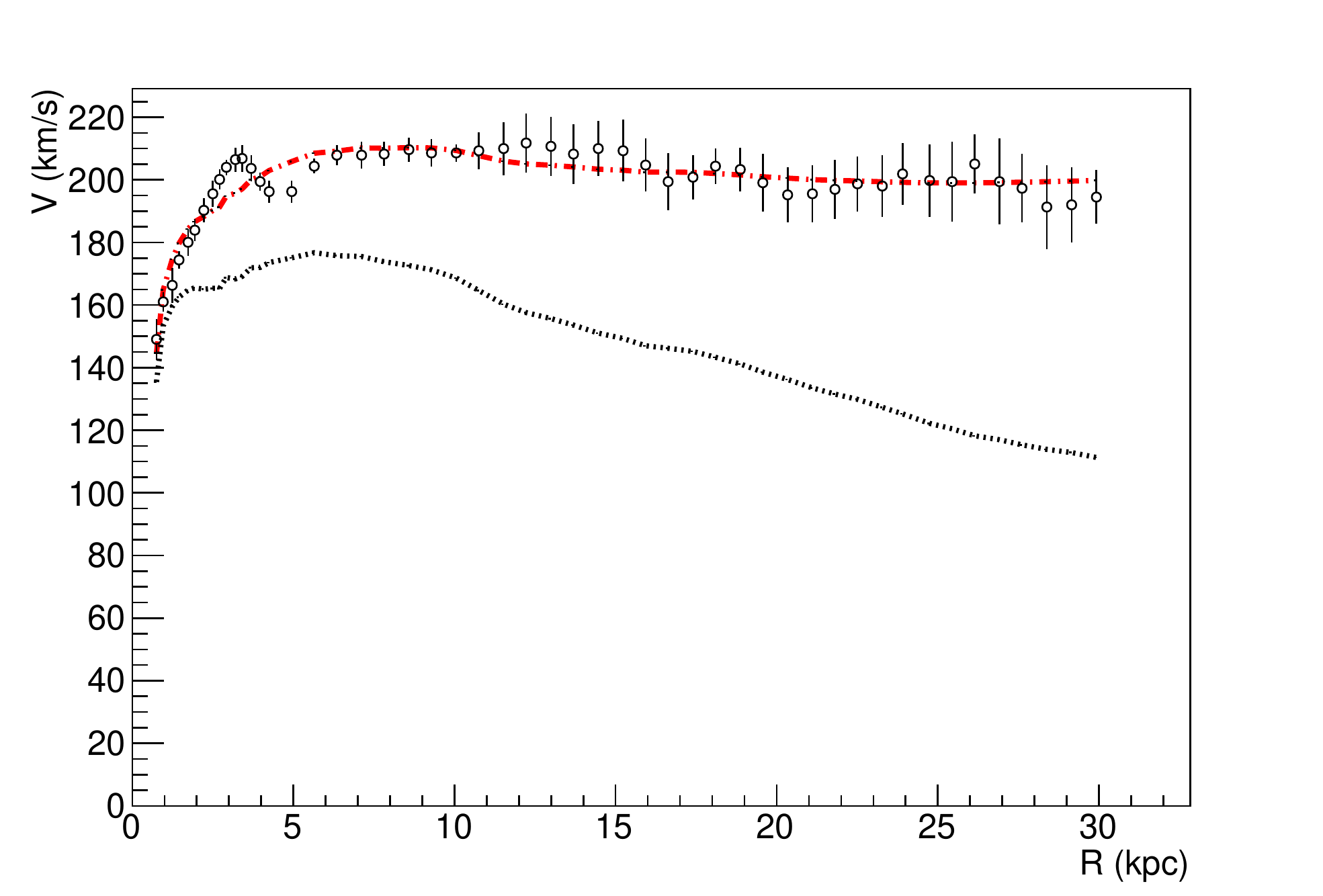}}
   \subfigure[NGC 5055, Ref.~7]{\includegraphics[width=0.33\textwidth]{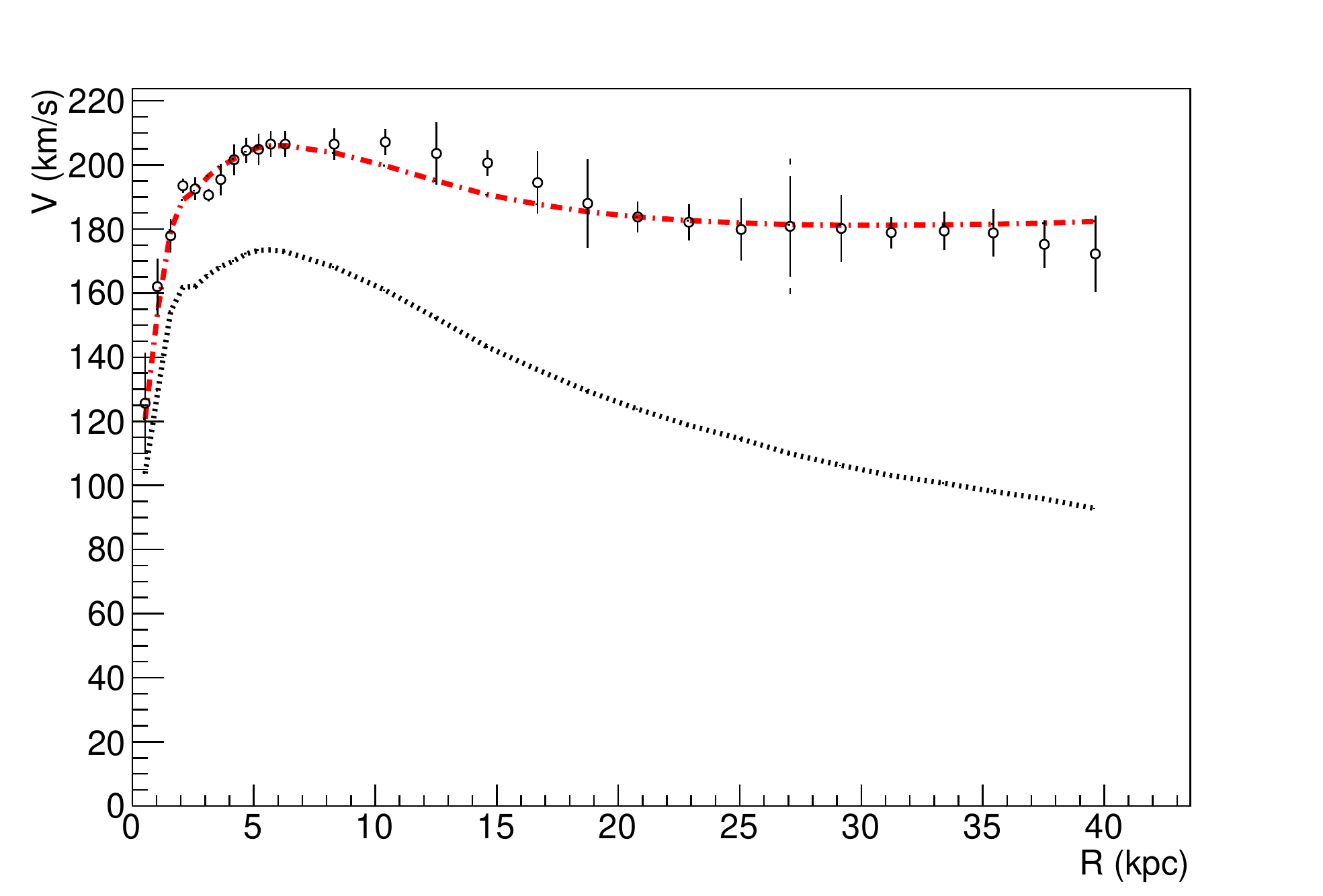}}
    \subfigure[NGC 5055, Ref.~ 8 ]{\includegraphics[width=0.33\textwidth]{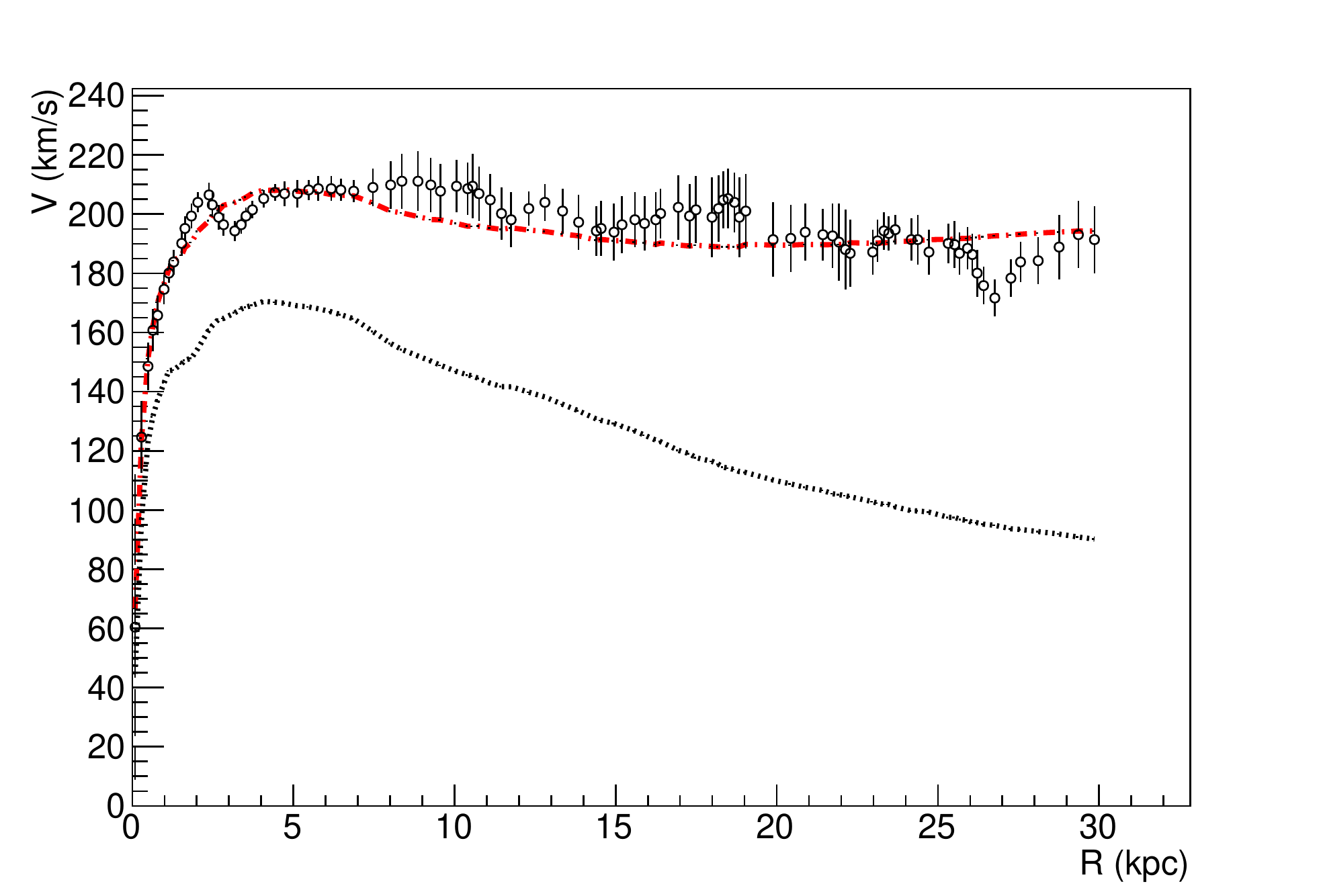}}
    
     \subfigure[ NGC 4138, Ref.~10]{\includegraphics[width=0.33\textwidth]{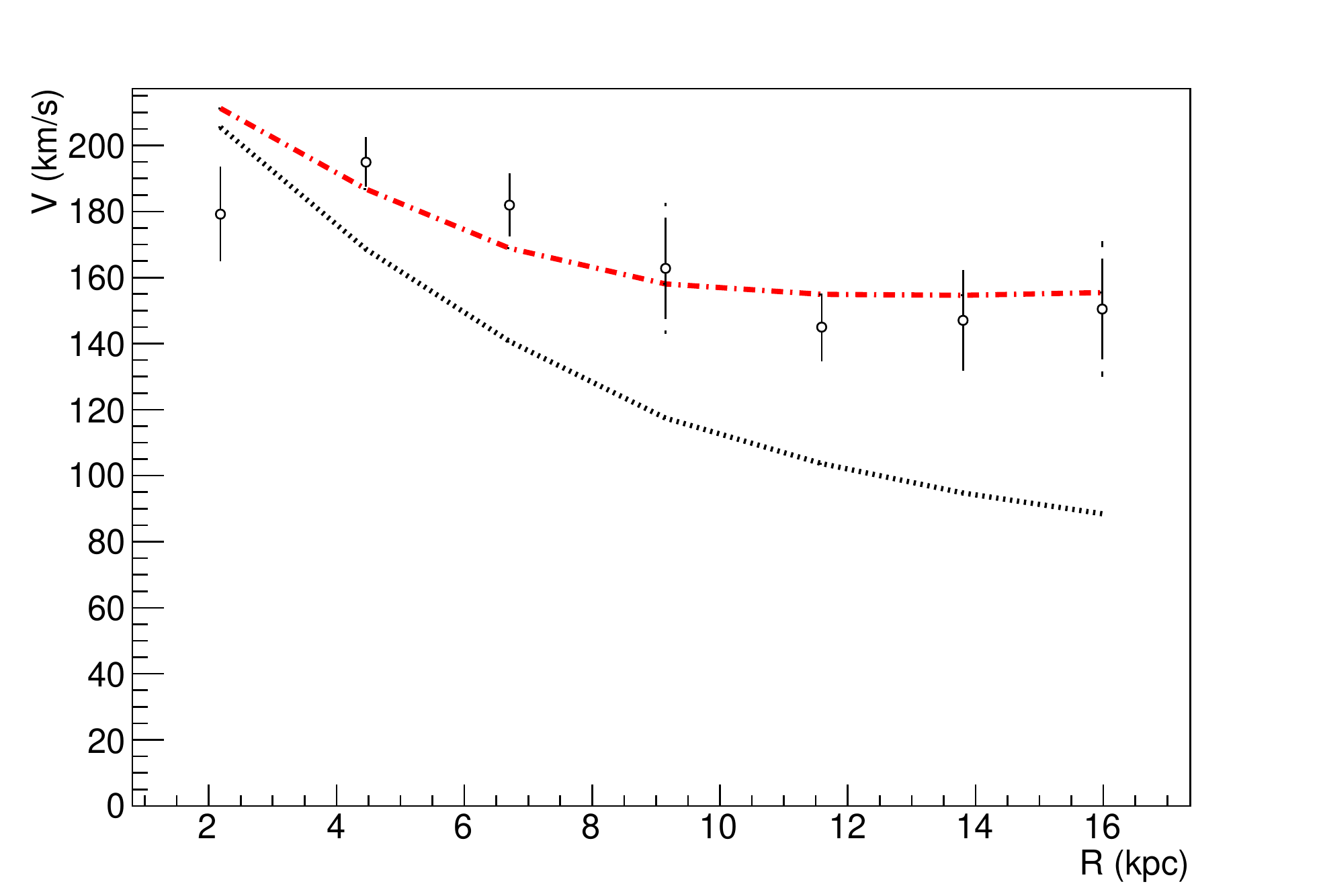}}
        \subfigure[ NGC 5907,  Ref.~10 ]{\includegraphics[width=0.33\textwidth]{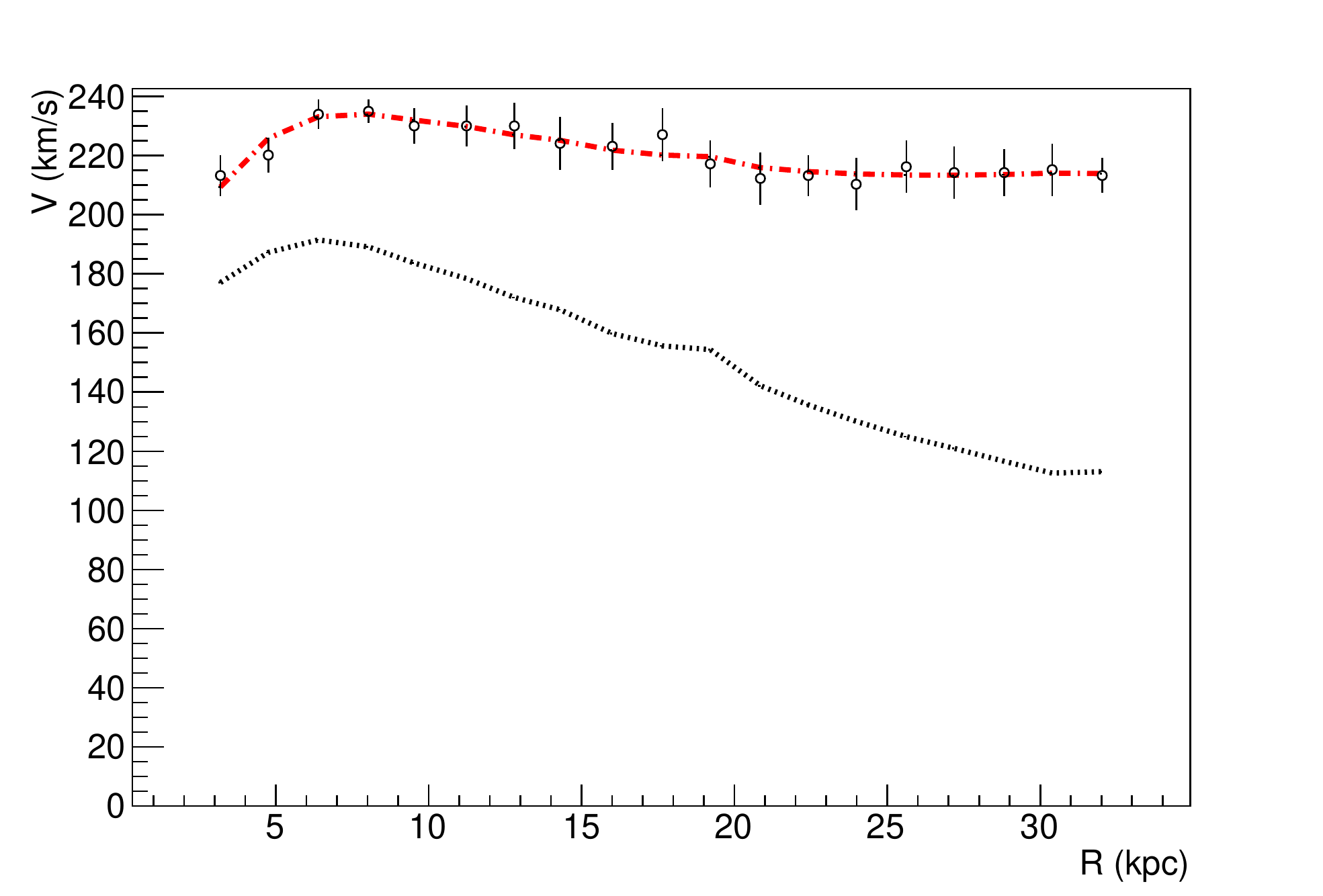}}  
            \subfigure[NGC 3992, Ref.~10 ]{\includegraphics[width=0.33\textwidth]{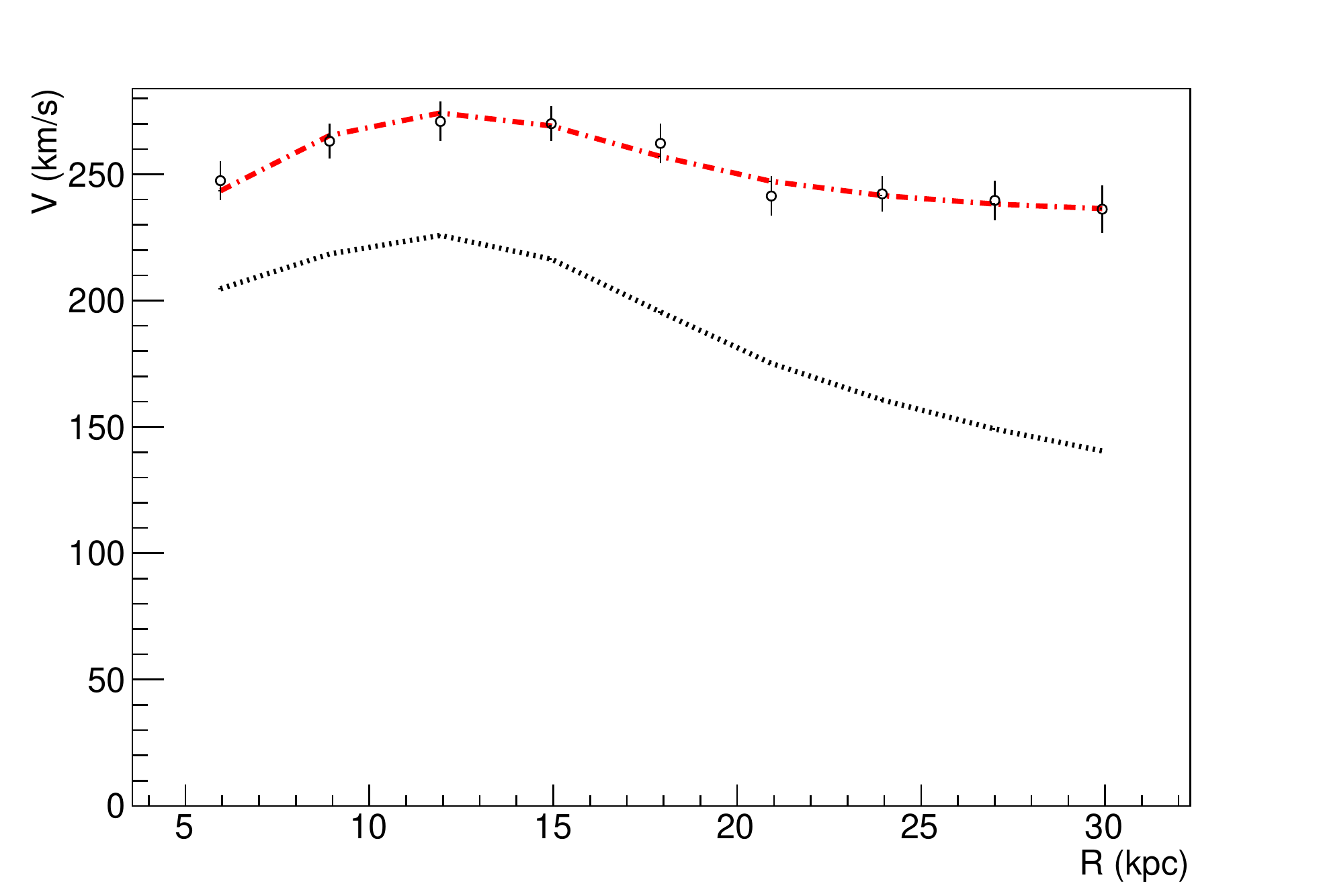}}

      \subfigure[ NGC 2903, Ref.~3 ]{\includegraphics[width=0.33\textwidth]{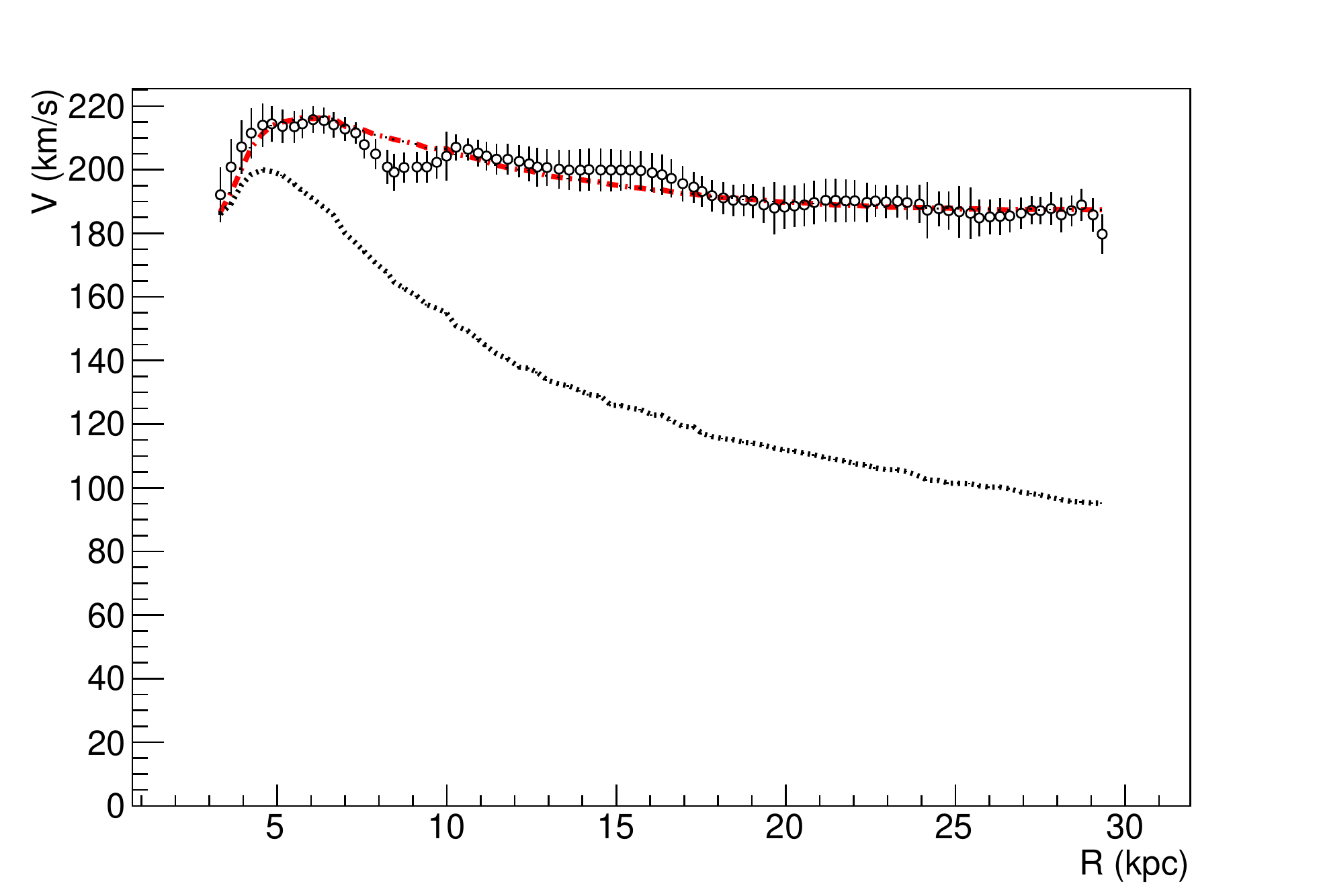}} 
      \subfigure[NGC 2903, Ref.~8]{\includegraphics[width=0.33\textwidth]{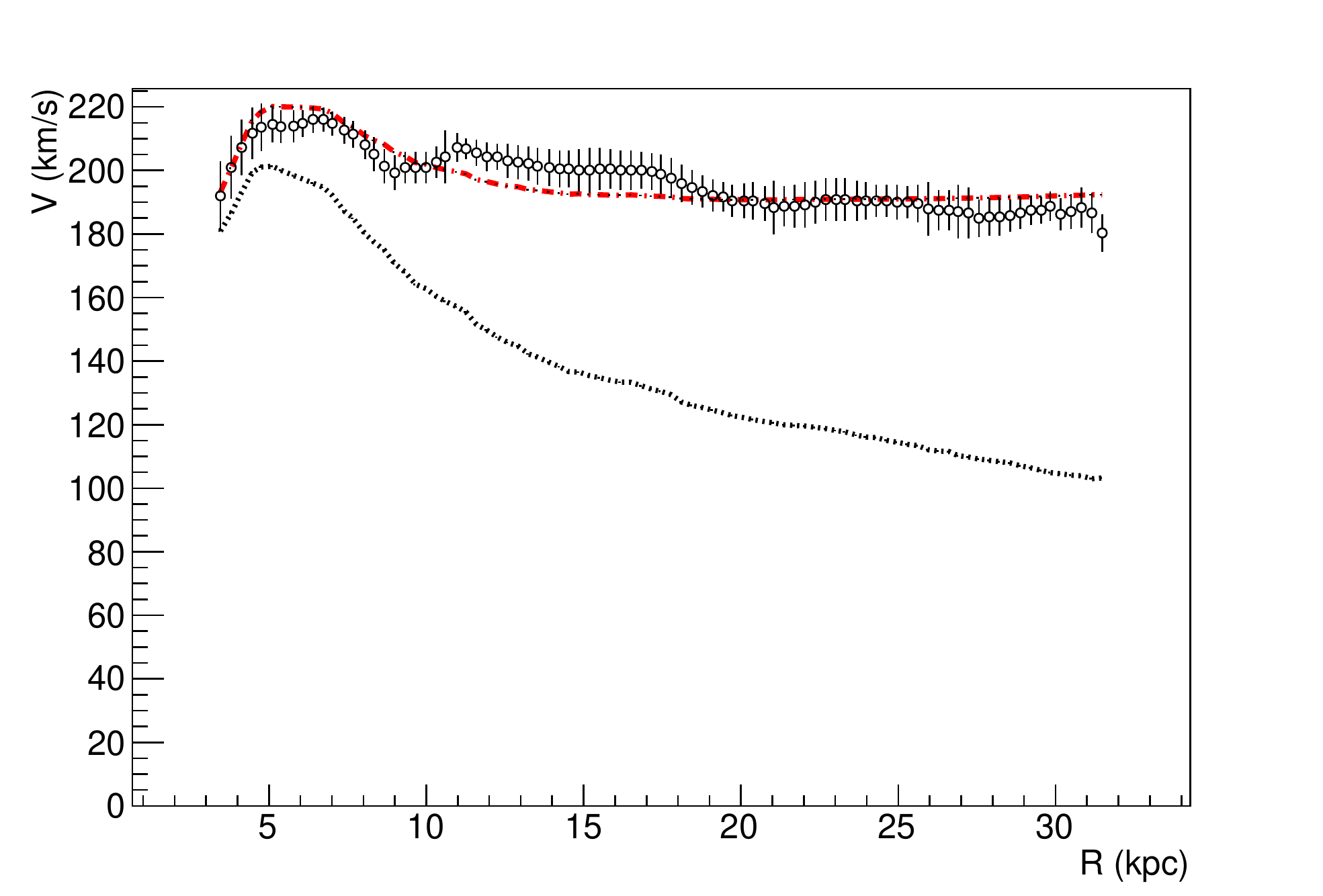}}
\subfigure[NGC 2903, Ref.~10 ]{\includegraphics[width=0.33\textwidth]{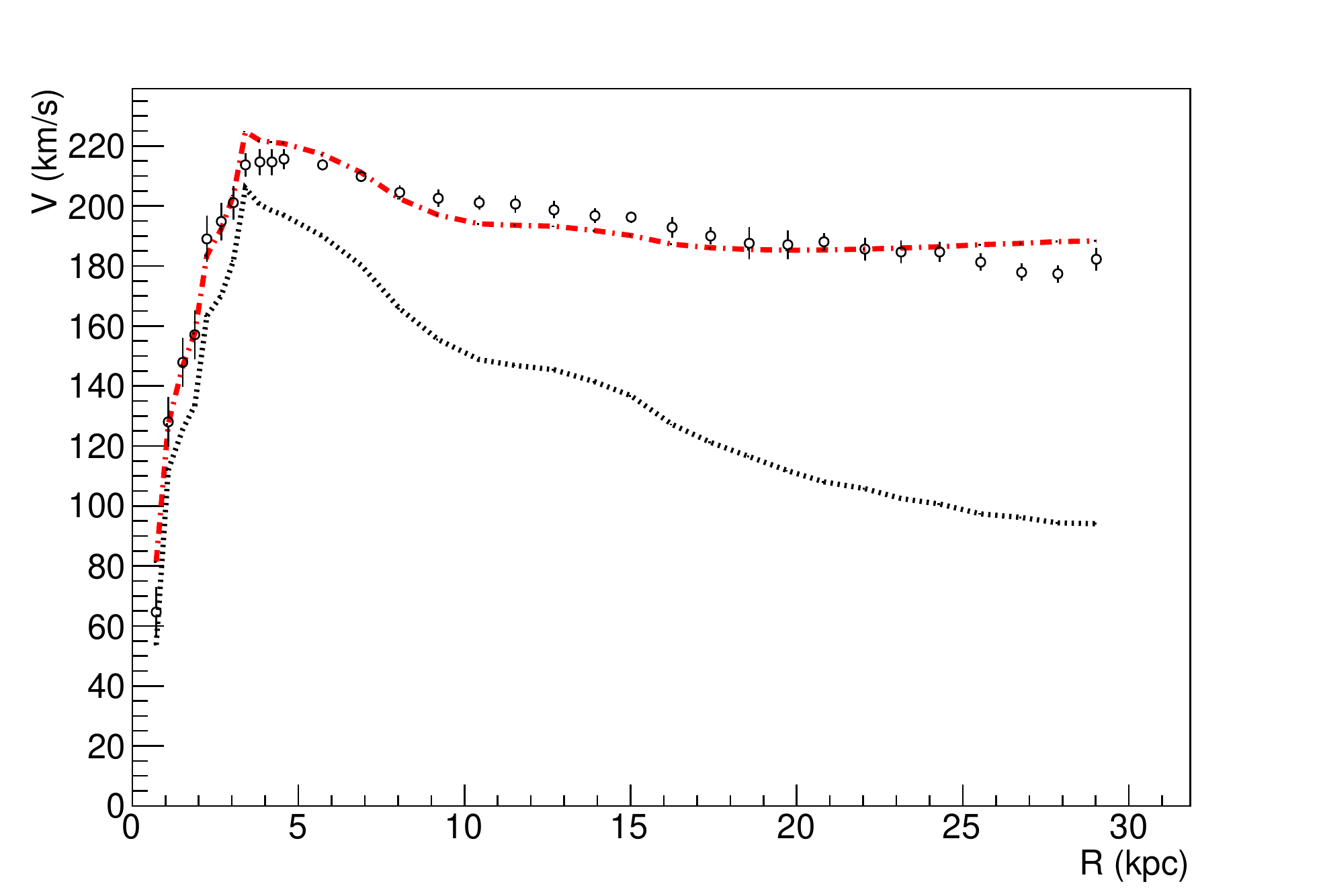}}\\

        \subfigure[NGC 6946, Ref.~3 ]{\includegraphics[width=0.33\textwidth]{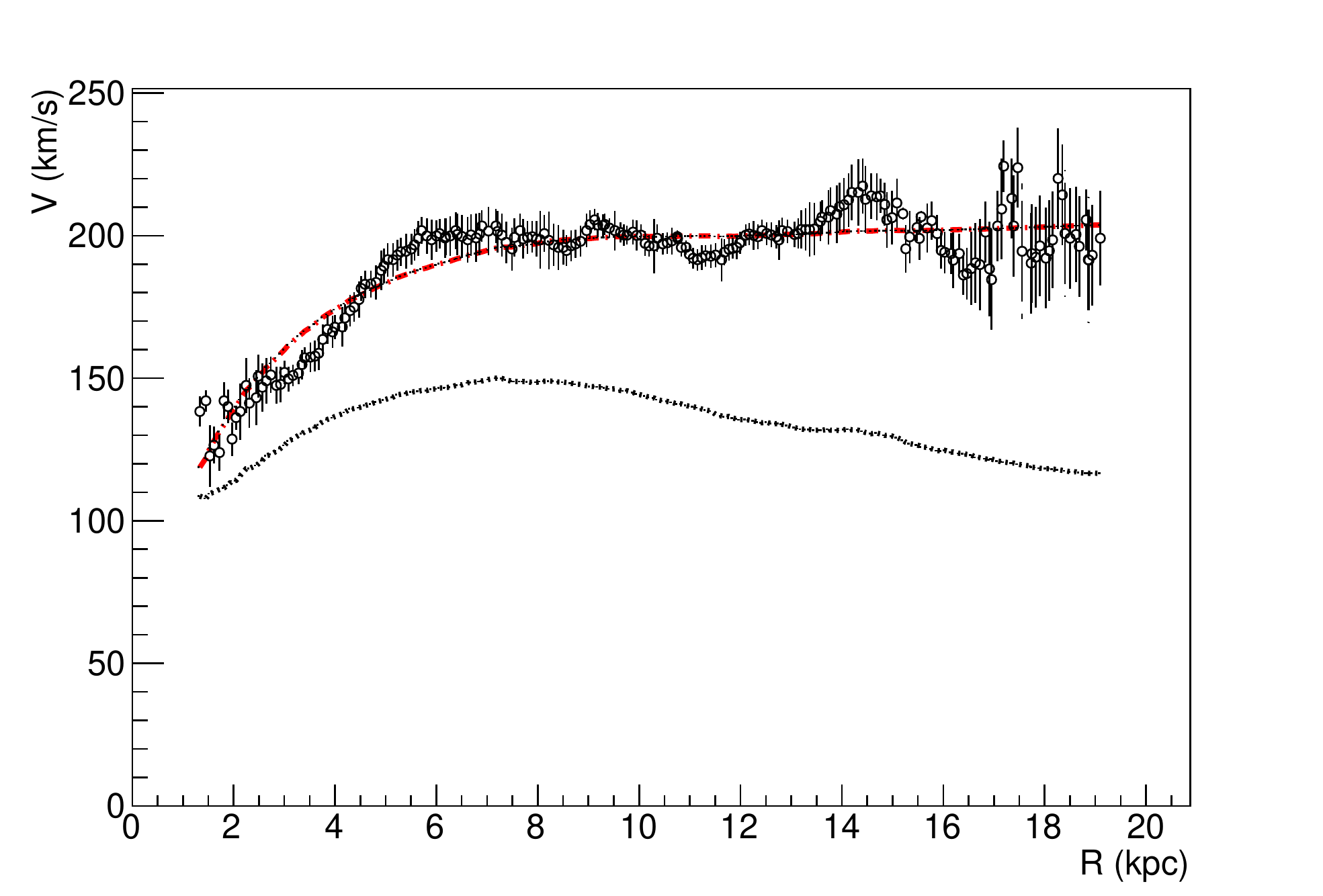}}
        \subfigure[NGC 6946, Ref.~10]{\includegraphics[width=0.33\textwidth]{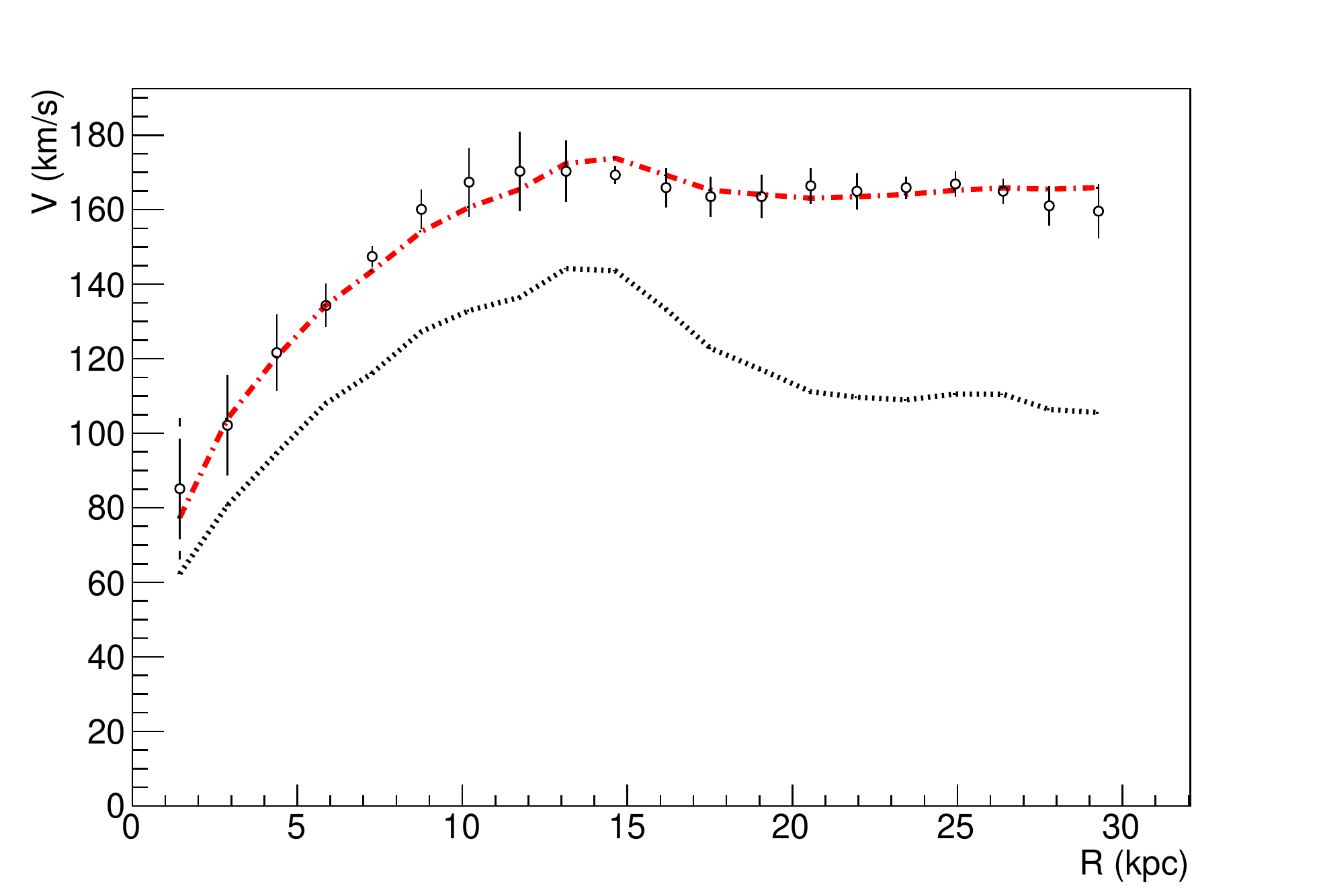}}
\caption{LCM rotation curve fits.   In all panels: lines are as in Fig.~\ref{fig:resultssmall}.   References  are as in Table~\ref{sumRESULTS}.}  
               \label{galaxiesSmaller}     
\end{figure*}

 The major distinguishing feature between the Ref.~3 luminous profile and those in Ref.~8 and Ref.~10 is that Ref.~3 includes a bulge. Since the errors and rotation curves from Ref.~3 and Ref.~8 are essentially identical, it is telling that the two separate luminous profiles returned essentially equivalent LCM $\chi^2_r$ values. However, visually the Ref.~3 luminous profile with a bulge, as compare to Xue Milky Way in Fig.~\ref{galaxiesSmaller} is slightly better than that for Ref.~8.  However, the Ref.~3 LCM results are unstable, as all the other Milky Way pairings returned a bulge M/L value of zero. This is an ambiguous result, because the chi-square for the Xue pairing is easily twice as good as any of the other pairs, see Table~\ref{sumRESULTSbigger}.  Additionally, this is the only pairing in the group of nine ($9$) pairs which returns a minimal gas scaling. 
 These three data sets do indicate consistency with the conjecture in  section~\ref{n891n7814} because the total light profile reported from Ref.~8 is  steeper in the density gradient in the inner $10$ kpc than either Ref.~3 or Ref.~10. This can be seen in the Keplerian profiles in   Fig.~\ref{galaxiesSmaller}.     

 \subsubsection{NGC 3953, UGC 6973,   NGC 4088 and NGC 3726}   

 The  rotation curves for   Ursa Major Galaxies  NGC 3953, UGC 6973   NGC 4088 and NGC 3726  are reported in Ref.~10~\citep{SanMcGa}, with luminous profiles resulting from   MOND fits.   Rotation curve data  is reported from  ~\citet{Ver98,San96}.  Rotation curve and modeling assumptions  are described in section~\ref{SanMc02}.   UGC 6973 and  NGC 4088 are reported as being kinematically disturbed. Fit results for all three galaxies are reported as consistent with PSM predictions.  Reported  gas masses,  in  H\,{\sevensize\bf I}  are: for N3953 $0 .27\times 10^{10}M_{\sun}$, for U6973  $0 .17\times 10^{10}M_{\sun}$,  for NGC 4088 $0 .79\times 10^{10}M_{\sun}$ and for  NGC 3726  $0 .62\times 10^{10}M_{\sun}$.

 These four galaxies are found to have    approximately the  same enclosed mass  at the limit of the data ($\kappa_\tau\approx 0.28$),  but   two distinct sets of $\tilde{a}$.  Averaged over the Milky Way pairings, these are: for 
  NGC 3953 \& UGC 6973   $\tilde{a}\approx 0.14$ and for  N4088 \&  N3726    $\tilde{a}\approx 0.08$.     As is clear in Fig.~\ref{fig:results3},   the more central mass distributions of    NGC 3953 and UGC 6973, as compared to      NGC 3953 and UGC 6973,   are consistent  with the conjecture in   section~\ref{n891n7814}.   
  
\subsubsection{NGC 6946} 
 
 NGC 6946 is a late-type spiral at the limit of low inclinations which can feasibly be studied kinematically.
The two   rotation curves we sample   are from Ref.~3~\citep{Blok} and  Ref.~10~\citep{SanMcGa}.   Their rotation curves differ enormously,  
both in magnitude and functional line shape with respect to $r$, see Fig.~\ref{galaxiesSmaller}. The rotation curve data in Ref.~10 comes from  ~\citet{Carig90} and they report  a MOND fit with luminous matter modeling estimates reported in  ~\citet{San96}. Ref.~3 notes that the differences in their reported rotation curves is most likely due to   the assumption of inclination angle; Ref.~3 uses  $i=32.6^{\circ}$ versus that  of Ref.~10  $i=38^{\circ}$.  They also use different distance estimates: Ref.~3 uses  $5.9$ Mpc ~\citep{Walter} and Ref.~10  uses $10.1$ Mpc
 ~\citep{Carig90}.   Ref.~3 notes for NGC 6946 the large galaxy size relative to the beam size. 
 
Since the  error budget on the Ref.~3  rotation curve is easily twice that of Ref.~10, it is significant that the LCM fits to from  Ref.~10  assumptions  are   three times better than those from the  Ref.~3 assumptions. Physically, Ref.~3 and Ref.~10 provide very different  NGC 6946s.  Ref.~3 gives a fixed bulge and disk M/L from PSM, and Ref.~10   gives only a stellar disk, with M/L free.    The LCM decreases the M/L reported in Ref.~3 by approximately $40\%$, and increases that reported by Ref.~10 by about the same margin.  The galaxy, as reported by Ref.~3, yields maximal gas scalings, whereas the galaxy as reported by Ref.~10 returns minimal gas scalings.  

 \subsubsection{NGC 3198}
  \label{n3198} 
 
\begin{figure*} 
 \centering
  \subfigure[ NGC 3953, Ref.~10 ]{\includegraphics[width=0.33\textwidth]{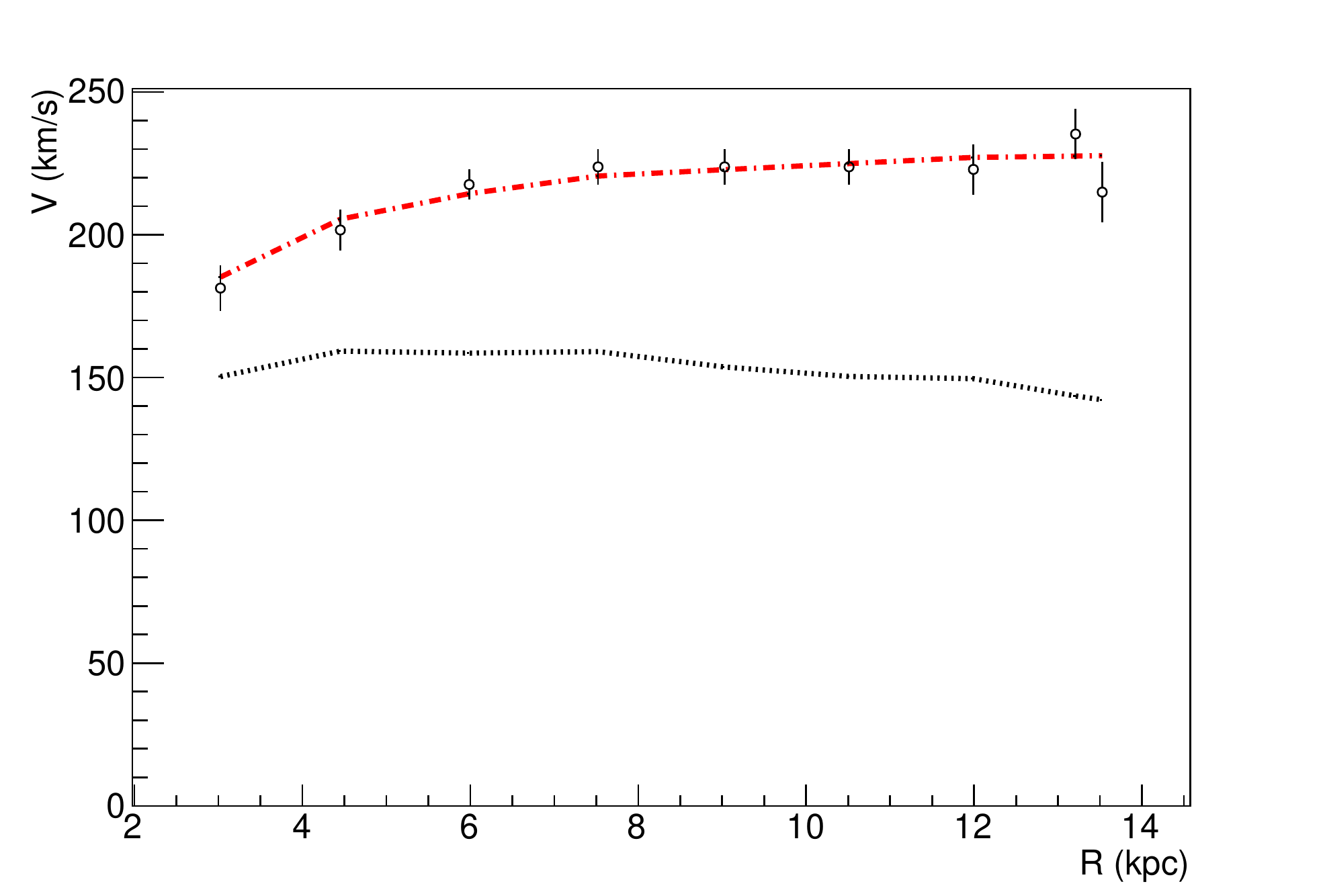}}
    \subfigure[ UGC 6973, Ref.~10 ]{\includegraphics[width=0.33\textwidth]{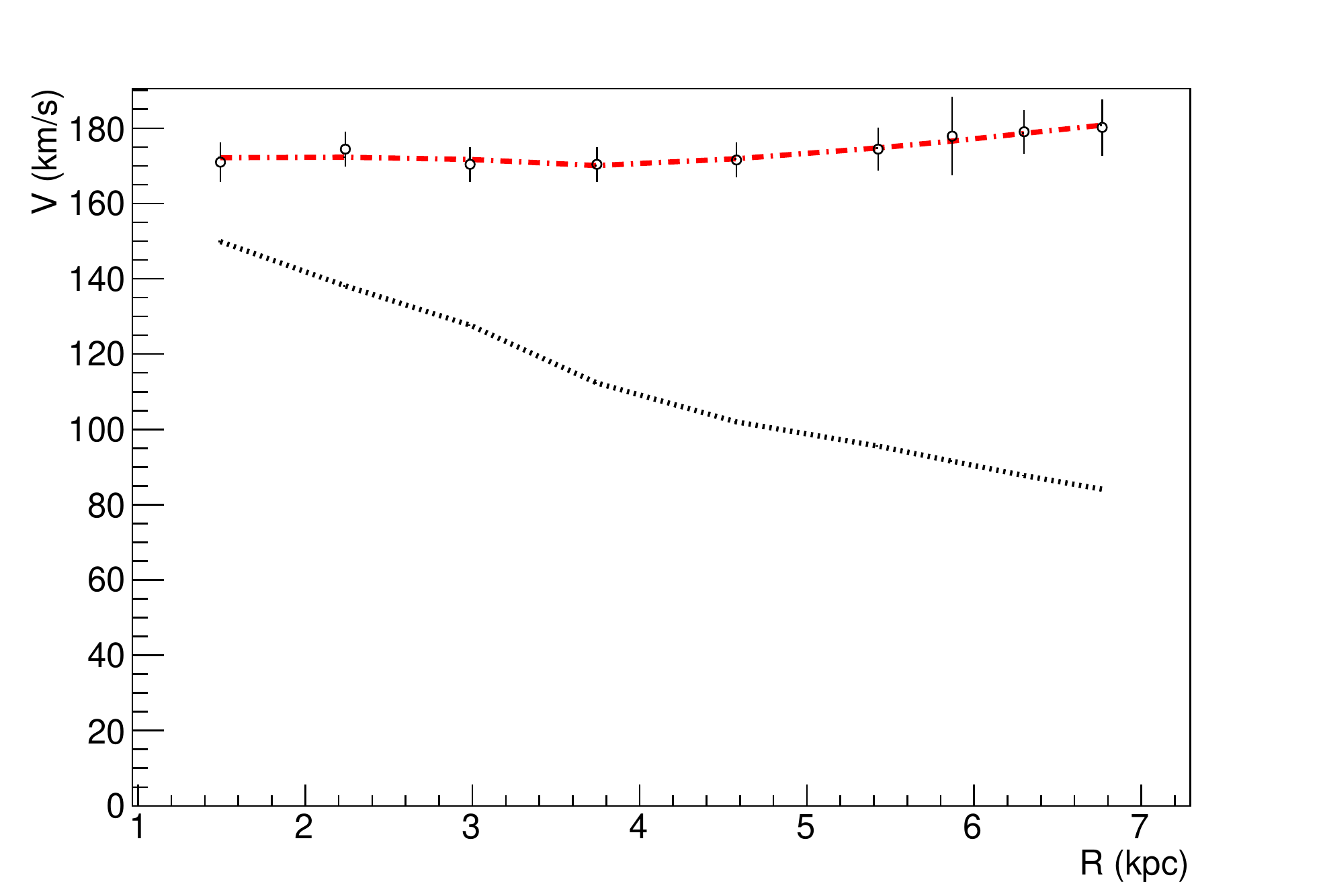}}
      \subfigure[ NGC 4088, Ref.~10]{\includegraphics[width=0.33\textwidth]{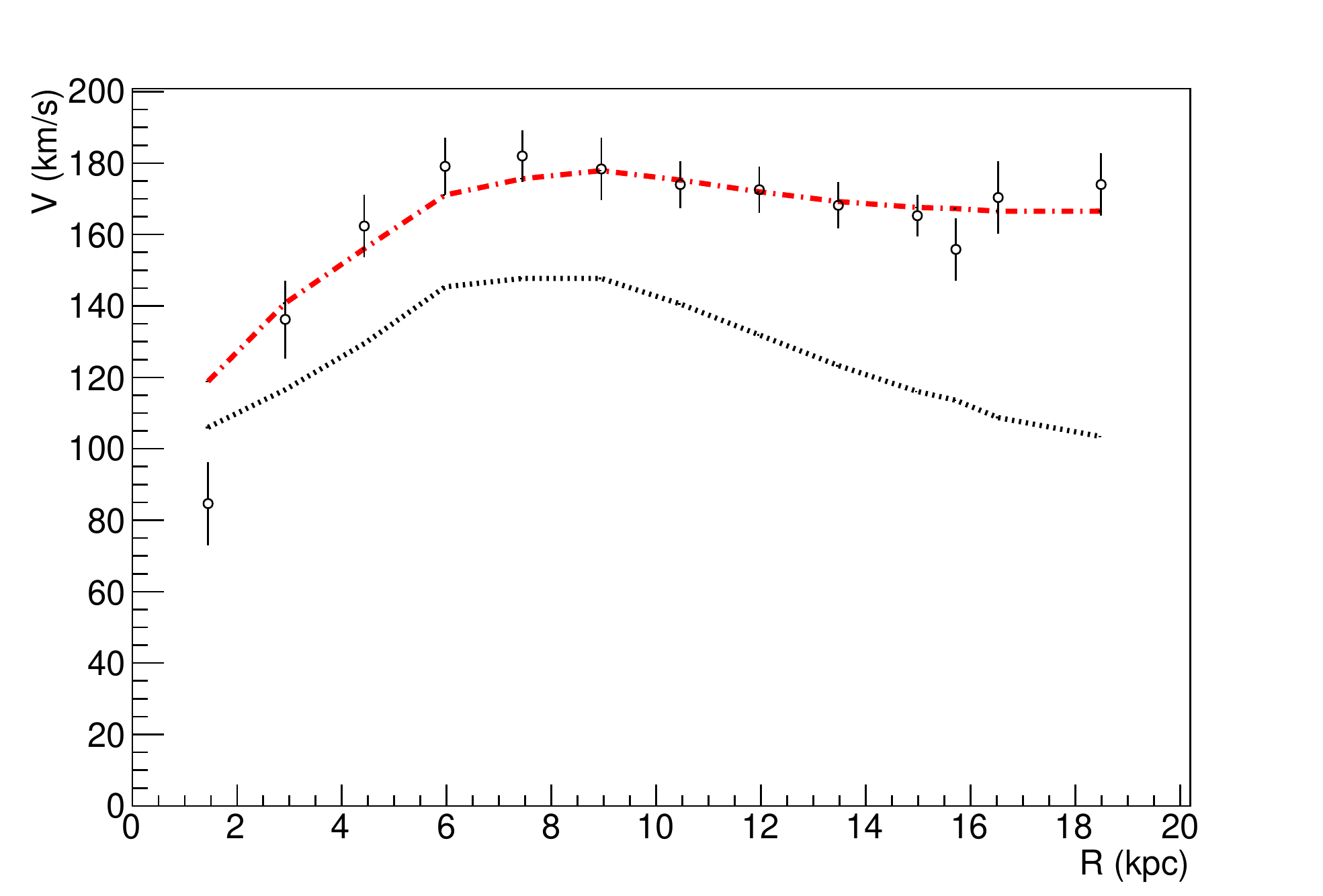}} \\
      
       \subfigure[ NGC 3726, Ref.~10]{\includegraphics[width=0.33\textwidth]{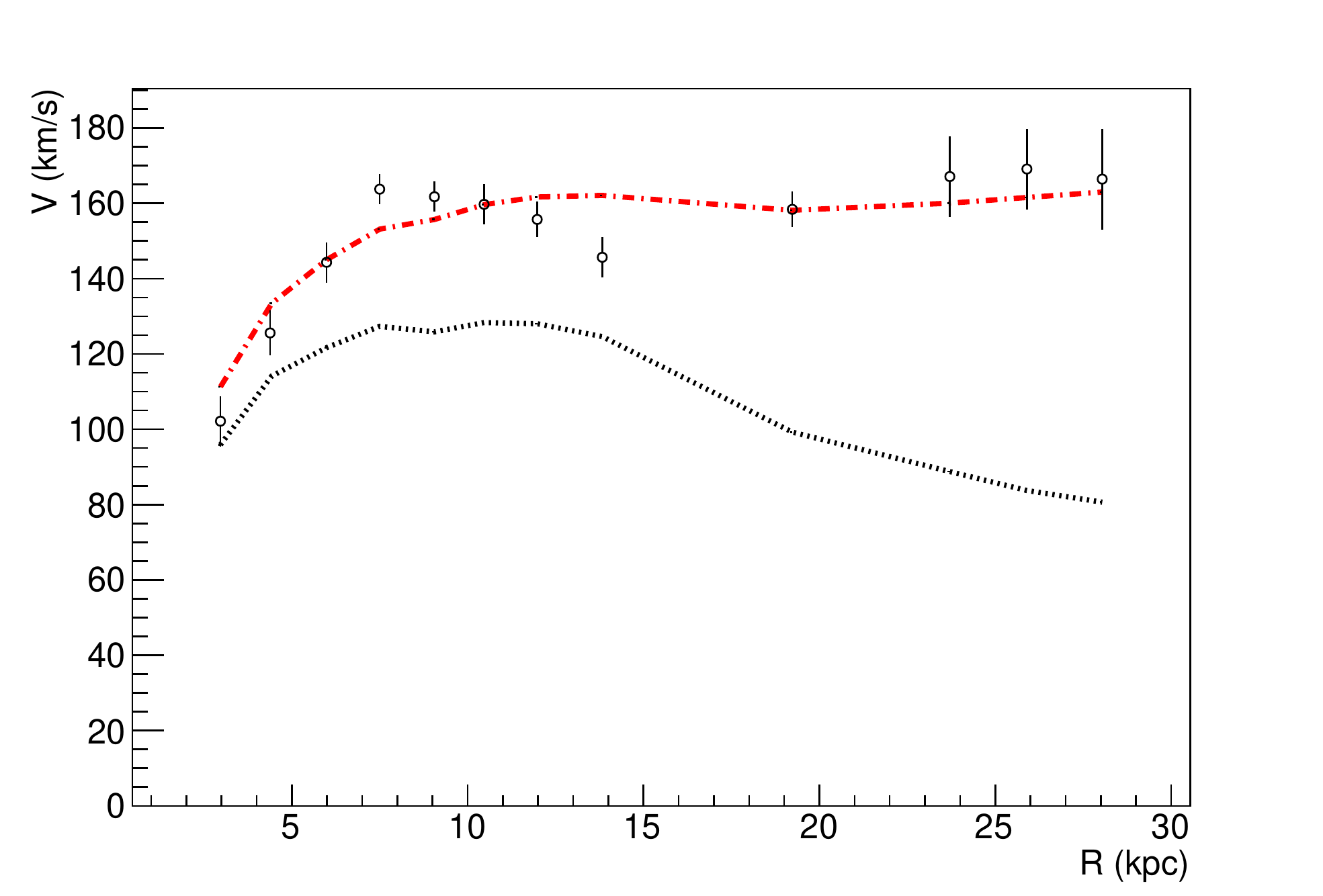}}
         \subfigure[NGC 3198, Ref.~1]{\includegraphics[width=0.33\textwidth]{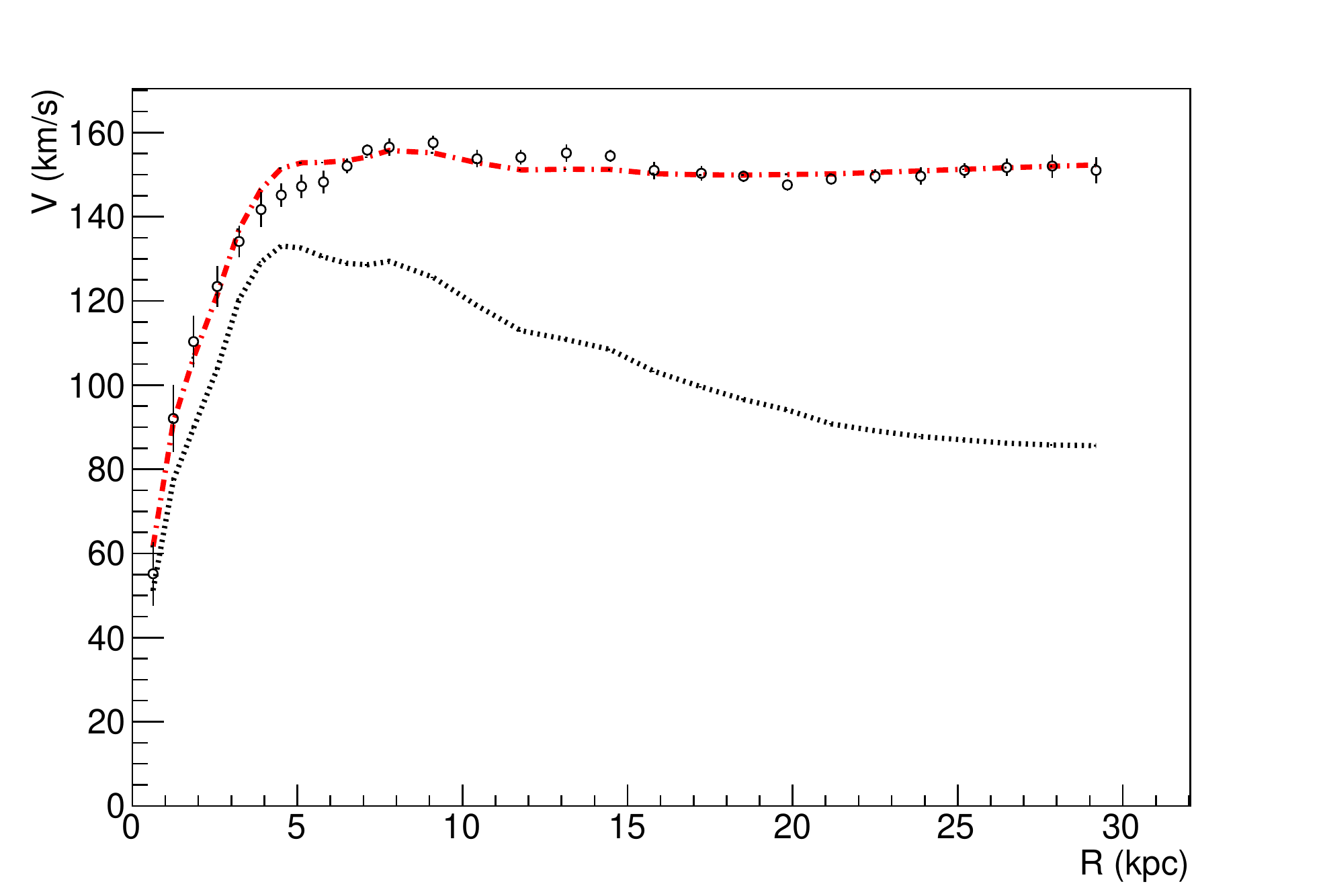}}
        \subfigure[NGC 3198, Ref.~2]{\includegraphics[width=0.33\textwidth]{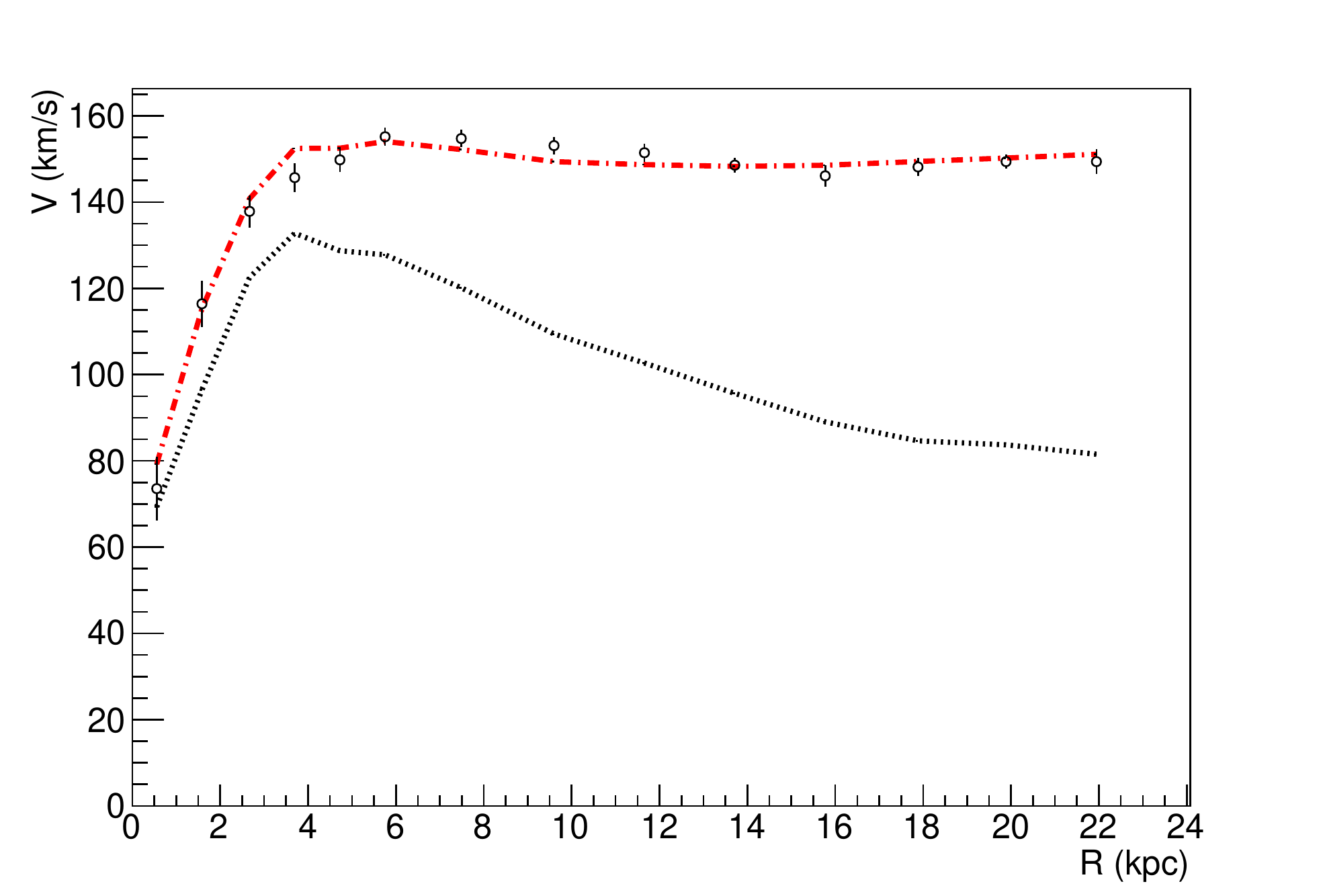}} \\
         
        \subfigure[ NGC 2403, Ref.~3]{\includegraphics[width=0.33\textwidth]{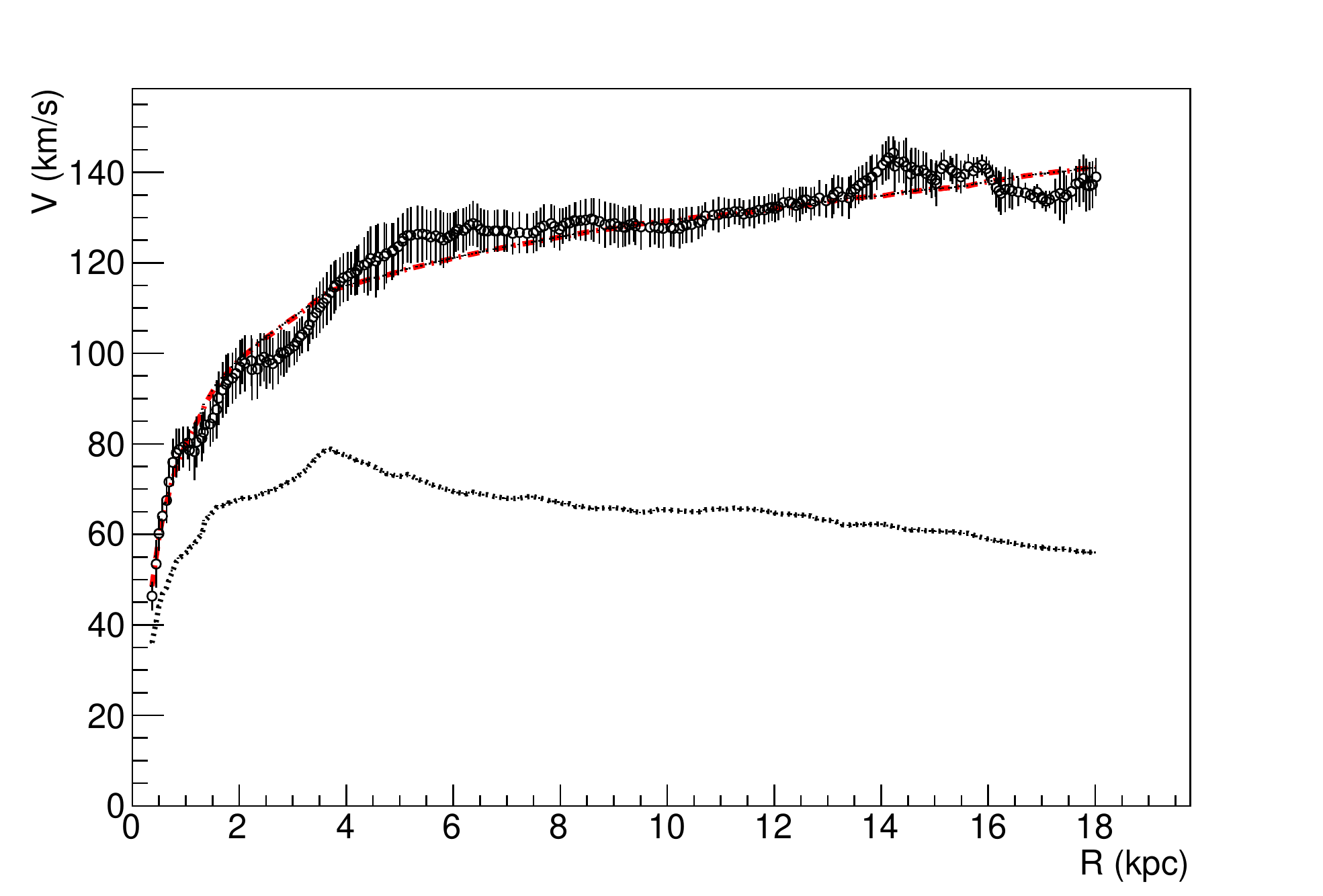}} 
         \subfigure[ NGC 3198, Ref.~3]{\includegraphics[width=0.33\textwidth]{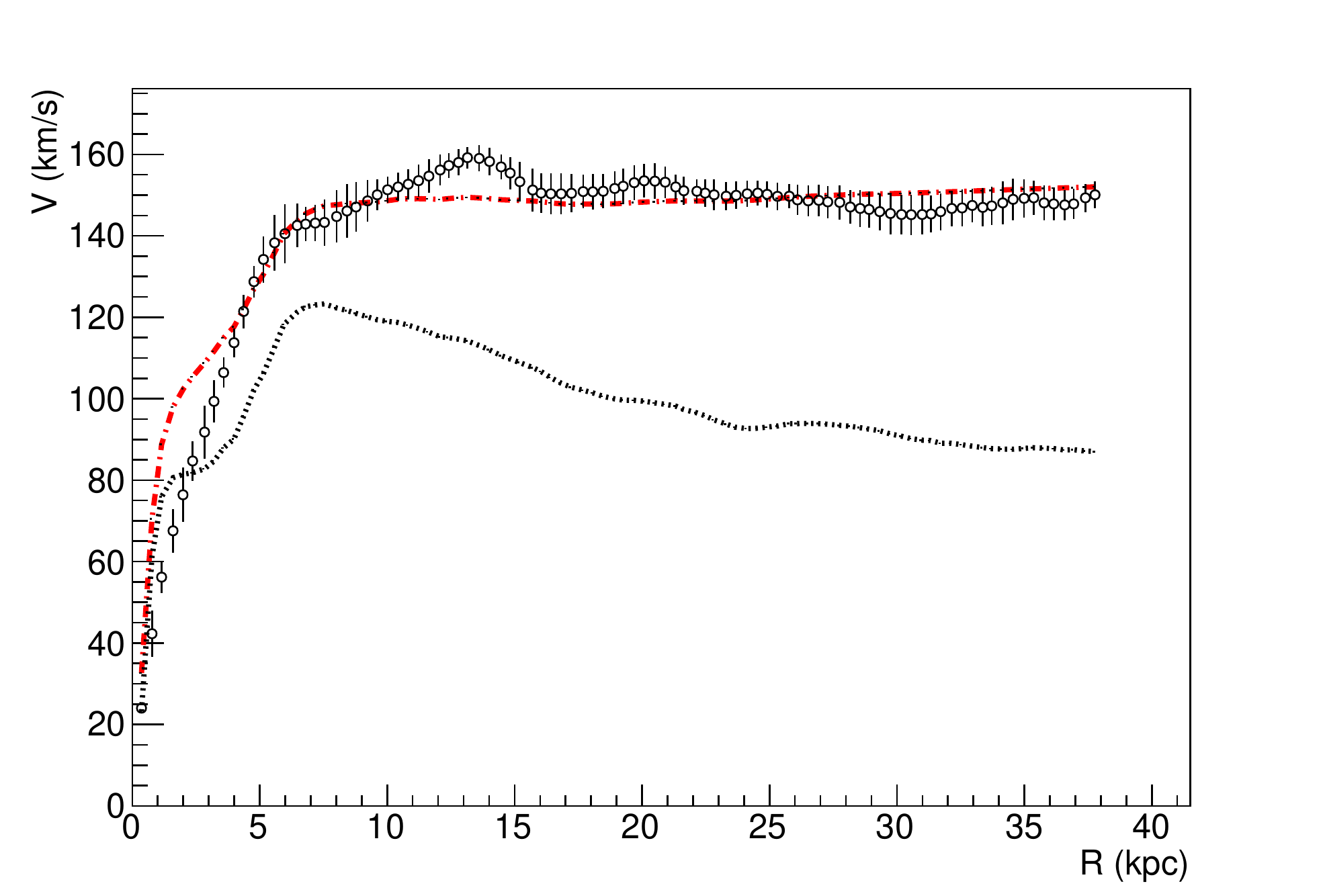}} 
          \subfigure[ NGC 3198, Ref.~4]{\includegraphics[width=0.33\textwidth]{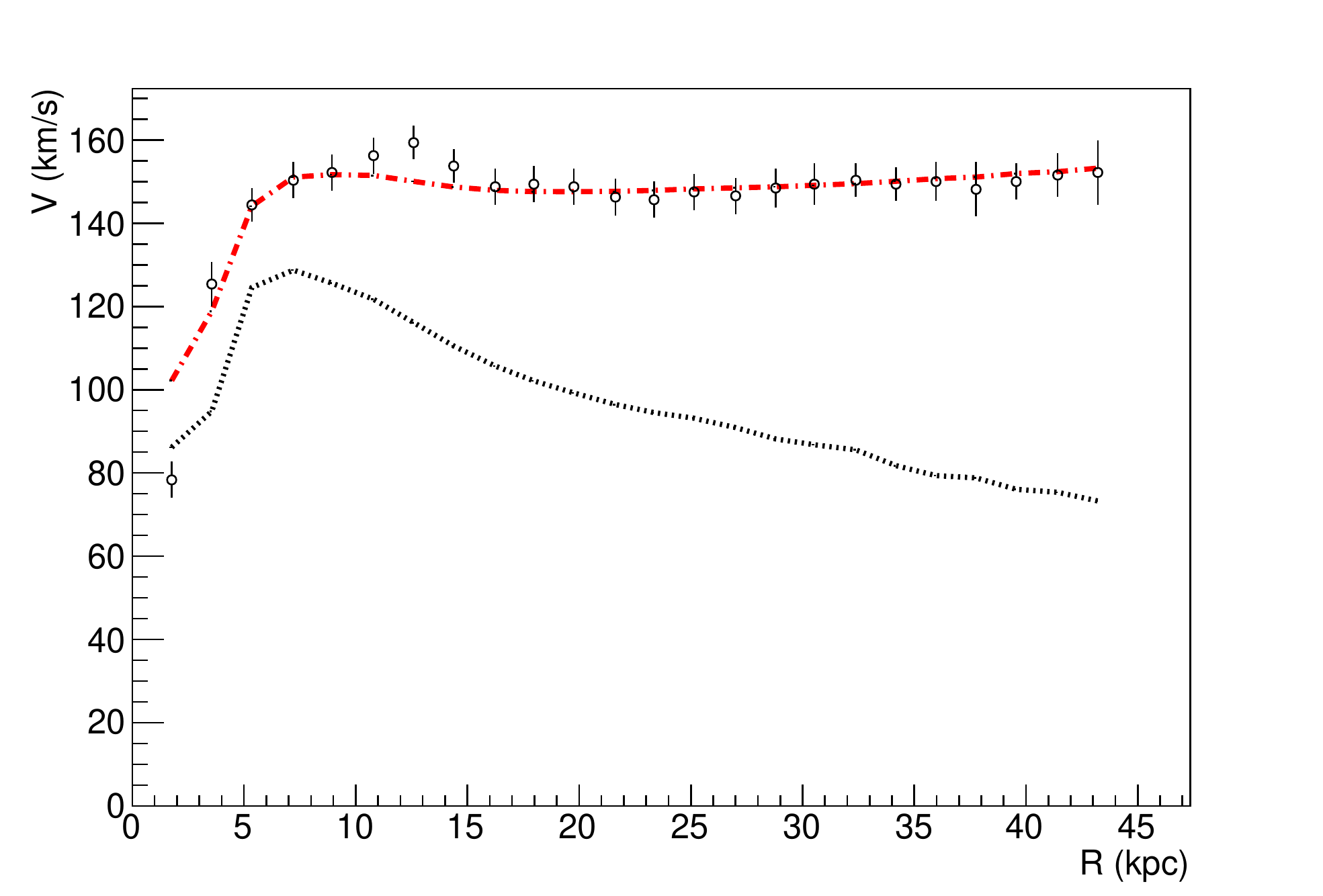}}\\
          
            \subfigure[ NGC 2403, Ref.~9]{\includegraphics[width=0.33\textwidth]{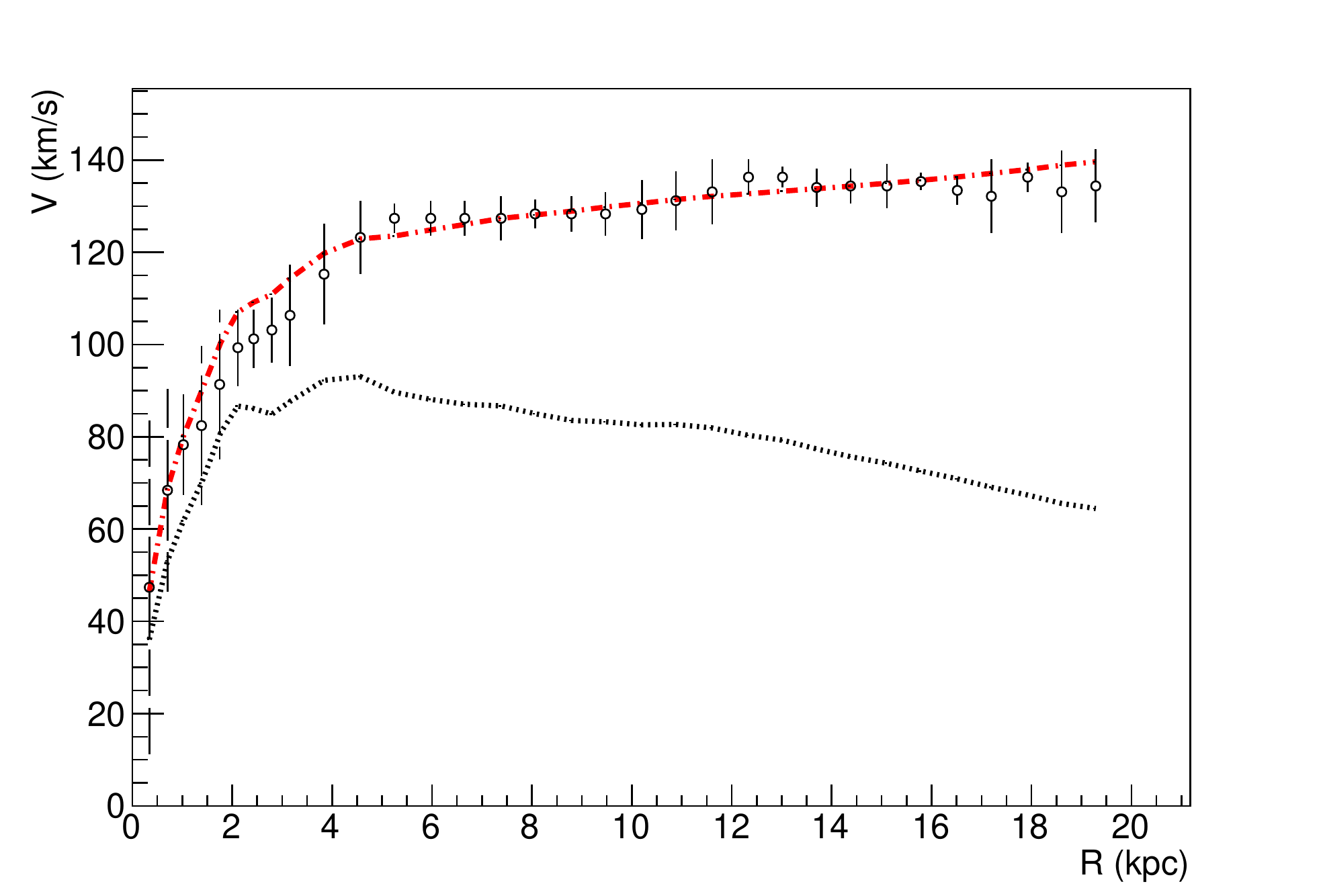}} 
             \subfigure[M 33, Ref.~5]{\includegraphics[width=0.33\textwidth]{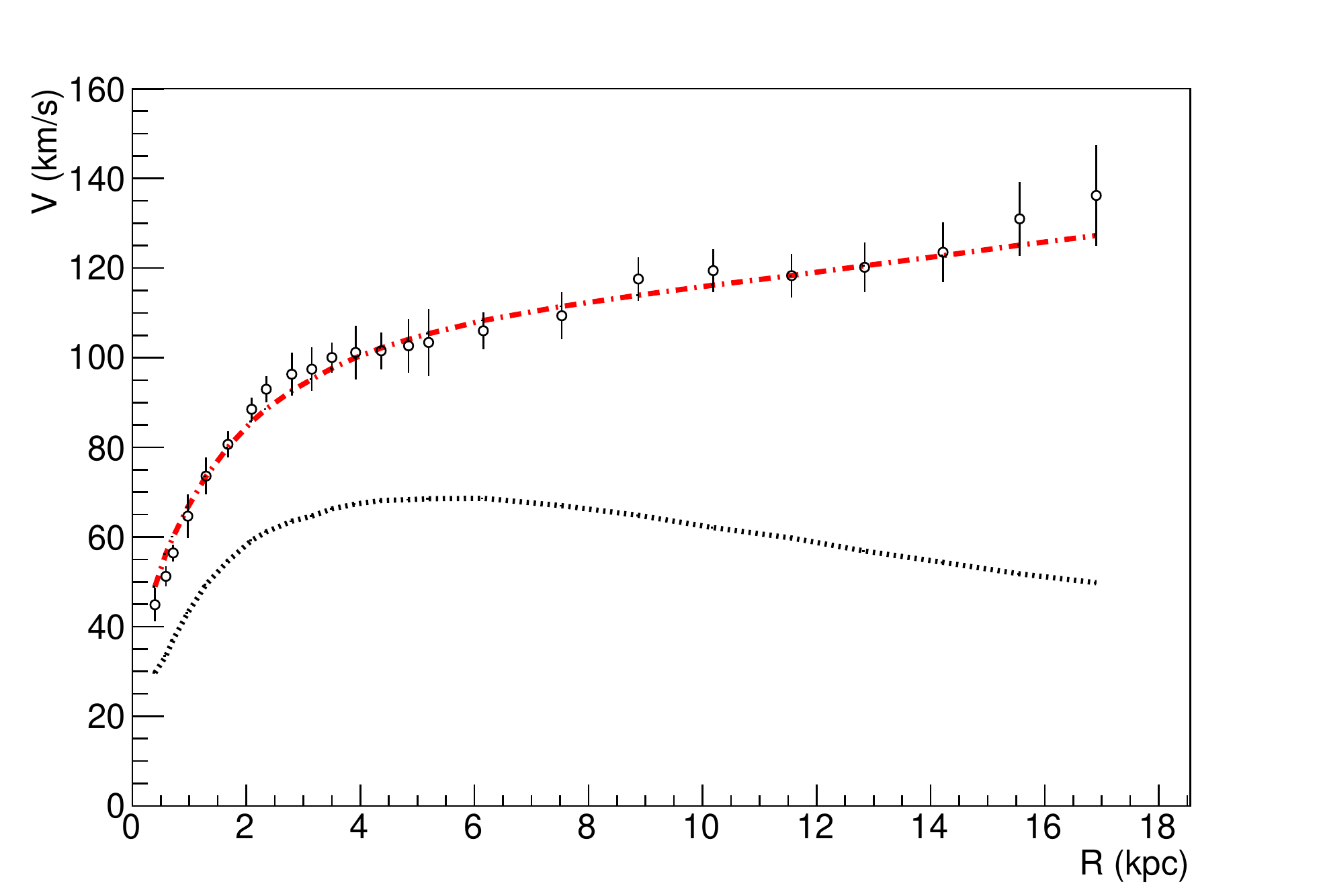}} 
              \subfigure[ M 33, Ref.~6]{\includegraphics[width=0.33\textwidth]{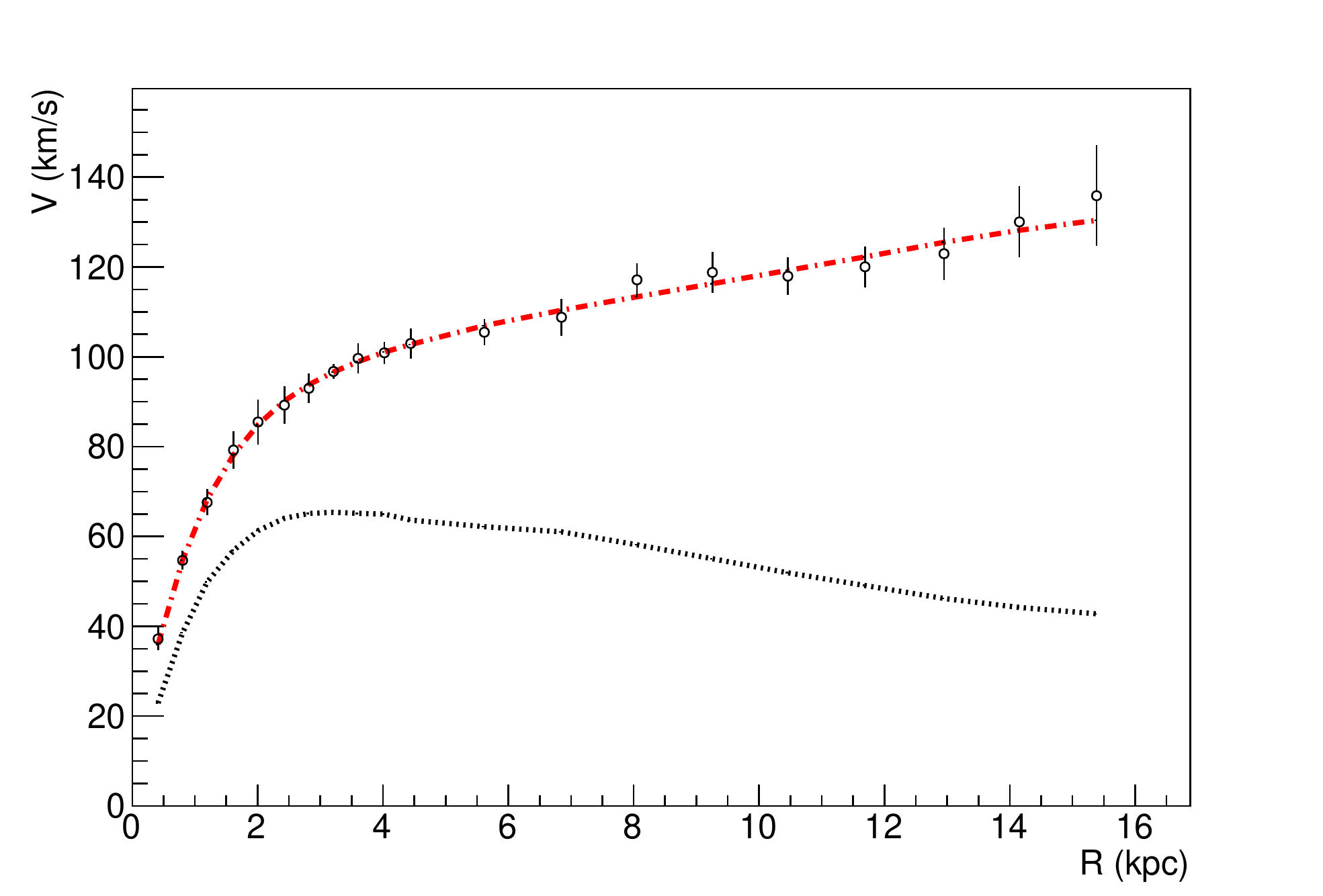}} 
 \caption{LCM rotation curve fits.   In all panels:lines are as in Fig.~\ref{fig:resultssmall}.    References  are as in Table~\ref{sumRESULTS}.}    
            \label{fig:results3}  
\end{figure*}

 For the high surface brightness spiral galaxy NGC 3198 we sample four different data sets.  
NGC 3198 is a moderately-inclined, late-type galaxy with   a small central bar and   is one of the standards for flat rotation-curve studies  due to its high symmetry and regular rotations.   We   compare the best gas measurements from Ref.~4~\citep{Maria},  the 
   best rotation curve data Ref.~ 3~\citep{Blok},  the older standard from  Ref.~1~\citep{Bege}, and   an alternate interpretation of the  Ref.~1 rotation curve in    Ref.~2~\citep{JNav}.   Ref.~1-3   report  NFW fits  and Ref.~4  a  MOND fit.    Ref.~1 and 2  use the distance   of  $9.4$ Mpc (Hubble Flow),      Ref.~3  uses $13.8$ Mpc (Cepheids), and    Ref.~4 uses  $12.3$ Mpc (constrained from the Cepheids).  Ref.~1  and Ref.~ 2  use the ~\citet{BegeTH}      rotation curve, based on   a tilted-ring model derivation.     The Ref.~3  rotation curve derivation is described in section~\ref{Blok}. The Ref.~4 rotation curve is from the HALOGAS ~\citep{Heald}  survey of the extra-planar gas in 22 spiral galaxies using very deep H\,{\sevensize\bf I}  observations, , augmented by $\rmn{ H}\alpha$ observations. The   rotation curve derivation reported in   Ref.~4 is  from a  tilted-ring model with intensity-weighted velocity fields, treating each side of the galaxy separately.   Ref.~4   uses a  3D model of the   H\,{\sevensize\bf I}   layers, giving attention to variable inclinations for the rings, position angles, and the features of variable rotation speeds as a function of distance from the plane.   They find previously unreported gas features and show a lag on the rotation velocities of the extra planar gas.   They analyze the H$\alpha$ emission to detect stellar light out to the end of the extra planar gas.       Ref.~4   reports   H\,{\sevensize\bf I}  gas mass of $1.08\times 10^{10} M_{\sun}$ which is $6 \%$  higher than reported in Ref.~3. They attribute the difference to amplitude calibrations and detection of extra planar gas.   Ref.~1  and 2 report     gas mass  of $5.0\times 10^9 M_{\sun}$.   All four  citations return  maximal LCM gas scalings of $1.5$.  In  Ref.~4,  the authors show   all  the   NGC 3198 rotation curves   compared here,  plotted on an arc second scale. On that scale, all the rotation curves are similar,  modulo the noise.  However, as can be seen in  Fig.~\ref{fig:results3},   the conversion   to the assumed    distance scales     causes   projection differences in the rotation curves.   
   
  As can be seen in Fig.~\ref{fig:massmodels18}, the reported luminous mass profiles vary significantly for this galaxy in geometry and magnitude. Modulo fixed geometry assumptions, the LCM returns   similar   luminous profiles for all four data sets,  as shown in Fig.~\ref{fig:results3}.  Ref.~4 uses the same geometry as Ref.~3, but  a lower M/L ratio.   
  Ref.~ 1  and Ref.~2 respectively report     M/L    of $3.8$ and $1.39$ in the r-band,  in solar mass units.  The resulting LCM average M/L   were respectively $3.3$ and $2.9$, demonstrating convergence.     Ref.~3   and Ref.~4 respectively report    M/L  of $0.8$ and $0.3$ in the $3.6 \,\umu \rmn{m}$  band,  in solar mass units.  The resulting LCM average M/L   were respectively $0.64$ and $0.61$, again demonstrating convergence.  Visually, the   LCM fit values for Ref.~2 are basically equivalent to those of Ref.~1 and those of    Ref.~4 are superior to  Ref.~3.  The Ref.~3 assumptions  over-predict  the  luminous profile at small radius. 
  
The  $\tilde{a}$ values are very consistent across all four of these posited luminous NGC 3198 profiles, despite differences in the posited distances.      

\subsubsection{NGC 2403}
Fig.~\ref{fig:results3}
NGC 2403 is a late-type Sc spiral in the M81 group,  dominated by  regular rotations. We sample two data sets:
 Ref.~3~\citep{Blok}  and
Ref.~9~\citep{Bot}.   The Ref.~3 luminous profile is fixed by PSMs, and  that of  Ref.~9 is  from a  MOND fit; both using  the  Cepheid based distance of  $ 3.22$ Mpc.   Ref.~3 reports an NFW fit to the rotation curve data.  The reported rotation curves are identical, modulo noise, as are the gas  profiles.   Ref.~9 reports a total gas mass of 
$0.4\,\times 10^{10}\,M_{\sun}$, which is   H\,{\sevensize\bf I}  multiplied by $1.3$  to account for primordial Helium abundances.      Ref.~3  notes  that the velocity in the inner kiloparsecs are   suspect to beam smearing effects,  given the steep increase in velocity.    

Both Ref.~3 and 9 use exponential disk geometries, respectively reporting stellar disk M/L in   Ref.~9 of $\rmn{M/L}_B=1.6$ and Ref.~3
 of  $\rmn{M/L}_{3.6\,\umu \rmn{m}}=0.41$. As can be seen in Fig.~\ref{fig:massmodels18}, this results in    higher     Keplerian velocities   due to the stellar disk in Ref.~9 than in Ref.~3,  by approximately $30\, \rmn{km\,  s^{-1}}$.    The LCM scales down the stellar disk M/L from   Ref.~9 and maintains that reported in     Ref.~3.  Both sets of fits return maximal gas scalings. 
 
    That the $\tilde{a}$ values  from Ref.~3 assumptions are slightly higher than those for Ref.~9, consistent with the conjecture in   section~\ref{n891n7814}, given that the geometry assumptions for the exponential disk in Ref.~3 indicates the same inner rise rate as   in Ref.~10, but a much faster decline in the light, and associated mass, profile -- and thereby a steeper density gradient.  
 
\subsubsection{M 33} 

The galaxy M 33 in our Local group  is used as a standard distance calibrator,  measured independent of the Hubble flow to be
at a distance of $0.84$ Mpc. It  is a  low-luminosity, very blue,  active star forming spiral galaxy with  major implications for galaxy evolution and formation models.    M33 is   smaller than the Milky Way and hence tends to be denoted as being dark matter dominated ~\citep{Salucci}.    The LCM successfully fits this galaxy without a dark matter halo in two reports: Ref.~5~\citep{Cor03}and  Ref.~6~\citep{CoSa00}.  These two rotation curve are respectively observed in:  molecular gas (Ref.~5) and  $21$-cm  (Ref.~6). The two rotation curves differ only in 
 fine detail and have essentially the  same    error budgets.    Ref.~5  reports their luminous profile in the context of an  NFW   fit, and    Ref.~6 in the context of a  model independent dark matter fit, consistent with CDM models. 
 
Including molecular gas in Ref.~5 results  in total gas mass of Ref.~5   $3.2\times 10^9 M_{\sun}$ as compared to that  previously reported in  Ref.~ 6 of  $1.8\times 10^9 M_{\sun}$).  This mass is predominantly   added  to M 33 past the radius  of  $8$\,kpc, which taken with the two different scale lengths used for the exponential disk geometries, results in slight modifications in the overall geometries implied in the two models. The
   LCM $\chi^2_r$ values  indicate that the     Ref.~6 luminous distribution  is approximately five times better than that of      Ref.~5. Since the gas observations are more complete  in Ref.~5, we assume the difference between these two fits is based upon     the choice of scale length.  Ref.~5 uses an  exponential scale length of $R_d\approx 1.3\pm0.2$ kpc and Ref.~6 uses $R_d\approx 1.2\,\pm\,0.2$\,kpc,  both   in the K-band.  
   
Consistent with the conjecture in section~\ref{n891n7814}, values for  $\tilde{a}$ are larger  for Ref.~6  assumptions with a  steeper density gradient.  Ref.~5 notes that  after subtracting the disk profile from the total light, there remains excess central emission which motivates the interpretation given here.  

\subsubsection{F 563-1 and NGC 925 and NGC 7793 }

The smallest galaxies in our sample are F 563-1, NGC 925 and NGC 7793.  For each galaxy we compare   two reports in the literature  (rotation curve and luminous assumptions).   
For F 563-1, we compare   NFW fits in    Ref.~2~\citep{JNav}  and Ref.~13~\citep{giraud2000universal}.
Ref.~2 is   rotation curve    data reported  by  ~\citet{deBlok:1997ut} and Ref.~13 is    rotation curve data reported by ~\citet{mcgaugh1998testing}.
 For  NGC 925,  we compare the fits from Ref.~3~\citep{Blok}, reported in the context of    NFW and   ISO models and  both reporting M/L fixed by PSM.  For  NGC 7793, we compare a MOND fit in Ref.~8~\citep{Gent}  and  an ISO fit in Ref.~14~\citep{Dicaire}.  
 
\begin{figure*} 
 \centering
 \subfigure[ F 563-1, Ref.~2]{\includegraphics[width=0.33\textwidth]{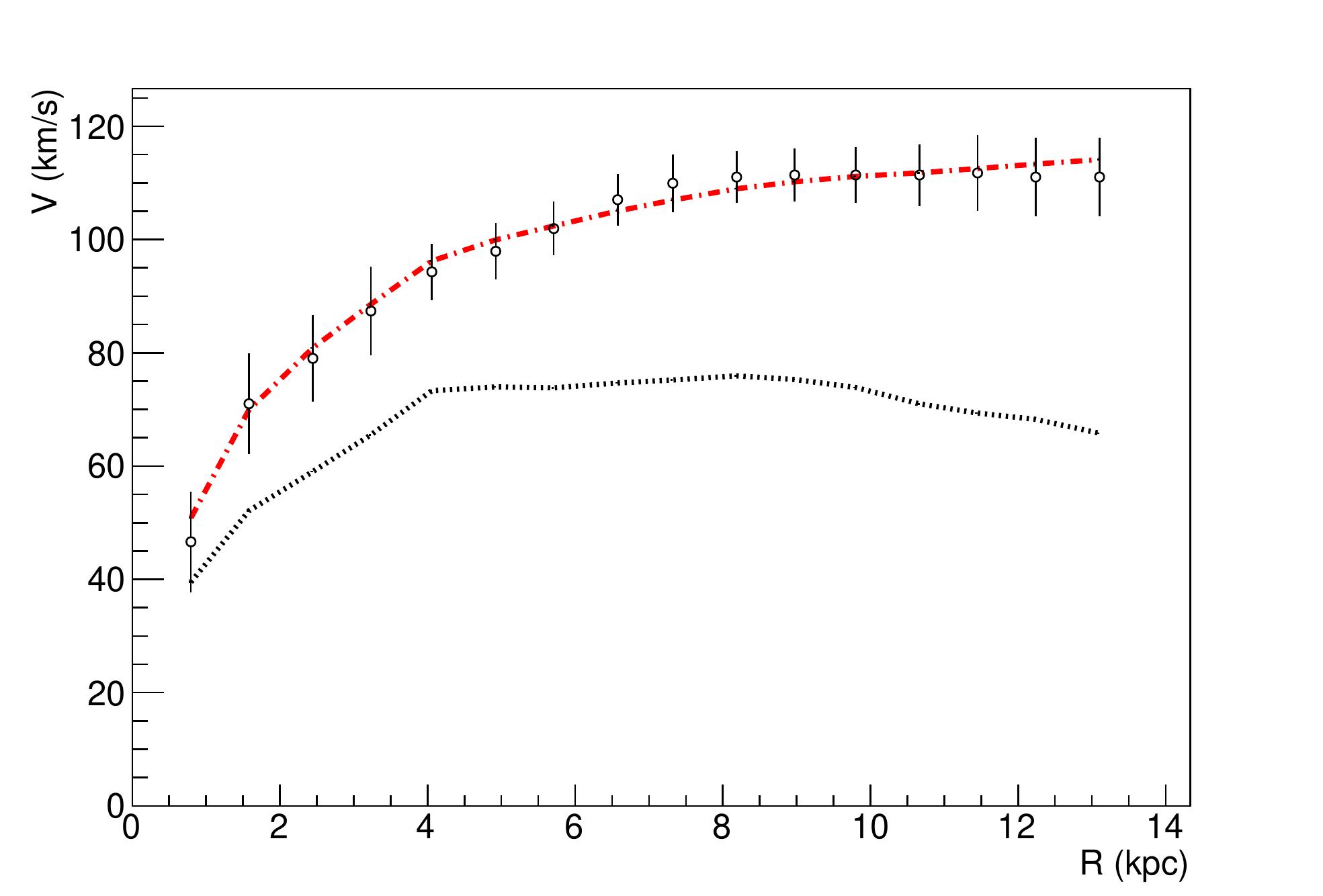}} 
  \subfigure[NGC 925, Ref.~3 (from NFW mass model)]{\includegraphics[width=0.33\textwidth]{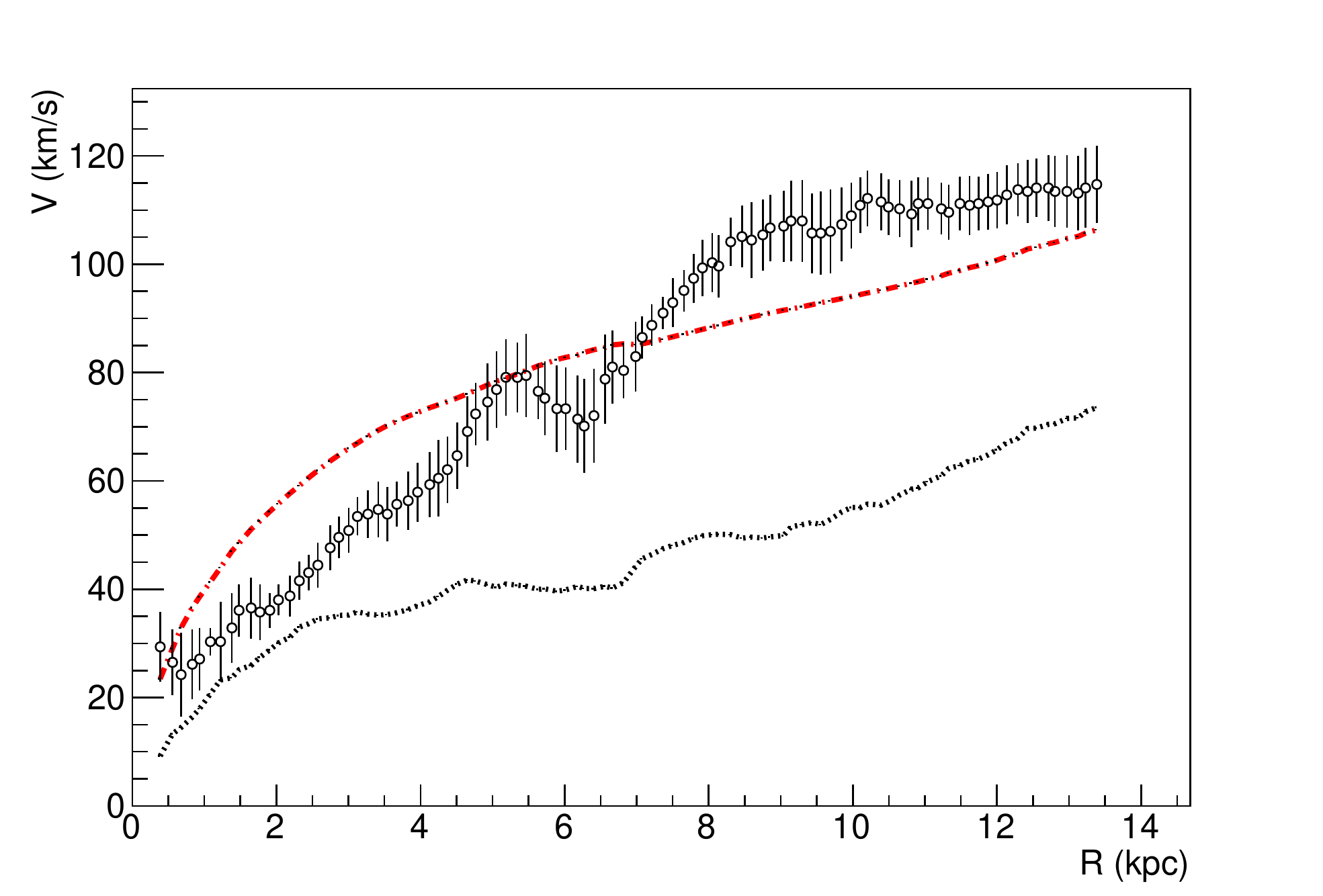}} 
   \subfigure[ NGC 7793, Ref.~8]{\includegraphics[width=0.33\textwidth]{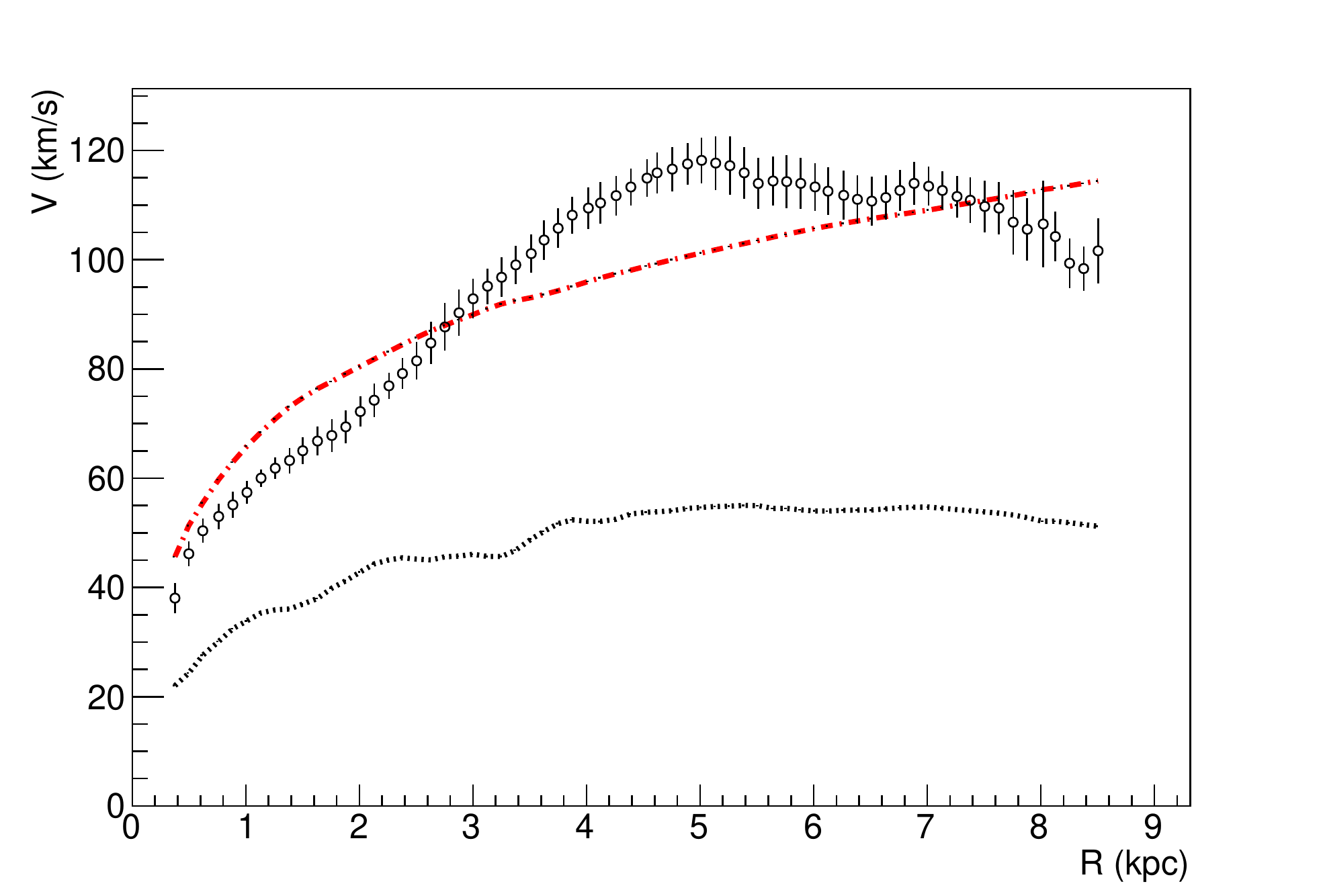}} 
    \subfigure[ F 563-1, Ref.~13]{\includegraphics[width=0.33\textwidth]{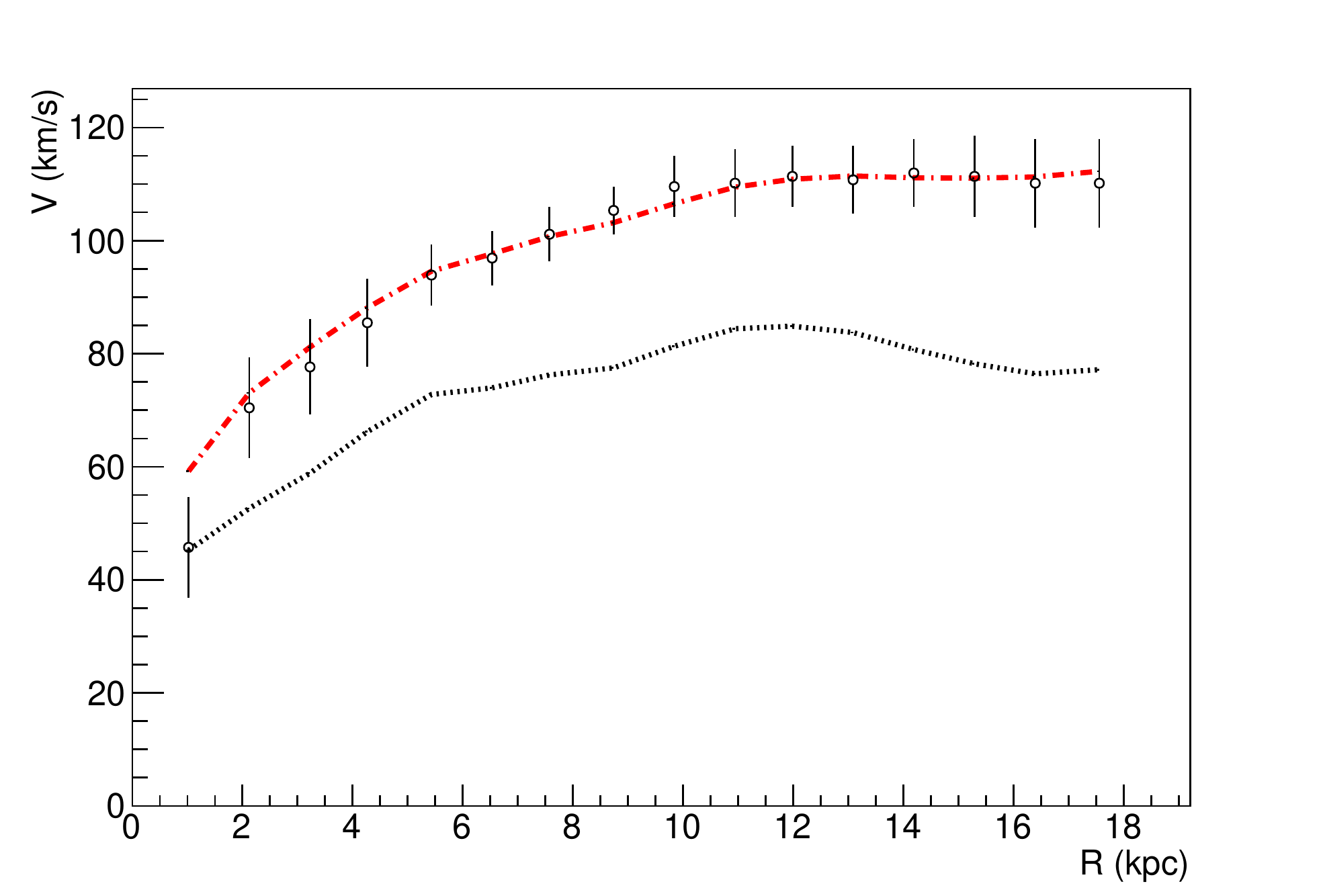}} 
     \subfigure[ NGC 925, Ref.~3  (from ISO mass model)]{\includegraphics[width=0.33\textwidth]{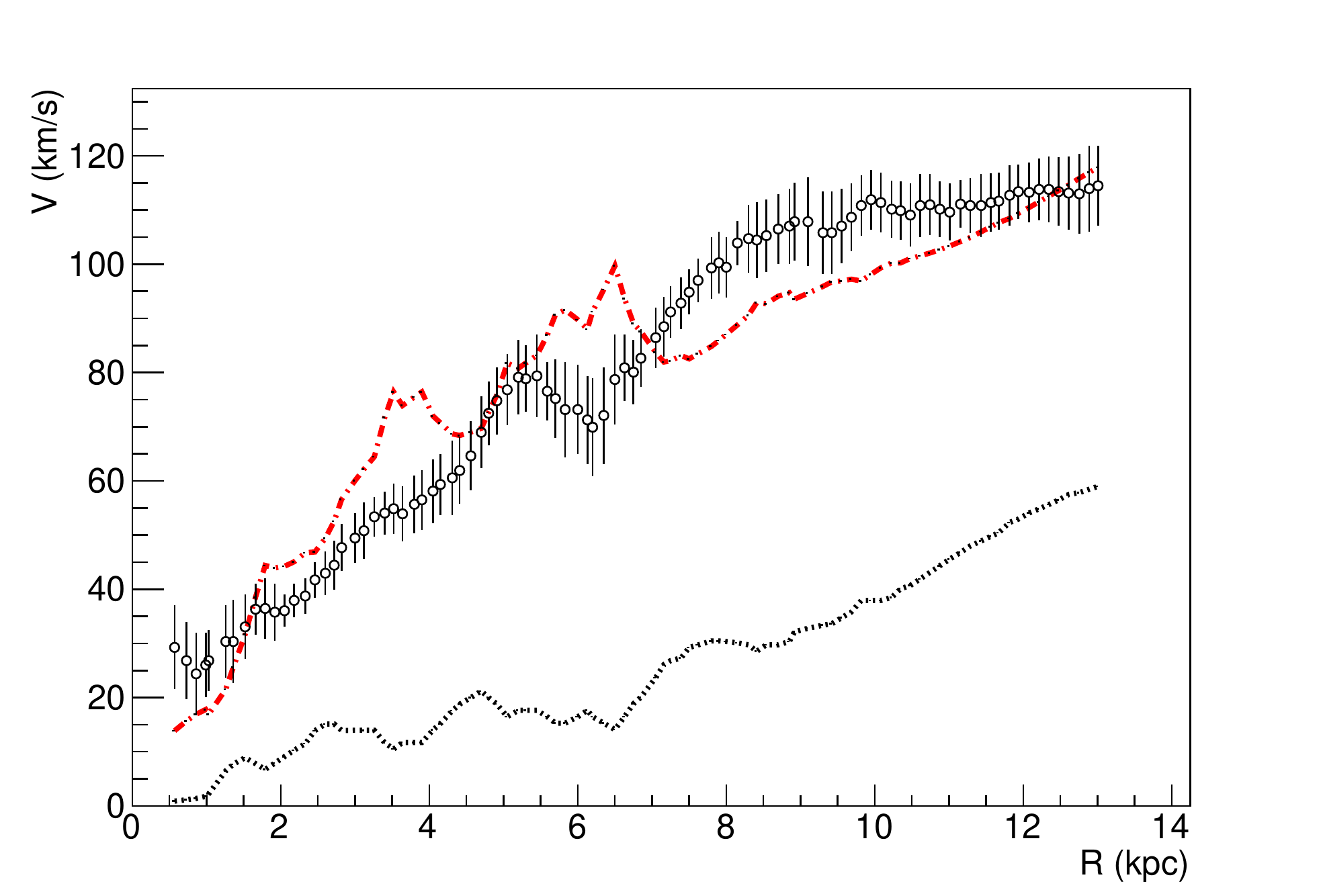}} 
      \subfigure[ NGC 7793, Ref.~14]{\includegraphics[width=0.33\textwidth]{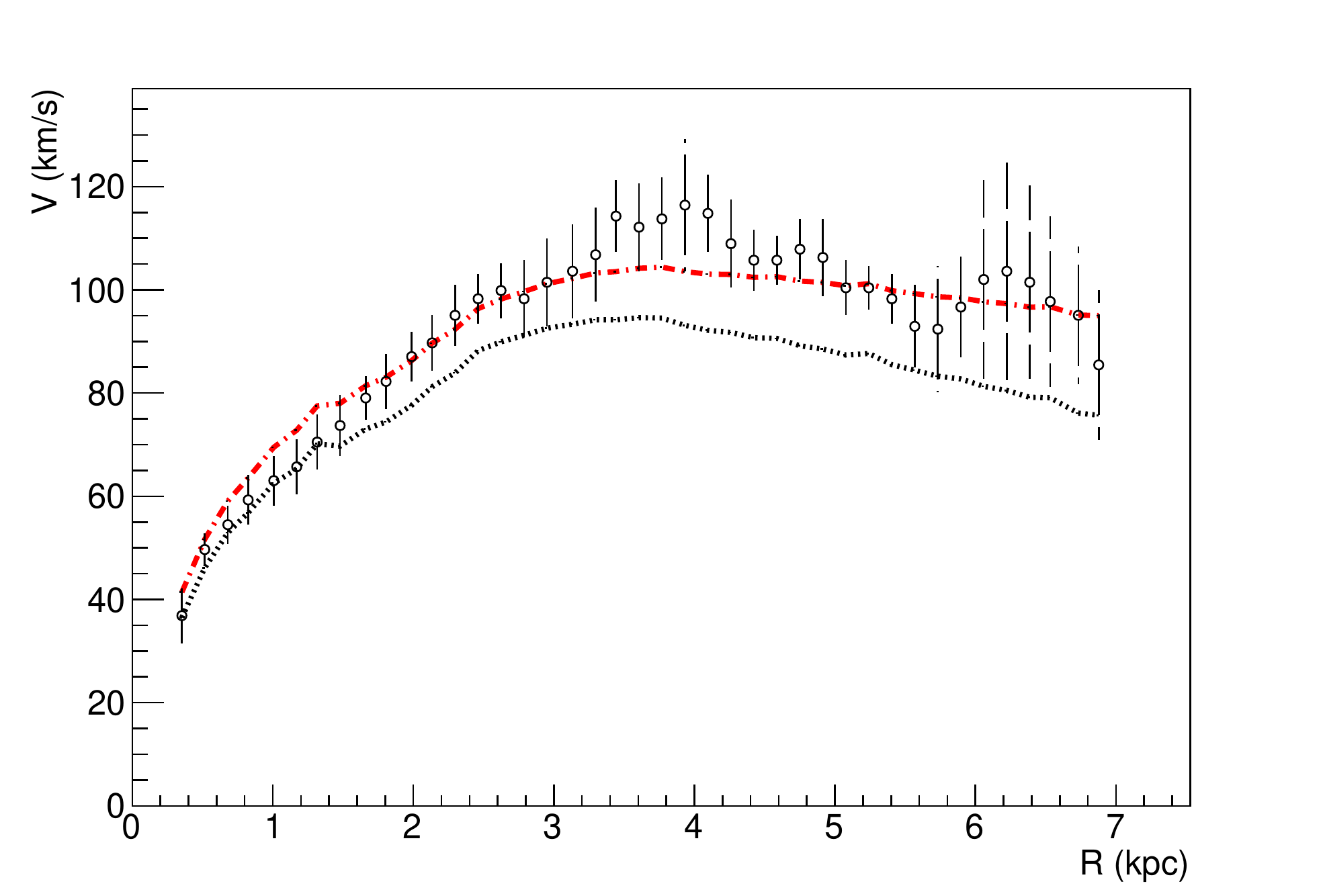}} 
\caption{LCM rotation curve fits.   In all panels:lines are as in Fig.~\ref{fig:resultssmall}.   References  are as in Table~\ref{sumRESULTS}.}  
             \label{galaxiesSmallest}   
\end{figure*}

The low surface brightness galaxy F 563-1   reported in    Ref.~2  and Ref.~13  is at  a distance of 45 Mpc.  Both references use an exponential disk. Visually,    in Fig.~\ref{galaxiesSmallest}, the LCM  fits using the luminous mass assumptions from  Ref.~2 are slightly  better than those from Ref.~13, but the resulting  M/L from Ref.~13 assumptions are   more physically acceptable. Ref.~13 uses a similar disk profile to that reported in Ref.~2, but on average resulting in associated orbital velocities which are about $30 \rmn{km\,  s^{-1}}$  higher(see Fig.~\ref{fig:massmodels18}).   Both sets of fits require maximal  LCM gas scalings.  
  
NGC 925,  late-type barred spiral at a reported  distance of $9.2$ Mpc, is a very   interesting example, as it indicates something physical about the LCM mapping.   From   Ref.~3 we sample    two very different luminous profiles,  consistent with  the ISO and NFW, bot fits for M/L fixed by PSM.    The ISO fit returns a significantly lower M/L than reported for    the NFW fits, based on observations of the rotation curve.  The LCM fit lowers the ISO M/L substantially, but does not set it to zero as is done in    Ref.~3 for their NFW fit  with M/L free.  The fascinating feature from  the ISO starting assumptions   is that the luminous mass features (in ISO)  are constructed to mimic the bumps in the rotation curve, but in the LCM fits  shows it is actually the mirror image of these features that  is   reflected in the rotation curve. We interpret this as a lens-type effect, where small luminous mass features in   galaxies that map mostly onto the bulge of the Milky Way are magnified and inverted.  This key feature could lead to  refinement of   the  luminous mass modeling of   low luminosity  and dwarf galaxies, based exclusively on the rotation curve data.   LCM fits using 
both the ISO and the NFW luminous profiles   failed  when paired to the \citet{Sofue} Milky Way.      

 NGC 7793 is a late-type spiral in the Sculptor group. The two rotation curves compared here come from $21$-cm observations reported in  Ref.~8 and  $\rmn{H}\alpha$ observations in Ref.~14. They differ outside of $4$ kpc, where Ref.~14 rotation velocities begin to decline faster than those reported in  Ref.~8.    The two luminous profiles use slightly different distance estimates, which may account for the differing rotation curves.   Ref.~8 uses  distance    of $3.91$ Mpc from the tip of the red giant branch indicator and Ref.~14 uses the distance of $3.38$ Mpc from ~\citet{Puche}.  Ref.~14 fails when compared to Milky Way ~\citet{Sofue}, but is otherwise very stable in the other three pairings.  Ref.~8 notes that  redder galaxies generally require higher stellar M/L, but they report    MOND best-fit
  $ (\rmn{M/L})_{disk}=0.28$.   The M/L  reported in Ref.~14   is  much larger, but in a different wavelength band. However, visually the LCM fits indicate that the  luminous assumptions in Ref.~14 are clearly preferable (Fig.~\ref{galaxiesSmallest}). 
  
 The  average $\tilde{a}$ values for  F 563-1, NGC 925 and NGC 7793 are consistent with  the conjecture in section~\ref{n891n7814}.
 
\section[]{Conclusions}
\label{sec:conclusion}
It has been  noted by ~\citet{Bot}  that   a credible, empirical alternative to dark matter   must: 
\begin{itemize}
\item   successfully predict rotation curves with reasonable estimates of the stellar mass-to-light ratios   and gas fractions,
\item  make sense within our physics framework,
\item  predict various other astrophysical observations,
\item   and  have one universal parameter. 
\end{itemize}  
 
 At this stage,  the LCM does the following:
  \begin{itemize}
 \item Successfully predict  rotation curves for  the sample of galaxies reported here  as evidenced by the average reduced $<\chi^2_r>$.   For the MOND subset of 21 data sets: the MOND $<\chi^2_r>=5.41$ vs. the LCM  $<\chi^2_r>=1.71$.   For the dark matter model subset of 19 data sets: the dark matter model 
  $<\chi^2_r>= 3.22$  whereas that for the LCM fits was $<\chi^2_r>= 2.07$. 	LCM fits resulted in  reasonable estimates of the luminous matter   reported in Table~\ref{sumRESULTS}.  
 \item Makes sense in our physical framework:    the  LCM is based on classical physics and luminous matter, and it fits galactic rotation curves based on the postulate that observed frequency shifts   include both  relative velocity and relative curvature components.   Both    effects are phrased in terms of the equivalent Lorentz Doppler-shifts, by the kinematics of the Lorentz group.  
\end{itemize}
In future investigations, the LCM can be extended to:
\begin{itemize}
\item Predict other astrophysical phenomena  by analytical extension to other metric geometries and dynamics; 
\item Include a physical  interpretation of  the free parameter  $\tilde{a}$.    There is some evidence of a correlation between  $\tilde{a}$ and the steepness of the density gradient (section~\ref{n891n7814}), but further theoretical developments and tests on a larger sample of galaxies are needed. This would  constitute the single  
  universal  parameter discussed by  ~\citet{Bot}.
   \end{itemize}
   
Our fits indicate that it may be possible to constrain population synthesis modeling directly from rotation curve data in the cases of   M 33 and NGC 925.  For M 33   the LCM  makes a direct constraint to the   two reported disk scale lengths, which differ by about   $8\%$.   In the case of NGC 925, the resulting LCM fits   demonstrate     the   galaxy's luminous profile appears to be    magnified and flipped in small spatial features  in the rotation curve, much as would be done by an optical lens.   In both cases, the defining feature that allows the LCM to    constrain population synthesis modeling is the compatibility of   rotation curves, uncertainties, distances, M/L bands, scale lengths, etc.  
  
  In many of the LCM fits   it is clear that the gas scalings are maximal or minimal, which is not to be taken as indication that the gas observations are at fault, but rather that the gas profiles are compensating for stellar models that are not appropriate.  In the view of the authors, once the stellar profiles and rotation curves have been constrained to best approximation, gas scalings will be indicative of real features.    
  This is already the case for  the M 33 fit using Ref.~6.
   
In the future, as the LCM sample size increases, trends in the fits will   indicate which  Milky Way luminous mass  model    most closely represents the true mass distribution.  At  this stage of the analysis there is not a  statistical preference in the fits for one model over another.  One particularly interesting  question for future work is    how to construct    the LCM for the rotation curve of the Milky Way itself; in this case the receiver frame is imbedded in the emitter frame.  Weak lensing problems are another place to which investigations can  easily be extended, as weak lensing has already been phrased in the geometric terms of Schwarzschild coordinate light speeds and effective indices of refraction by  \citet{Narayan}.
  
 Given the role that   the luminous matter of galaxies plays in the our understanding of  dark energy and   the Hubble flow,  indications from free parameters are critical to the big picture.  Initial investigations into relations between the free parameter and the  scale lengths are beyond the scope of the present paper.   Once the   physical significance of the     parameter $\tilde{a}$ has been   identified, it will be possible to directly report the luminous matter distribution  from the observed   shifted spectra.  
It remains possible that the LCM free parameter indicates scaling relations between the dark matter halo and the luminous matter distribution;  if that is so, the LCM offers an alternate method for luminous matter modeling with fewer parameter constraints on such a relationship. 

   \section[]{Acknowledgments}
The authors would like to thank R.\,A.\,M.\, Walterbos, V.\,P.\,  Nair,   M.\, Inzunza II,  J.\, Conrad,  V.\, Papavassiliou, T.\, Boyer, P.\, Fisher,   E.\, Bertschinger and I.\, Cisneros.   \\
 S.\,  Cisneros is supported by  the MIT Martin Luther King Jr. Fellowship, while J.\,A. Formaggio and N.\,A. Oblath are  supported by the United States Department of Energy under Grant No. DE-FG02-06ER- 41420.

\bibliography{LCM}{}

\bibliographystyle{mn2e}

 \bsp
 
\label{lastpage}

\end{document}